\documentclass[]{aa}

\usepackage{graphicx}
\usepackage{txfonts}
\usepackage{url}
\usepackage{natbib}
\bibpunct{(}{)}{;}{a}{}{,} 
\newcommand{\emm}[1]{\ensuremath{#1}}   
\newcommand{\emr}[1]{\emm{\mathrm{#1}}} 
\newcommand{\unit}[1]{\emm{\, \emr{#1}}}
\newcommand{\kms}{\unit{km\,s^{-1}}}
\newcommand{\changed}[1]{#1}   
\newcommand{\changedTwo}[1]{#1}   

\begin{document} 

   \title{Modelling clumpy PDRs in 3D}

   \subtitle{Understanding the Orion Bar stratification}

   \author{S.~Andree-Labsch
          \inst{1}
          \and
          V.~Ossenkopf{\changedTwo-Okada}
          \inst{1}
          \and
          M.~R\"ollig
          \inst{1}          
          }
   \institute{1. Physikalisches Institut, Universit\"at zu K\"oln,
              Z\"ulpicher Stra\ss e 77, 50937 K\"oln\\
              \email{sandree@ph1.uni-koeln.de}
             }

   \date{Received <date>; accepted <date>}

   \abstract
   {Models of photon-dominated regions (PDRs) still fail to fully 
   reproduce some of the observed properties, in particular the 
   combination of the intensities of different PDR cooling lines together
   with the chemical stratification, as observed e.g. for the Orion Bar
   PDR. }
   {We aim to construct a numerical PDR model, KOSMA-$\tau$~3D,
   to simulate full spectral cubes of line emission from arbitrary PDRs in
   three dimensions (3D). The model is to reproduce the intensity of the
   main cooling lines from the Orion Bar PDR and the observed
   layered structure of the different transitions.}
   {We build up a 3D compound, made of voxels (``3D pixels'') that contain
   a discrete mass distribution of spherical ``clumpy'' structures, 
   approximating the fractal ISM. To analyse each individual clump the new 
   code is combined with the KOSMA-$\tau$ PDR model. Probabilistic 
   algorithms are used to calculate the local FUV flux for each voxel as 
   well as the voxel-averaged line emissivities and optical depths, based 
   on the properties of the individual clumps. Finally, the computation of 
   the radiative transfer through the compound provides full spectral 
   cubes. To test the new model we try to simulate the structure of the 
   Orion Bar PDR and compare the results to observations from 
   HIFI/\emph{Herschel} and from the Caltech Submillimetre Observatory 
   (CSO). In this context new \emph{Herschel} data from the HEXOS 
   guaranteed-time key program is presented.
   }
   {Our model is able to reproduce the line integrated intensities within a 
   factor 2.5 and the observed stratification pattern within 0.016~pc for 
   the [C{\sc ii}] 158~$\mu$m and different $^{12/13}$CO and HCO$^+$ 
   transitions, based on the representation of the Orion Bar PDR by a 
   clumpy edge-on cavity wall. In the cavity wall, a large fraction of the
   total mass needs to be contained in clumps. The mass of the interclump 
   medium is constrained by the FUV penetration. Furthermore, the 
   stratification profile cannot be reproduced by a model having the same 
   amount of clump and interclump mass in each voxel, but dense clumps need 
   to be removed from the PDR surface.} 
   {}
 
   \keywords{photon-dominated region (PDR) -- ISM: structure -- ISM: clouds -- 
   submillimeter: ISM -- infrared: ISM -- radiative transfer}

   \maketitle
%
\section{Introduction}
\label{sec:Introduction}

\defcitealias{Hogerheijde_1995}{HJ95}
\defcitealias{Werf_1996}{Van der Werf et al. (1996)}
\defcitealias{Wiel_2009}{Van der Wiel et al. (2009)}

Stars form from the ISM, in it's dense and cold regions, inside molecular 
clouds. Hence, a better understanding of the chemical and physical 
processes taking place in molecular clouds, their internal structure, and 
the interaction between molecular clouds and the interstellar 
radiation field is an important step to constrain our knowledge on star 
formation processes. 

The energy which heats the different components of the ISM can originate 
from different sources, for instance from cosmic rays, from the 
dissipation of (magnetised) turbulence, or from the interstellar radiation 
field (including radiation from nearby stars). In photon-dominated 
(or photo-dissociation) regions (PDRs) the dominating energy input is 
provided by the interstellar radiation field. More precisely, a PDR is a 
region in interstellar space where the photon energies fall below the 
ionisation energy of hydrogen, but where the interstellar far-UV (FUV) 
radiation field still dominates the heating processes and the chemistry of 
the ISM (photon energies: 6~eV $<$ h$\nu$ $<$ 13.6~eV). Here, the lower 
threshold of 6~eV is an estimate of the work function of a typical 
interstellar dust grain\footnote{The estimate of the work function varies 
in literature. 6~eV are stated in \citet{deJong_1980}, more recent works 
discuss examples with work functions of 5~eV and of 7~eV 
\citep{Hollenbach_1999}, \citet{Weingartner_2001b} adopt 4.4~eV for 
graphite grains and 8~eV for silicates.}. Cooling of the gas is 
dominated by fine structure line emission by atoms and ions, especially 
[O{\sc i}] $63\,\mu$m and $145\,\mu$m, [C{\sc ii}] $158\,\mu$m and 
[C{\sc i}] $609\,\mu$m and $370\,\mu$m, by H$_2$ rovibrational, and by 
molecular rotational lines (mainly CO) \citep{Tielens_1985, 
Hollenbach_1997}. Far-infrared (far-IR) continuum emission by dust grains 
and the emission features of polycyclic aromatic hydrocarbons (PAHs) are 
observed. At high densities gas and dust are tightly coupled via collisions 
and the IR emission of the dust grains can contribute to the cooling of the 
gas. As PDR emission dominates the IR and sub-millimetre spectra of star 
forming regions and galaxies \citep{Roellig_2007} they are the subject of 
many observations and extensive modelling. PDRs can be found in many 
different astrophysical scenarios, however, here we focus on the 
transition zone between H{\sc ii}- and molecular regions illuminated by the 
strong FUV radiation from young stars.

Many different PDR models have been developed aiming to relate the 
observed line and continuum emission to the physical parameters of the 
emitting region and to understand the physical processes taking place 
in PDRs \citep[e.g.][]{Tielens_1985, Sternberg_1989, Koester_1994}. The 
models focus on different key aspects and exploit different geometries. 
An overview, emphasizing advantages and disadvantages of the different 
PDR models, can be found in the comparison study by 
\citet{Roellig_2007}. {\changedTwo Since then most of the codes have 
been significantly improved 
\citep[see e.g.][]{Roellig_2013,LeBourlot_2012,Ferland_2013}. A major
new step was provided by the extension to fully three-dimensional
configurations, which allows for the modelling of PDRs with arbitrary 
geometries, by \citet{Bisbas_2012}.}

In the molecular clouds the FUV field is attenuated, mainly due to 
absorption by dust grains. The decreasing FUV field strength causes a 
layered structure of different 
chemical transitions, referred to as \emph{chemical stratification}. Chemical
stratification can be observed in many different PDRs and within different scenarios,
for instance in the H{\sc ii} region and molecular cloud M17 
\citep[][]{Stutzki_1988, Pellegrini_2007, Perez_2012}, the Horsehead Nebula \citep{Pety_2007},
planetary nebulae \citep[for example NGC~7027, see][]{Graham_1993} or within protoplanetary disks 
\citep[see for instance][]{Kamp_2010}. Furthermore, it is observed in the Orion Bar PDR as discussed in
Sect.~\ref{Sect:OriBar}.

In other PDRs we find a spatial coexistence of different PDR tracers
that can be explained by a clumpy or filamentary cloud structure 
\citep{Stutzki_1988, Stutzki_1990, Howe_1991}. Actually, most
observations of molecular clouds show filamentary, 
turbulent structures and substructures 
on all scales observed so far. Such clouds can be described by fractal scaling laws.
Fractal structures contain surfaces everywhere throughout the cloud, hence, 
a large fraction of the molecular material 
is located close to a surface. Combined with a low volume filling 
factor (VFF) of the dense condensations this implies that surfaces inside 
the clouds are exposed to the interstellar radiation field - i.e.~form PDRs 
\citep{Burton_1990, Ossenkopf_2007}.

Several attempts have been made to model the 3D and inhomogeneous structure of PDR gas. For instance, \citet{Stutzki_1998}
have proven that the fractal properties
can be mimicked by an \emph{ensemble} of clumps with 
an appropriate mass spectrum. Based on this approach
\citet{Cubick_2008} have shown that an ensemble of such clumps, 
immersed in a thin inter-clump medium, can be used to simulate the 
large scale fine structure emission from the Milky Way. 
More recent, \citet{Glover_2010} developed 3D simulations of the turbulent interstellar
gas with coupled thermal, chemical and dynamical evolution.
\citet{Levrier_2012} use the Meudon PDR code to compare the 
chemical abundances in a homogeneous cloud to the chemical abundances 
in a cloud with density fluctuations.

However, a distribution of spherical clumps of different sizes 
that enables modelling of arbitrary 3D geometries has not yet been 
described. For the Orion Bar PDR, one of the most prominent PDRs in the 
solar neighbourhood, a match between observations and simulation results of 
the high-$J$ CO line intensities, combined with the observed stratification 
profile is still pending. Plane-parallel PDR models fail in this context,
because a match of the high-$J$ CO line intensities always requires
high densities which imply a very sharp and dense 
PDR structure\footnote{For example in \citet{Roellig_2007} (their Fig.~11) 
the C$^+$-to-C-to-CO transition has been simulated using many different PDR 
codes (for a gas density of $10^{5.5}$~cm$^{-3}$ and an FUV field strength 
of 10$^5$ times the mean interstellar radiation field \citep{Draine_1978}).
For all models the transition takes place at optical depths 
$A_{\rm V}\lesssim 4$ and using 
$A_{\rm V}/N_{\rm H}=6.289\times 10^{-22}$~cm$^{-2}$ \citep{Roellig_2007} 
we find that the stratified layers do not cover more than 0.0065~pc.} 
that is not consistent with the observed stratification covering, in the 
case of the Orion Bar PDR, at least 0.03~pc 
\citep[see for example][or the data presented in this work, Sect.~\ref{Sect:Observations}]{Pellegrini_2009}.
More sophisticated models are necessary to reproduce the observed 
line intensities as well as the observed chemical stratification. 
%
To overcome this deficiency we have set up an extension of the KOSMA-$\tau$ 
PDR code, denoted KOSMA-$\tau$ 3D, which enables us to model clumpy PDRs in 
3D. The code supports a spatial variation of PDR parameters, like the mean 
density, the clump-size distribution, or the strength of the impinging FUV 
field. Furthermore, to exploit the copious information contained in 
observed line profiles, the new code analyses a region at arbitrary 
velocities and hence the simulations of full line profiles.

In Sect.~\ref{Sect:Code} we discuss the extension of the KOSMA-$\tau$ PDR 
model to a clumpy 3D PDR model. To test the new code we use selected 
observations of the Orion Bar PDR which are presented in 
Sect.~\ref{Sect:OriBar}. The 3D model of the Orion Bar PDR is discussed in 
Sect.~\ref{Sect:OriModel}. In Sect.~\ref{Sect:ParameterScans} we present 
the fitting process: first we discuss the parameters that are varied within 
our model set-up and define functions of merit that are used for the 
evaluation of different models. We do then present the simulation outcome 
for many different models and provide a discussion. The results are 
summarised in Sect.~\ref{Sect:summary}.

\section{3D PDR modelling}
\label{Sect:Code}
In this section we discuss the extension of the KOSMA-$\tau$ PDR model to a 
clumpy 3D PDR model. First the properties of the KOSMA-$\tau$ PDR model are
summarised and modelling of the inhomogeneous ISM based on fractal 
structures is discussed. Afterwards, the 3D model set-up is described including 
all steps which are necessary to simulate maps and spectra, comparable to 
astronomical observations.
%
%
\subsection{The KOSMA-$\tau$ PDR model}
\label{Sect:KOSMA}

The KOSMA-$\tau$ PDR model\footnote{\url{http://www.astro.uni-koeln.de/kosma-tau}} \citep{Roellig_2006} has been developed at the University of 
Cologne in collaboration with the Tel-Aviv University. Contrary to many 
other models (see \citealt{Roellig_2007} and references therein), which 
are based on plane-parallel geometries, the KOSMA-$\tau$ model utilises a 
spherical geometry, \emph{clumps}, to model the structure of a PDR. 

A single clump is parameterised by its total hydrogen mass
$M_{\rm{cl}}$, the surface hydrogen density 
$n_{\rm s}=n_{\rm H,s}+2\, n_{{\rm H}_2,{\rm s}}$ and the strength of 
the incident FUV field. The FUV flux is assumed to be isotropic
{\changedTwo (see discussion in Sect.~\ref{Sect:Discussion_FUV})} and 
is measured in units of the Draine field integrated over the FUV 
range \citep[$\chi_0=2.7\cdot 10^{-3}$~erg\, s$^{-1}$\, cm$^{-2}$,][]{Draine_1978}. In addition, the model accounts for cosmic ray 
primary ionisations at a constant rate. In this work a rate of 
$2\cdot10^{-16}$~s$^{-1}$ per H$_2$ molecule is used 
\citep{Hollenbach_2012}. In the model the radial density distribution 
$n(r)$ of the clumps is divided into a core and an outer region:
\begin{equation}
n(r)=n_{\rm s}\begin{cases}
(\frac{r}{R_{\rm{cl}}})^{-a}, & \text{ for } x\,R_{\rm{cl}}\le r\le R_{\rm{cl}}\\
x^{-a}, & \text{ for } r<x\,R_{\rm{cl}}
\end{cases}
\label{Eq:clumpDensityDistribution}
\end{equation}
where $R_{\rm cl}$ is the radius of the clump and $x\, R_{\rm{cl}}$ 
with $x\in [0,1]$ is the radius of the clump core. The exponent, $a$, and 
the size of the core are input parameters of the KOSMA-$\tau$ code, 
$a=0$ and $x\ne 0$ for example enforces a constant density sphere. 
In many studies \citep[][]{Stoerzer_1996, Cubick_2008} and also in this 
work, $a=1.5$ and $x=0.2$ are chosen with the aim to generate clumps that 
approximate Bonnor-Ebert spheres\footnote{Bonnor-Ebert spheres are 
isothermal spheres in hydrostatic equilibrium embedded in a pressurised 
medium with a finite density at the position of the clump centre.}.
Consequently, the averaged density of one clump is given by
\begin{equation}
\overline{n_{\rm{cl}}}=\frac{1}{\frac{4}{3}\pi R_{\rm cl }^3}\int 4\pi r^2\, n(r)\, {\rm d}r \approx 1.91\, n_{\rm s}\, .
\end{equation}

To analyse such a clump the frequency dependent mean (averaged over the 
full solid angle) FUV intensity is derived at different positions between 
clump centre and clump surface (for different radii 
$r\in [0,R_{\rm cl}]$). This calculation is based on the multi component 
dust radiative transfer (MCDRT) code 
\citep[see][]{Yorke_1980,Szczerba_1997,Roellig_2013}
and includes isotropic scattering. The same code accounts for the IR 
continuum radiative transfer inside the clump. The KOSMA-$\tau$ PDR code 
includes H$_2$ self-shielding based on the results from 
\citet{Draine_1996}, furthermore, CO photodissociation is computed based 
on \citet{Visser_2009}.

To derive the physical conditions and the chemical composition of the 
clump, the KOSMA-$\tau$ code iteratively solves the following steps: The 
chemical network, which can be assembled from a modular chemical network 
\citep{Roellig_2013}, is used to derive local abundances, based on the local 
conditions. Using line of sight integrated escape probabilities for the main 
cooling lines the local energy balance, i.e.~heating and cooling 
processes are evaluated in steady state. After sufficient iterations 
(when a pre-defined convergency criterion is met; here: when the calculated column densities vary less than 1\% between subsequent iterations), ray tracing through the clump is solved for lines of 
sight at different impact parameters $p$ from the centre point of the 
spherical clump. For details on the ``ONION'' radiative transfer model 
see \citet{Gierens_1992}. 
The emission from the spherical clumps is not sensitive to their internal
density structure, but fully parametrized by their surface density. A
parameter study by \citet{Mertens2013} showed that modifications of the
internal density profile hardly change the chemical abundance profiles
as long as the surface density is kept constant.

In the first part of the KOSMA-$\tau$ \emph{3D} PDR code the averaged 
attenuation of the FUV flux caused by clumps with different masses and 
densities is needed. The second part of the code uses clump-averaged line 
intensities and optical depths of atomic and molecular transitions. 
To derive the averaged FUV attenuation of a clump we calculate the 
hydrogen column density along a line of sight through the clump, 
depending on the impact parameter, i.e.
\begin{equation}
 N_{\rm H}(p)=2\int_{0}^{\sqrt{R_{\rm cl}^2-p^2}} n\left(\sqrt{p^2+x^2}\right) {\rm d}x\, ,
 \label{Eq:NHp}
\end{equation}
where $n(...)$ is the density profile as given by 
Eq.~\ref{Eq:clumpDensityDistribution}. $N_{\rm H}(p)$ can be used to 
calculate the attenuation in the FUV range, 
$\tau_{j,\, {\rm FUV}}(p)$\footnote{The index $j$ is related to 
the mass of the clump (see Sect.~\ref{Sect:discrete}).}, assuming that 
both quantities are proportional to each other (see 
Sect.~\ref{Sect:FUV_pixel}). Furthermore, the attenuation of line 
intensities is proportional to the factor $\exp(-\tau (p))$ (see for 
example Eq.~\ref{Eq:case2}), where $\tau(p)$ denotes the optical depth 
for a line of sight with impact parameter $p$. Therefore, the factor 
$\exp(-\tau (p))$ needs to be averaged over the projected surface of the 
clump, i.e. for each clump we numerically solve the integral
\begin{equation}
\overline{\tau_{j,\, {\rm FUV}}}= -\ln\left[ \frac{2}{R_{{\rm cl}}^2}\int_{0}^{R_{{\rm cl}}}{\rm e}^{-\tau_{j,\, {\rm FUV}}(p)}\, p \,{\rm d}p \right] \, .
\label{Eq:tauClAvFUV}
\end{equation}
The line intensities $I_{j,\,{\rm line}}(p)$ of different atomic and 
molecular transitions have been averaged correspondingly, i.e.
\begin{equation}
\overline{I_{j,\, {\rm line}}}= \frac{2}{R_{{\rm cl}}^2}\int_0^{R_{{\rm cl}}}I_{j,\, {\rm line}}(p)\, p\, {\rm d}p
\label{Eq:IClAv}
\end{equation}
and the optical depths of the different transitions, 
$\tau_{j,\, {\rm line}}(p)$, are processed analogously to Eq.~\ref{Eq:tauClAvFUV}.
The $\overline{\tau_{j,\, {\rm FUV}}}$, $\overline{I_{j,\, {\rm line}}}$ 
and $\overline{\tau_{j,\, {\rm line}}}$ have been derived on a parameter 
grid of surface densities, clump masses and impinging FUV fluxes. The 
KOSMA-$\tau$ 3D code introduced in this work imports such a model grid 
and, if necessary, interpolates between gridpoints to derive the 
intensities and optical depths needed in the simulations. Details on the 
grid used for the presented Orion Bar simulations are summarised in 
Table~\ref{Tab:Grid_parameter}.

\begin{table*}
\caption{Overview of the most important model parameter. The 
numbers in parentheses indicate powers of ten.}             
\label{Tab:Grid_parameter}      
\centering          
\begin{tabular}{l l l c}  
\hline\hline   
{\bf Parameter}                 &{\bf  Value(s)}                          & {\bf  Comments} & {\bf Reference} \\ \hline
\multicolumn{4}{c}{Gridpoints}\\ \hline
$\lbrace M_{\rm cl}\rbrace$     & $10^i$~M$_\odot$ with $i=-3,-2,..., 3$  & clump mass & \\
$\lbrace n_{\rm s}\rbrace$      & $10^i$~cm$^{-3}$ with $i=3, 4,..., 7\tablefootmark{a}$   & clump surface density & \\
$\lbrace I_{\rm UV} \rbrace$    & $10^i\chi_0$ with $i=-1,0,...,6$        & FUV scaling factor & \\ \hline
\multicolumn{4}{c}{Abundances relative to the total hydrogen abundance}                        \\ \hline
He/H                            & 0.0851                                  & & (1)\\
O/H                             & $4.47(-4)$                              & & (2)                                     \\
C/H                             & $2.34(-4)$                              & & (2)  \\
$^{13}$C/H                      & $3.52(-6)$                              & based on a $^{12}$C/$^{13}$C ratio of about 67 in Orion & (3) \\
S/H                             & $7.41(-6)$                              & & (2) \\ \hline
\multicolumn{4}{c}{Others}                                                                                 \\ \hline
$\alpha$                        & 1.8                                     & clump-mass power law index  & (4)\\
$\gamma$                        & 2.3                                     & mass-size power law index   & (4)\\
$Z$                             & 1                                       & solar metallicity\\
$\zeta_{CR}$                    & $2(-16)$~s$^{-1}$                       & cosmic ray primary ionisation rate per H$_2$ & (5)\\
$R_\mathrm{V}$                  & 5.5\tablefootmark{b}                    & ratio between visual extinction and ``reddening'' & (6) \\
                                &                                         &  for dense clouds &\\
$\sigma_\mathrm{g}$             & $8.41(-22)$~cm$^2$                      & FUV dust cross section per H & (7)\\
$b$                             & 1~km~s$^{-1}$                           & Doppler broadening parameter\tablefootmark{c}\\
$A_\mathrm{V}/N_{\rm H}    $    & $5.3(-22)$~cm$^2$                       & normalisation for extinction curve & (8) \\
\hline           
\end{tabular}
\tablebib{\changed
(1) \citet{asplund_2005}; (2) \citet{Simon-Diaz_2011}; (3) \citet{Langer_1990}; (4) \citet{Heithausen_1998}; (5) \citet{Hollenbach_2012}; 
(6) \citet{Draine_1996}; (7) \citet{Roellig_2013}; (8) \citet{Weingartner_2001}.
}
\tablefoot{
\tablefoottext{a}{{\changedTwo For densities higher than 
$10^7$~cm$^{-3}$ the steep chemical gradient and short reaction time 
scales can cause numerical problems.}}
\tablefoottext{b}{Correspondingly, we use an averaged, normalised extinction 
$k_{\rm FUV}=\langle A(\lambda)/A(V)\rangle_\lambda=1.722$ with 
$\lambda=912...2066$~\AA\ \citep{Roellig_2013}.}
\tablefoottext{c}{${\it b}={\rm FWHM}/(4\ {\rm ln} 2)^{1/2}$, i.e.~$b=1$~\kms{} corresponds to FWHM=1.67~\kms{} \citep{Draine_1996}.}
}
\end{table*}

\subsection{Modelling the fractal ISM}
\label{Sect:FractalISM}
The fractal structure of molecular clouds can be mimicked by a superposition of spherical clumps following a
well-defined \emph{clump-mass spectrum}, building up a  
\emph{clumpy ensemble} \citep{Stutzki_1998, Cubick_2008}. The clump-mass spectrum can be described by 
a power-law
\begin{equation}
\frac{{\rm d}N_{\rm{cl}}}{{\rm d}M_{\rm{cl}}}=AM_{\rm{cl}}^{-\alpha}
\label{Eq:CMS}
\end{equation}
giving the number of clumps ${\rm d}N_{\rm cl}$ in the mass bin ${\rm d}M_{\rm cl}$. In addition
the masses of the clumps are related to their radii $R_{\rm{cl}}$ by the \emph{mass-size relation}
\begin{equation}
M_{\rm{cl}} = C R_{\rm{cl}}^{\gamma}\, .
\label{Eq:MassSize}
\end{equation}
{\changed The power-law exponents $\alpha$ and $\gamma$ have been subject to many studies.
\citet{Kramer_1998} present clump mass spectra, 
derived using the square-fitting procedure {\sc gaussclump} \citep{Stutzki_1990},
of seven different molecular clouds,
covering a wide range of physical properties and cloud sizes. 
They test and discuss the reliability of the mass spectra by studying the dependence on the control parameter 
of the decomposition algorithm. For all clouds from their sample 
they find that $\alpha$ lies between 1.6 and 1.8 implying that small clumps are more numerous.
No turnover of the power-law index is observed especially not for small, gravitationally unbound objects.

The power-law exponent $\gamma$ has for instance been discussed by \citet{Elmegreen_1996}. They analyse 
different cloud surveys from literature (based on different methods of clump identification)
and find an exponent $\gamma=2.4-3.7$
for single cloud surveys and an ``all-cloud slope'' in the range $2.2-2.5$. Hence, smaller clumps are 
expected to be denser. Using a second method they derive a fractal dimension $D=2.3\pm 0.3$ for the 
same surveys which theoretically is expected to be equal to the exponent $\gamma$.}

\citet{Heithausen_1998} {\changed combine and analyse large and small scale data of the Polaris Flare 
to derive the power law slopes over a mass range of at least 5 orders of magnitude, from several 
10~$M_\sun$, down to masses less than that of Jupiter (about 10$^{-3}$~$M_\sun$). Using the CO~$1-0$ 
and $2-1$ lines they find $\alpha =1.84$ and $\gamma=2.31$, values which are comparable to the ranges 
stated above and which we adapt for this work.}

We note that in some more recent works 
\citep[see review by][and references therein]{Offner_2014}
a turnover in the core-mass function has been reported for low-mass 
clumps. Such a deviation from the power-law is not included in the 
KOSMA-$\tau$ 3D PDR code. The influence of the very small clumps on
the {\changedTwo simulation} outcome is investigated in 
Sect.~\ref{Sect:MassPoints}.

\subsubsection{Continuous description}
\label{Sect:continuous}

Assuming that the masses of the clumps in an ensemble lie between a lower and an upper cut-off mass, 
$m_{\rm{l}}$ and $m_{\rm{u}}$, one can derive the number of clumps $N_{\rm{ens}}$ in the ensemble \citep[see][]{Cubick_2008}: 
\begin{equation}
N_{\rm{ens}}=\frac{A}{\alpha - 1}\left( m_{\rm{l}}^{1-\alpha}-m_{\rm{u}}^{1-\alpha} \right)\quad\mbox{for }\alpha\ne 1
\label{Eq:ClumpsPerInterval}
\end{equation}
and the total ensemble mass 
\begin{equation}
M_{\rm{ens}}=\frac{A}{2 - \alpha}\left(m_{\rm{u}}^{2-\alpha}-m_{\rm{l}}^{2-\alpha} \right)\quad\mbox{for }\alpha\ne 2
\label{Eq:MassPerInterval}
\end{equation}
relating the constant $A$ to the ensemble mass.
For the observed values of $\alpha$ below two {\changed an ensemble contains more low-mass than
high-mass clumps, still, the high-mass clumps provide a larger fraction of the ensemble mass}. 
The constant $C$ in Eq.~\ref{Eq:MassSize} depends on the averaged ensemble density 
$\rho_{\rm ens}$ and the cut-off masses:
\begin{equation}
C=\Big(\frac{4\pi}{3}\frac{2-\alpha}{1+3/\gamma-\alpha}\frac{m_{\rm{u}}^{1+3/\gamma-\alpha}-m_{\rm{l}}^{1+3/\gamma-\alpha}}{m_{\rm{u}}^{2-\alpha}-m_{\rm{l}}^{2-\alpha}}\rho_{\rm ens}\Big)^{\gamma/3} .
\label{Eq:C}
\end{equation}

\subsubsection{Discrete description}
\label{Sect:discrete}

We use a discrete description for a simplified numerical treatment of the 
\emph{clumpy ensemble} \citep[see][]{Cubick_2005}. Here, the mass 
spectrum of the clumps is not continuous, but represented by clumps at 
discrete \emph{mass points} $\lbrace M_{\it j}\rbrace_{j=1...n_M}$. We 
use a logarithmic parameter scale, 
i.e.~$\frac{M_{{\it j} + 1}}{M_{\it j}}=B$ with $B=10$. Indices are 
ordered with increasing masses. For the Orion Bar simulations we used
$M_{n_M} = 1\,M_{\sun}$ (\citealt{Lis_2003}, see 
Sect.~\ref{Sect:OriModel}) whereas the simulations of the whole Milky Way 
by \citet{Cubick_2005} rather correspond to $M_{n_M} = 100\,M_{\sun}$.
We assume that the number of clumps $N_j$ with mass $M_j$ is given by the 
power law
\begin{equation}
N_j = A_{\rm d} \cdot M_j^{1-\alpha}
\label{Eq:Nj}
\end{equation}
with a constant $A_{\rm d}$ (${\rm d}$=discrete) similar to Eq.~\ref{Eq:ClumpsPerInterval}. This
yields for the total mass $M_J$ of clumps with mass $M_j$
\begin{equation}
M_J=M_jN_j = A_{\rm d} \cdot M_j^{2-\alpha}\, .
\end{equation} 
For each ensemble the total mass of the ensemble $M_{\rm{ens}}$ and the averaged ensemble density $\rho_{\rm ens}$ 
are input parameters which can be fixed if the physical parameters of the PDR are known or they can be used as fitting 
parameters otherwise.
The total ensemble mass is given by $M_{\rm{ens}} = \sum_j N_{\it j} M_{\it j}$. 
Inserting $N_j$ from Eq.~\ref{Eq:Nj} we find
\begin{equation}
A_{\rm{d}}=\frac{M_{\rm{ens}}}{\sum_j M_{\it j}^{2-\alpha}}\, .
\label{Eq:Ad}
\end{equation}
The density of the individual clumps in the ensemble deviates from the ensemble averaged density $\rho_{\rm ens}$ 
according to the mass-size relation (Eq.~\ref{Eq:MassSize}), depending on their specific masses.
For given {\changed mass points} $\lbrace M_j\rbrace_{j=1...n_M}$ the volumes $\lbrace V_j\rbrace_{j=1...n_M}$ of individual clumps 
can be calculated using Eq.~\ref{Eq:MassSize} which yields
\begin{equation}
V_j=\frac{4}{3}\pi R_j^3=\frac{4}{3}\pi\left(\frac{M_j}{C}\right)^{3/\gamma}
\label{Eq:VolumeMass}
\end{equation}
and consequently the averaged density of a clump is found to be
\begin{equation}
\rho_j=\frac{M_j}{V_j}=\frac{3}{4\pi}C^{3/\gamma} M_j^{1 - 3/\gamma} .
\label{Eq:clumpDensity}
\end{equation}
The ensemble averaged density $\rho_{\rm ens}$ is equal to the total ensemble mass, divided by the total ensemble volume
\begin{equation}
\rho_{\rm ens}=\frac{\sum_j N_j M_j}{\sum_j N_j V_j}
\label{Eq:rhoEns}
\end{equation}
and inserting Eq.~\ref{Eq:Nj} and Eq.~\ref{Eq:VolumeMass} we derive 
\begin{equation}
\rho_{\rm{ens}}=\frac{3}{4\pi}C^{3/\gamma}\frac{\sum_j M_j^{2 - \alpha}}{\sum_j M_j^{1 + 3/\gamma - \alpha}}.
\label{Eq:rhoEnsDiscrete2}
\end{equation}
Inserting Eq.~\ref{Eq:clumpDensity} yields an expression for the density of clumps with mass $M_j$ as a function of the average 
ensemble density{\changed ,}
\begin{equation}
\rho_j=\frac{ M_j^{1 - 3/\gamma}  \sum_k M_k^{1 + 3/\gamma - \alpha}}{\sum_k M_k^{2 - \alpha}}\rho_{\rm{ens}}\, .
\label{Eq:rho_cl}
\end{equation}
In addition, the number of clumps $N_j$ with mass $M_j$, as a function of the total ensemble mass,
is found by combining Eqs.~\ref{Eq:Nj} and~\ref{Eq:Ad}:
\begin{equation}
N_{j}=\frac{M_j^{1-\alpha}}{\sum_k M_{\it k}^{2-\alpha}}M_{\rm{ens}}\, .
\label{Eq:Nbinned2}
\end{equation}
$N_{j}$ as given by Eq.~\ref{Eq:Nbinned2} and $\rho_j$ given via Eq.~\ref{Eq:rho_cl} uniquely define the parameters of 
the overall ensemble.

In the 3D model the clumps of an ensemble are randomly distributed in a 
voxel (``3D pixel'') with a known volume $\Delta s^3$ (see 
Sect.~\ref{Sect:EmissivityOpacity}). The VFF, i.e.~the 
fraction of the volume filled by clumps is given by
\begin{equation}
f_V=\frac{\sum_j N_j V_j}{\Delta s^3}\, .
\end{equation}
In principle for the discrete description artificial cut-off masses can 
be chosen in such a way that the parameters of the discrete description 
match those of the continuous description. However, one should note that 
it is not possible to conserve the total mass \emph{and} the number of 
clumps within a mass interval when switching from the continuous to the 
discrete description (while using $B=10$). Here, we have fixed the total 
ensemble mass which is assumed to be a known quantity. As the continuous 
description is not needed for the 3D PDR model, we will stick to the 
discrete description as an independent model.

\subsection{Three-dimensional set-up}
\label{Sect:3Dset_up} 

In irradiated molecular clouds we find position dependent conditions: 
the FUV field strength will decrease with increasing depth into the 
clouds due to extinction; furthermore, the average density and 
composition of the cloud may change. To model PDRs we set up a 3D 
model which can {\changedTwo replicate} arbitrary geometries using 
voxels. Each voxel contains at least one clumpy ensemble. 
Furthermore, for each mass point and for each voxel a velocity 
dispersion between the individual clumps is applied. The radiative 
transfer (Sect.~\ref{Sect:radTrans}) enables the simulation 
of line integrated maps as well as the modelling of full line 
profiles.

\subsubsection{Ensemble statistics: Area filling and clumps intersecting one line of sight}
\label{Sect:EnsembleStatistics}

In the 3D set-up each ensemble is contained in a 3D voxel having a projected 
surface area $\Delta s^2$ perpendicular to the line of sight between the observer and the voxel\footnote{The shape of the 
projected surface is arbitrary, but the volume should be spanned by the product of this surface area with the voxel depth.
For the presented algorithm for example a cuboid or a cylinder could be used and give the same results. 
A different viewing angle to the same geometry, therefore needs a {\changedTwo re-sampling} of the density structure
into new voxels where the $z$ axis is parallel to the line of sight.}.
The clumps, building up the ensemble, are randomly positioned in the voxel resulting in 
a {\changed number surface density $N_j/\Delta s^2$} for each mass point $j$.

Consider one arbitrary line of sight, perpendicular to the projected area $\Delta s^2$, through the ensemble. 
The probability distribution describing with how many randomly positioned clumps of mass $M_j$ the line of sight intersects 
{\changed is given} by the binomial distribution 
\begin{equation}
 B(k_j \ | \ p_j,\ N_j)=\binom{N_j}{k_j}p_j^{k_j}(1-p_j)^{N_j-k_j} 
 \label{Eq:Binomial}
\end{equation}
where $k_j$ is the number of clumps pierced by the line of sight, $p_j$ is the probability 
that the line of sight intersects with a specific clump of mass $M_j$ and $N_j$ is the total 
number of clumps with mass $M_j$. The intersection probability $p_j$ is given by 
$p_j=\pi R_{\rm cl, \it{j}}^2/\Delta s^2$.

In the following Sects.~\ref{Sect:FUV_pixel} and \ref{Sect:EmissivityOpacity} 
binomial distributions (Eq.~\ref{Eq:Binomial}) are used to calculate 
ensemble-averaged quantities, namely the ensemble-averaged FUV attenuation as 
well as ensemble-averaged line intensities and optical depths. As
binomial distributions are discrete probability distributions, the numbers of 
clumps, $N_j$, need to be integer values. This is not automatically provided by 
Eq.~\ref{Eq:Nbinned2}, however, scaling of the surface size $\Delta s^2$ (projected surface 
of the voxel = pixel) does not change the results for the ensemble-averaged quantities
as long as the number surface density, $N_j/\Delta s^2$, is kept constant for 
each mass point. Therefore, we rather consider a scaled "superpixel`` of area 
$(\Delta s')^2 = \Delta s^2\cdot c$ with a constant $c > 0$. Consequently, the numbers of 
clumps, $N_j$, need to be scaled accordingly: $N_{j}^{'}= N_j\cdot c$. The constant $c$ is 
chosen in a way that the following conditions are met:
\begin{itemize}
\item[a)] The projected clump areas of the largest clumps need to be smaller than 
{\changed a superpixel area: $\pi R_{{\rm cl},n_M}^2<(\Delta s')^2$, i.e.~$p^{'}_{n_M}<1$.}
\item[b)] $N_{n_M}^{'}$, i.e.~the number of clumps with mass $M_{n_M}$, is always 
an integer value.
\item[c)] $N_{n_M}^{'}$ is chosen to be the smallest value possible 
that does not contradict a) or b) to optimise for computing speed. {\changed This 
typically\footnote{\changed In this work scaling to $N_{n_M}^{'}=1$ is only performed for 
the ensemble representing the dense clumps. For the interclump medium, which contains 
only one type of clumps, $N_{n_M}^{'}$ is chosen to be larger (typically $N_{n_M}^{'}=100$). 
For details see Appendix~\ref{appendix:bruteForce}.} implies $N_{n_M}^{'}=1$}.
\end{itemize}
After clump numbers and pixel surface area have been scaled the numbers of clumps, $N_{j\ne n_M}^{'}$, 
{\changed are} rounded to integer values. As the numbers of low-mass clumps {\changed are} significantly 
higher then the number of high-mass clumps, $N_{j\ne n_M}^{'}>N_{n_M}^{'}$,
due to the \emph{clump-mass-spectrum} (see Eq.~\ref{Eq:CMS}) these rounding errors are negligible. 

In general, rounding errors can always be decreased by scaling to larger surface areas and 
consequently larger numbers of clumps. However, we found that the error made by rounding after 
step c) is already negligible and therefore optimised for computing speed. Furthermore, to 
increase {\changed the} computing speed, the binomial distribution can be approximated by a 
{\changed normal distribution} if the expected value $\mu_j=N_j\, p_j\gg 1$ {\changed and the 
number of clumps $N_j\rightarrow\infty$.} In the {\changed code} this simplification {\changed is 
implemented for the case $\mu_j>5$ and $N_j>1000$}. {\changed However, in the presented Orion 
Bar set-up we usually find $\mu_j< 5$ due to the low number surface densities.}

\subsubsection{Voxel dependent FUV field strength}
\label{Sect:FUV_pixel}

The line intensities and optical depths of the clumps contained in the 
voxels at different positions in the 3D model depend on the local FUV 
field strength. The FUV flux, coming from a known direction, is 
attenuated inside the PDR.

The FUV attenuation is proportional to the total hydrogen column density 
along the line of sight between the FUV source and the respective voxel. 
\citet{Roellig_2013} discussed the FUV extinction in terms of the FUV-to-V 
color, $k_{\rm FUV}=\langle A(\lambda)/A(V)\rangle_\lambda$, where the 
averaging is performed over an energy range from 6 to 13.6~eV. 
$k_{\rm FUV}$ is derived based on different grain-size distributions for 
the interstellar dust. For the model of the Orion Bar we adapt 
$k_{\rm FUV}\approx 1.7$ based on the grain-size distribution from
\citet{Weingartner_2001} for $R_{\rm V}=5.5$ {\changed ($R_{\rm V}=A_{\rm V}/(E(B-V))$ 
is the ratio between visual extinction and ''reddening``)}, the highest 
carbon abundance {\changed (for $R_{\rm V}=5.5$) in the small-grain 
populations} and a constant grain volume per hydrogen {\changed atom 
during their fitting process}\footnote{The grain-size distribution 
denoted with ``WD01-25'' in \citet{Roellig_2013} from line~25 in Table~1 
in \citet{Weingartner_2001} was used.}. $R_{\rm V}=5.5$ {\changed has 
been observed towards $\Theta^1$~Ori~C and is potentially 
representative for very dense clouds \citep{Draine_1996}. The high 
carbon abundance in small ``grains'' is justified by the observation 
of strong PAH features in the Orion Bar region \citep{Pilleri_2012}.
Combining $k_{\rm FUV}$ with} the normali{\changed s}ation for the 
extinction curve $A_{\rm V}/N_{\rm H}=5.3\cdot 10^{-22}$~cm$^{2}$ from 
\citet{Weingartner_2001}, {\changed hydrogen column densities that 
have been derived for different lines of sight through individual 
clumps (see Eq.~\ref{Eq:NHp}) can be transformed into the related FUV 
attenuations.}

{\changed For a given ensemble, let $X$ be the set of all possible 
combinations of clumps intersecting one line of sight, 
i.e.~$X=\lbrace k_{j}\rbrace_{j=1,...,\,n_M}$ with $0\le k_{j}\le N_j^{'}$. 
Consequently, one specific combination of clumps is described by 
an element $x\in X$, which is a set of $n_M$ numbers that provide
the number of clumps that intersect at each mass point $j$. For 
an element $x$ the FUV attenuation along the line of sight} is given by 
\begin{equation}
\tau_x=\sum^{n_M}_{j=1} k_{j}\, \overline{\tau_{j,\,{\rm FUV}}}\, .
\label{Eq:taux}
\end{equation}
The probability to find this 
combination of clumps intersecting the line of sight is the product of
binomial distributions, Eq.~\ref{Eq:Binomial}, 
\begin{equation}
p_x=\prod^{n_M}_{j=1} B(k_{j} \ | \ p_j,\ N_j)\, .
\label{Eq:px}
\end{equation}
In principle Eq.~\ref{Eq:px} has to be evaluated for each 
{\changed possible} combination, however, some combinations are highly 
improbable and can be neglected within the algorithm to increase 
{\changed its} computing speed. This is done by only accounting for 
numbers of clumps $k_j$ which lie in {\changed a $n_{\rm cut}\times\sigma_j^{'}$} 
interval\footnote{\changed For the binomial distribution the standard 
deviations, $\sigma_j^{'}$, are given by $(\sigma_j^{'})^2=N_j^{'}\ p_j^{'}(1-p_j^{'})$.}  around the expected 
value of the respective binomial distribution. 
Calculations presented in this paper have been performed with $n_{\rm cut}=3$.
In the presented simulations we found 
\begin{equation}
\sum_{\changed{x\in\lbrace \mu_j\pm 3\sigma_j^{'}\rbrace}} p_x>0.998\, ,
\end{equation} 
confirming the low error of our approximation. Finally, the ensemble-averaged 
FUV attenuation can be derived using
\begin{equation}
\langle {\rm e}^{-\tau_{\rm FUV}}\rangle_{\rm ens}=\sum_{x\in\lbrace \mu_j\pm 3\sigma_j^{'}\rbrace} {\it p}_x\cdot{\rm e}^{-\tau_x}
\label{eq:averagedtauFUV}
\end{equation} 
where $\langle\rangle_{\rm ens}$ denotes the ensemble-averaged value.
The {\changed F}UV attenuation in each voxel can then be described by the
effective optical depth  
$\langle \tau_{\rm FUV}\rangle_{\rm ens} = -\ln\left(\langle \rm{e}^{-\tau_{\rm FUV}}\rangle_{\rm ens}\right)$. 
{\changed 
If a voxels contains two ensembles to represent interclump medium and dense clumps, the algorithm 
presented above needs to be run for both ensembles separately. Finally, the ensemble-averaged 
contributions from clump and interclump medium are summed up, 
i.e.~$\langle \tau_{\rm FUV}\rangle_{\rm tot} = \langle \tau_{\rm FUV}\rangle_{\rm ens, cl}+\langle \tau_{\rm FUV}\rangle_{\rm ens, inter}$.}

\begin{figure}
  \resizebox{\hsize}{!}{\includegraphics{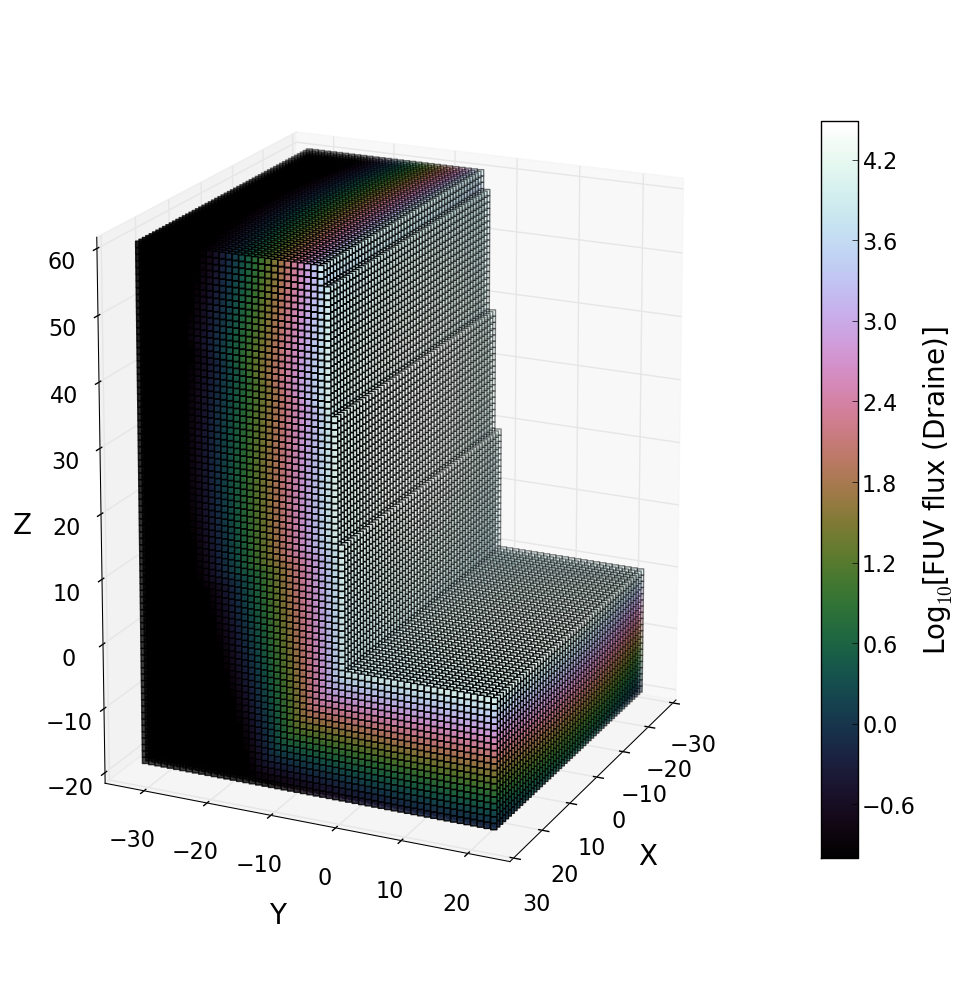}}
  \caption{A 3D compound {\changedTwo replicating} a possible 
  geometry (model 1m, see Table~\ref{Tab:models}) of the Orion Bar
  PDR. Each cube represents one voxel filled with at least one clumpy
  ensemble. Coordinates are given in voxel sizes, corresponding to 
  0.01~pc. The ensemble parameters can be varied between different voxels. 
  The colour scale \citep{Green_2011} shows the impinging FUV flux,
  calculated for each voxel, for a FUV source located at 
  $\lbrack 0, 22.3, 30\rbrack$. The direction 
  to earth corresponds to the positive $z$ direction.}
  \label{Fig:OriBarGeometry_FUV}
\end{figure}

\subsubsection{Ensemble-averaged emissivities and opacities}
\label{Sect:EmissivityOpacity}

Similar to the FUV attenuation, we need the ensemble-averaged line 
intensities 
and {\changed optical depths} of each voxel to compute the emission of 
the PDR. Clump averaged line intensities $\overline{I_{j,\, {\rm line}}}$ {\changed and 
optical depths $\overline{\tau_{j,\, {\rm line}}}$} are provided by the 
KOSMA-$\tau$ model ({\changed see Sect.~\ref{Sect:KOSMA})}. {\changed In 
principle, the ensemble-averaged emissivities and optical depths can be 
derived based on the same algorithm as the ensemble-averaged FUV 
attenuation in Sect.~\ref{Sect:FUV_pixel}. However, there is 
one major difference: while FUV absorption is a continuum process, 
line absorption and emission is velocity dependent and hence
we have to account for the intrinsic velocities of single clumps.
An algorithm accounting for a velocity distribution between \emph{identical} 
cloud fragments has already been discussed by \citet{Martin_1984}.
The algorithm presented here is similar but capable to treat ensembles 
made of different clumps. 

In the KOSMA-$\tau$ 3D code the velocity-space is discretised into velocity 
bins of width $\Delta {\rm v}$ around the centre velocities 
${\rm v}_i$ ($i=1,...,i_{\rm max}$) and the velocity of each clump 
from a bin ${\rm v}_i\pm \Delta{\rm v}/2$ is approximated by ${\rm v}_i$. 
Furthermore, we assume a Gaussian velocity distribution with standard deviation $\sigma_{{\rm ens},j}$ 
for the clumps at mass point $j$, accounting for random motions of the single clumps.
If $N_j$ is the total number of clumps at mass point $j$ then the number of clumps at 
mass point $j$ whose centre velocity lies inside the bin around centre velocity ${\rm v}_i$ 
is given by
\begin{equation}
\label{Eq:Nji}
\Delta N_{j,i}=\frac{N_j}{\sqrt{2\pi}\sigma_{j,\,{\rm ens}}}\exp\Bigg\lbrack -\frac{1}{2}\left(\frac{{\rm v}_i-{\rm v_{sys}}}{\sigma_{j,\,{\rm ens}}}\right)^2\Bigg\rbrack\Delta {\rm v}
\end{equation}  
where ${\rm v_{sys}}$ denotes the systematic velocity of the whole region 
(or of the voxel). The same discussion as in Sect.~\ref{Sect:EnsembleStatistics} 
applies here, i.e.~it is necessary to ensure that the numbers of clumps 
$\Delta N_{j,i}$ are integer values which can be {\changed achieved} by scaling of 
the pixel size. To convert the $\Delta N_{j,i}$ into integer values 
$\Delta N_{j,i}^{'}$ the algorithm presented in Sect.~\ref{Sect:EnsembleStatistics} 
is used. Note that in general the scaling factor (factor $c$ in 
Sect.~\ref{Sect:EnsembleStatistics}) will be different for each velocity bin.

If we consider a specific emission line and an observing velocity 
${\rm v_{obs}}$\footnote{In the current set-up the arrays containing the 
centre velocities ${\rm v}_i$ and the sampling velocities ${\rm v_{obs}}$ are 
chosen to be identical.} we are interested in the contributions to the line 
intensity and optical depth provided by the clumps at all centre velocities 
$\{{\rm v}_i\}$ around velocity ${\rm v_{obs}}$. To derive the contribution from 
the clumps at \emph{a single} centre velocity ${\rm v}_i$, we investigate 
(analogously to Sect.~\ref{Sect:FUV_pixel}) combinations\footnote{\changed 
In general possible combinations $x_i$ are different for each velocity bin. 
The case where the $\sigma_{j,\,{\rm ens}}$ are identical for all $j$ 
yields an exception. This case is treated separately by the KOSMA-$\tau$ 3D code 
to improve the computing speed.} of clumps $x_i \in X_i$ with 
$X_i=\lbrace k_{j,i}\rbrace_{j=1,...,n_M}$ and with $0\leq k_{j,i}\leq\Delta N_{j,i}^{'}$.
Furthermore, we can calculate the line intensities and optical depths of each combination 
$x_i$, which are given by\footnote{In Eqs.~\ref{Eq:Ixi} and \ref{Eq:tauxi} is would 
be more precise to average over the velocity bin, i.e.~replace the exponential function by
\begin{equation}
 \int_{{\rm v}_i-\Delta {\rm v}/2}^{{\rm v}_i+\Delta {\rm v}/2}\exp\Bigg\lbrack -\frac{1}{2}\left(\frac{{\rm v-v}_{\rm obs}}{\sigma_{j,\,{\rm line}}}\right)^2\Bigg\rbrack {\rm d v}\, . \nonumber
\end{equation}
However, we find that the effect of the integral is small. In a test run with 
11 velocity bins with $v_i=6.3, 7.3,... , 16.3$ we compared the calculated 
ensemble-averaged quantities with and without integrating over the exponential 
function for different ensembles (clumps and interclump or only interclump medium), 
different $v_{\rm obs}$, and for the different transitions analysed in this paper. 
The worst case relative deviation is about 3\%, for most other constellations 
the deviation is orders of magnitude smaller. As the integral significantly 
increases the computing time it is only optional in the code and not included in 
the presented simulations.}
\begin{align}
\label{Eq:Ixi}
 I_{x_i}({\rm v_{obs}})  = &\sum_{j=1}^{n_M} k_{j,i}\, \overline{I_{j,\, {\rm line}}}\exp\Bigg\lbrack -\frac{1}{2}\left(\frac{{\rm v}_i-{\rm v}_{\rm obs}}{\sigma_{j,\, {\rm line}}}\right)^2\Bigg\rbrack \\
\label{Eq:tauxi}
 \tau_{x_i}({\rm v_{obs}})= &\sum_{j=1}^{n_M} k_{j,i}\, \overline{\tau_{j,\, {\rm line}}}\exp\Bigg\lbrack -\frac{1}{2}\left(\frac{{\rm v}_i-{\rm v}_{\rm obs}}{\sigma_{j,\,{\rm line}}}\right)^2\Bigg\rbrack
\end{align}
where $\sigma_{j,\,{\rm line}}$ is the intrinsic line width (standard 
deviation) of a single clump at mass point $j$ {\changedTwo and
$\overline{I_{j,\, {\rm line}}}$ and $\overline{\tau_{j,\, {\rm line}}}$
are the clump-averaged (Eq. \ref{Eq:tauClAvFUV} and \ref{Eq:IClAv}) line-centre
(peak) intensities and optical depths computed by the KOSMA-$\tau$ PDR code.}
As discussed in Sect.~\ref{Sect:EnsembleStatistics} the probability to 
find a specific combination of clumps (a specific element $x_i \in X_i$) 
on a line of sight through the voxel is given by 
\begin{equation}
\label{Eq:pxi}
 p_{x_i}=\prod_{j=1}^{n_M} B( k_{j,i}\, | p_{j,i},\, N_{j,i})\, .
\end{equation}
As a next step the effective line intensity and optical depth \emph{of each velocity bin} 
is calculated by averaging over all combinations of clumps found in one bin, i.e.
\begin{eqnarray}
 \langle I\rangle_i({\rm v_{obs}}) &=& \sum_{x_i\in\lbrace \mu_{j,i}\pm 3\sigma_{j,i}^{'}\rbrace} p_{x_i}\,  I_{x_i}({\rm v_{obs}}) \\ 
 \label{Eq:IAvi}
 \langle \tau \rangle_i ({\rm v_{obs}}) &=&-\ln\lbrack \langle {\rm e}^{-\tau}\rangle_i ({\rm v_{obs}}) \rbrack \\ \nonumber
 &=&-\ln\Big\lbrack\sum_{x_i\in\lbrace \mu_{j,i}\pm 3\sigma_{j,i}^{'}\rbrace} p_{x_i}\, {\rm e}^{-\tau_{x_i}({\rm v_{obs}})}\Big\rbrack\, .
 \label{Eq:tauAvi}
 \end{eqnarray}
 Note that the expected values, $\mu_{j,i}$, and the corresponding standard derivations, 
$\sigma_{j,i}$, do now depend on mass point and velocity bin. Finally, the contributions 
from the different bins are summed up to give the complete ensemble-averaged intensity 
and optical depth
\begin{eqnarray}
 \label{Eq:Iens}
 \langle I\rangle_{\rm ens}({\rm v_{obs}}) &=& \sum_{i=1}^{i_{\rm max}}  \langle I\rangle_i({\rm v_{obs}})\\ 
 \label{Eq:tauens}
 \langle \tau \rangle_{\rm ens} ({\rm v_{obs}}) &=& \sum_{i=1}^{i_{\rm max}}  \langle \tau\rangle_i({\rm v_{obs}})\, .
\end{eqnarray}
{\changedTwo In Eqs.~\ref{Eq:Ixi} and \ref{Eq:Iens}, where we sum
up the line intensities from different clumps and velocity bins, we 
have assumed that the line is locally optically thin. By choosing
the voxel size $\Delta s$ sufficiently small, this condition 
can always be met.} The ``probabilistic approach'' presented 
in this section for the calculation of the ensemble-averaged 
quantities was verified by comparison to a second method 
(see Appendix~\ref{appendix:bruteForce}). If a voxel contains two 
ensembles the line intensities and optical depths of both ensembles 
are calculated separately and summed up, as described in 
Sect.~\ref{Sect:FUV_pixel}.}


\subsubsection{Radiative transfer}
\label{Sect:radTrans}

In the 3D PDR simulations the geometry of the PDR is 
{\changedTwo replicated} using voxels having the volume 
$(\Delta s)^3$. To derive maps and spectra that are comparable to 
observations the radiative transfer through the 3D model needs to be 
calculated. An example for a 3D set-up is shown in 
Fig.~\ref{Fig:OriBarGeometry_FUV} ({\changedTwo replicating} the 
Orion Bar PDR, which will be introduced in Sect.~\ref{Sect:OriBar}). 
Each small cube in Fig.~\ref{Fig:OriBarGeometry_FUV} represents one 
voxel and the colour scale shows the FUV flux at the different voxels 
within the compound, for an FUV source located at the position 
$\lbrack 0, 22.3, 30\rbrack$.

The attenuation of the FUV photons inside the PDR is given by the sum of 
the optical depths, $\langle \tau_{\rm FUV}\rangle_{\rm ens}$, of all 
voxels between the FUV source and the voxel of interest. The 
KOSMA-$\tau$ 3D code accounts for the absorption of photons and for 
isotropic scattering {\changedTwo within the individual clumps} 
(see Sect.~\ref{Sect:KOSMA}). The code uses voxels with the shape of a 
cuboids, which are oriented in a way that one side of the cuboid is 
{\changedTwo perpendicular} to the line of sight between observer and 
voxel ($z$-axis). The voxels need to be sufficiently small to trace 
all relevant changes of different quantities within the compound, but 
some compromise has to be made to reduce the overall computational 
effort. Therefore, the KOSMA-$\tau$ 3D code can make use of a set-up 
where the  {\changedTwo FUV attenuation} is calculated {\changedTwo 
for} different 
positions within the voxels, i.e.~on a 3D Cartesian grid at sub-voxel 
scale, {\changedTwo and is averaged over the voxel afterwards. For 
each sub-voxel the code derives the line of sight that connects the 
centre-point of the sub-voxel and the source of the FUV radiation and 
evaluates which voxels intersect with this line of sight (i.e.~shield 
FUV radiation). The FUV attenuation is weighted by the distance that is
effectively crossed by a FUV photon within each voxel.} The
simulations presented in this work have been performed using a 
$3\times 3\times 3$ voxel sub-grid. For the calculation of the 
ensemble-averaged line 
intensity and optical depth (see Sect.~\ref{Sect:EmissivityOpacity}), 
and consequently also for the radiative transfer, the sub-voxel grid is 
not used. This has two reasons, which are (a) while the lines of
sight between the observer and a voxel are oriented perpendicular to the 
voxel surface, this is in general not the case for the lines of sight 
between \emph{the FUV source} and the voxels. If we would calculate the 
FUV attenuation between FUV source and voxels by just summing up the 
$\langle \tau_{\rm FUV}\rangle_{\rm ens}$ between the position 
of the FUV source and the midpoints of the voxels we would introduce 
unnecessarily large errors in the case where a voxel is \emph{partly} 
shielded by another voxel in the foreground. (b) The sub-voxel treatment 
is too costly for the calculation of the ensemble-averaged line 
intensity and optical depth where calculations need to performed at 
different (here: $i_{\rm max}=21$) velocities.

{\changed We perform the radiative transfer for the same velocities ${\rm v_{obs}}$ 
that have already been used in Sect.~\ref{Sect:EmissivityOpacity}.}
The ensemble-averaged volume emissivity and absorption coefficient, 
$\langle \epsilon \rangle_{\rm ens}({\rm v_{obs}})$ and $\langle \kappa \rangle_{\rm ens}({\rm v_{obs}})$,
are calculated based on
the results from the last section, {\changed Eqs.~\ref{Eq:Iens} and \ref{Eq:tauens}, i.e.}
\begin{eqnarray}
\langle \epsilon \rangle_{\rm ens}({\rm v_{obs}})=&\frac{1}{\Delta s}\, \langle I \rangle_{\rm ens}({\rm v_{obs}})\\
\langle \kappa \rangle_{\rm ens}({\rm v_{obs}})=&\frac{1}{\Delta s}\, \langle \tau\rangle_{\rm ens}({\rm v_{obs}})\, ,
\label{Eq:kappaAv}
\end{eqnarray}
where $\Delta s$ denotes the depth of the voxel along the line of sight.
In the following we will omit the velocity dependence (${\rm v_{obs}}$) 
and the ``ensemble-averaged'' brackets $\langle ... \rangle_{\rm ens}$ 
in our formulae for readability reasons.

For radiation travelling {\changed a distance ${\rm d}s$ along a straight path} 
the change in intensity is given by the equation of radiative {\changed transfer}, 
which {\changed (omitting the dependence on frequency) reads}
\begin{equation}
{\rm d}I= -I \kappa\, {\rm d}s +  \epsilon\, {\rm d}s.
\label{Eq:radtrans1}
\end{equation}
Integration along a straight path length, between 0 and $\Delta s$, yields 
\begin{eqnarray}
I = {\rm e}^{-\int_{0}^{\Delta s}  \kappa\,  {\rm d} x}\left[ \int_0^{\Delta s} \epsilon\, {\rm e}^{\int_{0}^{s'}  \kappa\,  {\rm d} x}\, ds' + I_{\rm bg}\right]
\label{Eq:radtrans2}
\end{eqnarray}
{\changed where $I_{\rm bg}=I(0)$ is the background intensity of radiation travelling along the same path}.
For radiative {\changed transfer} from voxel $p-1$ to the neighbouring voxel $p$ \citep{Ossenkopf_2001} $\epsilon$ and 
$\kappa $ are linearly interpolated, {\changed i.e.~we define $ \epsilon = e_0+e_1 s'$ and 
$ \kappa = k_0+k_1 s'$} with $s'\in \left[0,\Delta s\right]$ and with 
\begin{eqnarray}
k_0 &=&\kappa_{p-1}\nonumber\\
k_1 &=&\big(\kappa_{p}- \kappa_{p-1}\big)/\Delta s\nonumber\\
e_0 &=&\epsilon_{p-1}\nonumber\\
e_1 &=&\big(\epsilon_{p}-\epsilon_{p-1}\big)/\Delta s\, ,
\label{Eq:approx}
\end{eqnarray}
which can be inserted into Eq.~\ref{Eq:radtrans2} yielding
\begin{equation}
I=\frac{1}{{\rm e}^{k_0\Delta s+\frac{1}{2}k_1(\Delta s)^2}}\left[ \int_0^{\Delta s}(e_0+e_1\cdot s'){\rm e}^{k_0 s'+\frac{1}{2}k_1 (s')^2}{\rm d}s' + I_{\rm{bg}}\right]\, .
\label{Eq:linearApproxRadTrans}
\end{equation}
Eq.~\ref{Eq:linearApproxRadTrans} is solved numerically and tabulated for the simulations (see Appendix~\ref{appendix:radTrans}).

\section{The Orion Bar PDR}
\label{Sect:OriBar}

The Orion Bar PDR is a prominent feature located in the Orion Nebula (M42, NGC~1976). 
In observations of typical cooling lines from the UV down to radio wavelength and the 
IR continuum the bar appears as a bright rim. With a distance of $414\pm 7$~pc 
\citep{Menten_2007} the Orion Bar PDR is one of the nearest and hence brightest PDRs 
to the terrestrial observer. Consequently, a large amount of observations of the Orion 
Bar PDR has been performed providing us with an excellent test case for PDR models.
 
\emph{Chemical stratification} has been observed for the Orion Bar PDR by different {\changed groups 
\citep{Tielens_1993, Werf_1996, Simon_1997, Marconi_1998, Walmsley_2000, Wiel_2009, Pellegrini_2009, Bernard-Salas_2012}. 
For example} \citet{Wiel_2009} discuss a layered structure with C$_2$H emission peaking close to the ionisation front (IF),
followed by H$_2$CO and SO, while other species like C$^{18}$O, HCN and $^{13}$CO peak deeper into the cloud.

Nowadays, i{\changed t} has become clear that a ``simple'' homogeneous 
{\changed (i.e.~non-clumpy) b}ar is an insufficient description of the Orion Bar PDR.
High angular resolution observations show that the {\changed b}ar breaks down into substructure.  
{\changed The commonly accepted picture is that the {\changed b}ar 
includes an extended gas component of $n_{\rm H}=10^{4-5}$~cm$^{-3}$ that causes the 
chemical stratification and is the dominating origin for low-$J$ molecular line emission. 
Embedded in this ``interclump medium'' a clumpy high-density ($n_{\rm H}=10^{6-7}$~cm$^{-3}$)
component is needed to provide the emission of the ``high-density tracers'', among others 
the lines of high-$J$ CO isotopologues, 
CO$^+$, and the observed H$_2$ or OH \citep[for a summary and additional references see][]{Goicoechea_2011}.
The low filling factor of the dense clumps ensures that the FUV field can penetrate deep into the cloud.}
We cannot list all observations that have dealt with the spatial structure of the Orion Bar.
Just to name a few, \citet{YoungOwl_2000} presented combined single-dish and 
interferometric data of HCO$^+$ and HCN $J=1-0$ which show a clumpy NE and SW {\changed b}ar, 
\citet{Lis_2003} showed interferometric data of the Orion Bar PDR in H$^{13}$CN and H$^{13}$CO$^+$ 
and identify at least 10 dense condensations in the H$^{13}$CN image, and individual
clumps have also been resolved by \citet{Werf_1996} who showed that a PDR surface can be
found on each clump inside the Orion Bar. {\changed More recent studies on the structure of the Orion Bar PDR
have been performed by
\citet{Goicoechea_2011, Cuadrado_2014}.}

\subsection{Geometry}
\label{Sect:geometry}

A common explanation for the existence of the {\changed b}ar is the ``Blister model'':
{\changed t}he Orion Nebula embeds a cluster of bright and young stars
which ionise their surrounding medium creating an H{\sc ii}-region inside 
the molecular cloud. At the side of the nebula facing earth this ``H{\sc ii}-bubble'' 
has broken out of the cloud, enabling observations of the cavity and of the Orion 
Bar PDR which forms one of the edges of the cavity, illuminated by the strong FUV 
radiation from the young star cluster (see for example \citealt{Wen_1995} and 
references therein).

The dominating ionising source and most massive star is $\Theta^1$~Ori~C which produces 
$\sim\ $80\% of the H-ionising photons. $\Theta^1$~Ori~D, 
the second most massive star of the ``Trapezium'' system, accounts 
for another $\sim\ $15\% \citep{Draine_2011}. 
The IF, as marked for example by the peak position of the [O{\sc ii}] or [Fe{\sc ii}] emission \citep{Walmsley_2000}, 
[S{\sc ii}] \citep{Pellegrini_2009},
or [N{\sc ii}] \citep{Bernard-Salas_2012}
is located at 111'' (corresponding to 0.223~pc) projected distance from  $\Theta^1$~Ori~C.

The flux at the IF has been estimated to correspond to an enhancement over the 
average interstellar radiation field{\changed , $\chi_0$, by} a factor $\approx4.4\cdot10^4$ 
\citep{Hogerheijde_1995, Jansen_1995} {\changed (the series of papers by 
\citet{Hogerheijde_1995} and \citet{Jansen_1995} is hereafter abbreviated 
\citetalias{Hogerheijde_1995})}. We have verified that this value lies in {\changed the} 
probable range (see Appendix~\ref{appendix:FUVatIF}).

\begin{figure}
  \resizebox{\hsize}{!}{\includegraphics{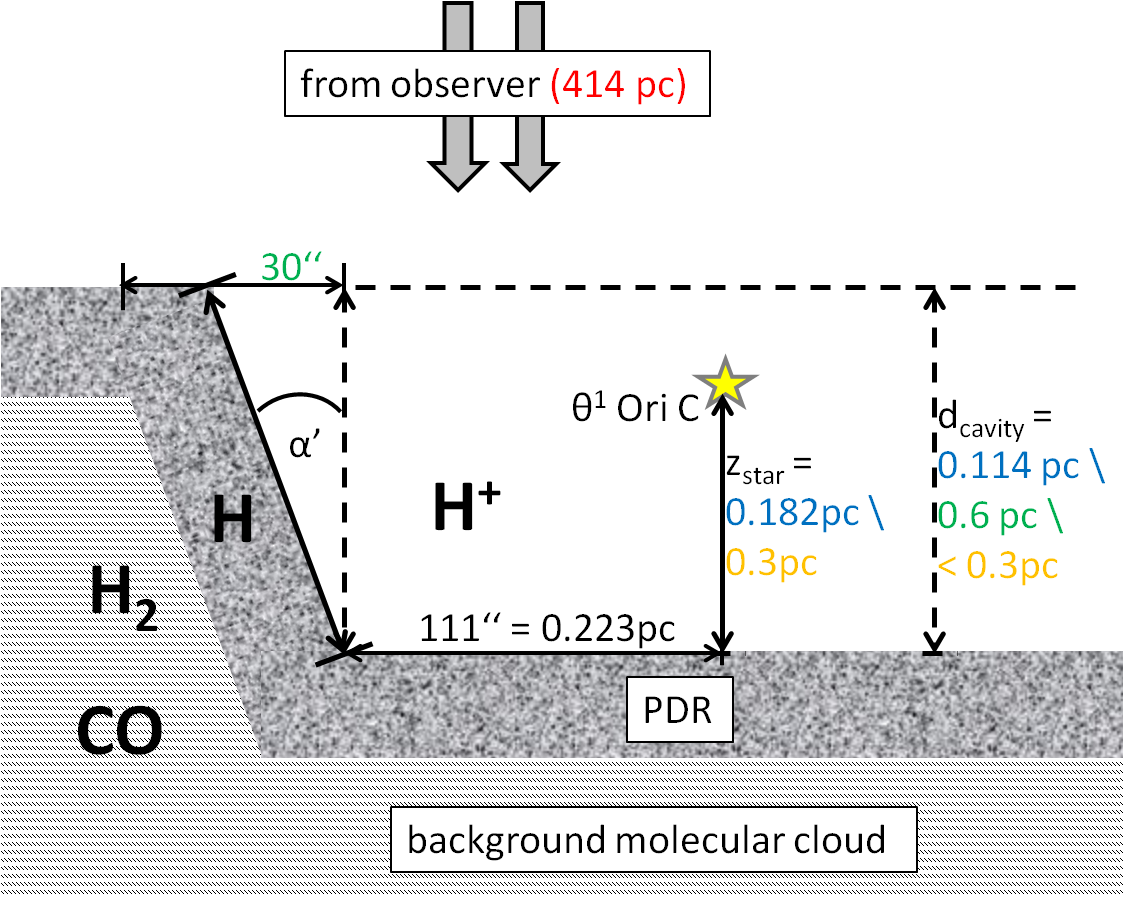}}
  \caption{``{\changed Face-on/edge-on/face-on}'' Orion Bar Geometry as proposed by \citetalias{Hogerheijde_1995}. 
              Values are taken from:  
              green: \citetalias{Hogerheijde_1995}; red: \citet{Menten_2007}; blue: \citet{Pellegrini_2009}; 
              and orange: \citet{Werf_2013}. For the inclination angle, $\alpha'$, values {\changed 
              between less than $3\degr$} and 15$\degr$ have been discussed \citep{Jansen_1995, Melnick_2012}.}
  \label{Fig:Ori_geometry}
\end{figure}

Different geometries have been proposed to model the Orion Bar, the dominating idea 
is a slightly inclined face-on/edge-on/face-on geometry first introduced by 
\citetalias{Hogerheijde_1995}. A schematic picture of this geometry is shown in 
Fig.~\ref{Fig:Ori_geometry}.
Many other workgroups have used adoptions of this  geometry to model observations 
\citep[see e.g.][]{Pellegrini_2009}. Due to the increased column density 
along the line of sight, this geometry naturally explains the observed intensity peak.
The depth of the cavity, the inclination angle of the {\changed b}ar ($\alpha'$, not to 
be confused with the power-law exponent $\alpha$ from Eq.~\ref{Eq:CMS}) and the 
``$z$-position'' (position on the line of sight to the observer) of the illuminating 
cluster have been subject to discussions. Different possibilities are indicated in 
Fig.~\ref{Fig:Ori_geometry}.

{\changed The face-on/edge-on/face-on} geometry is consistent with all the FIR 
and submm observations, but an indication that this geometry needs at least some 
modifications stems from optical observations \citep{McCaughrean2002} that show some 
shadowing at the very edge of the Orion Bar. This would be explained by a configuration 
where the Orion Bar is not the edge of a cavity but rather a filament as proposed by 
\citet{Walmsley_2000, Arab_2012}.

\subsection{Observations}
\label{Sect:Observations}

A tremendous amount of data is available for the Orion Bar PDR.
Recent observations of the Orion Bar PDR, observed with the 
\emph{Herschel} Space Observatory \citep{Pilbratt_2010}, can for 
instance be found in \citet{Habart_2010, Goicoechea_2011, Nagy_2013, 
Nagy_2014}. Recently, the whole Orion molecular cloud 1 region, which includes the Orion Bar PDR, has been mapped velocity resolved by \citet{Goicoechea_2015}.

As the aim of this paper focuses on the description and the testing of the
{\changed KOSMA-$\tau$ 3D code}, we selected only observations of abundant and simple species:
CO isotopologues, HCO$^+$ and the [C{\sc ii}] cooling line.
An expansion including many more species is of course possible.

We use [C{\sc ii}], CO~$10-9$, CO~$16-15$, $^{13}$CO~$5-4$, $^{13}$CO~$10-9$ and HCO$^+$~$6-5$ 
line observations of the Orion Bar PDR observed with the Heterodyne Instrument for 
the Far-Infrared \citep[HIFI,][]{Graauw_2010} on-board the \emph{Herschel} Space
Observatory \citep{Pilbratt_2010}. The 
observations have been performed as part of the EXtra-Ordinary Sources (HEXOS) 
guaranteed-time key program \citep{Bergin_2010}. Combined with low-$J$ CO and HCO$^+$ 
rotational lines (see below) these lines are well suited to trace the chemical 
stratification observed in the Orion Bar PDR. 

{\changed The [C{\sc ii}] observations have already been discussed in \citet{Ossenkopf_2013}.
Furthermore, \citet{Nagy_2014} show an HCO$^+$ map. All other \emph{Herschel} data is 
presented here for the first time. Further analysis of the data will be provided in 
subsequent papers (\citealt{Choi_2014}, Nagy et al., in prep).}

\begin{figure}
  \resizebox{\hsize}{!}{\includegraphics{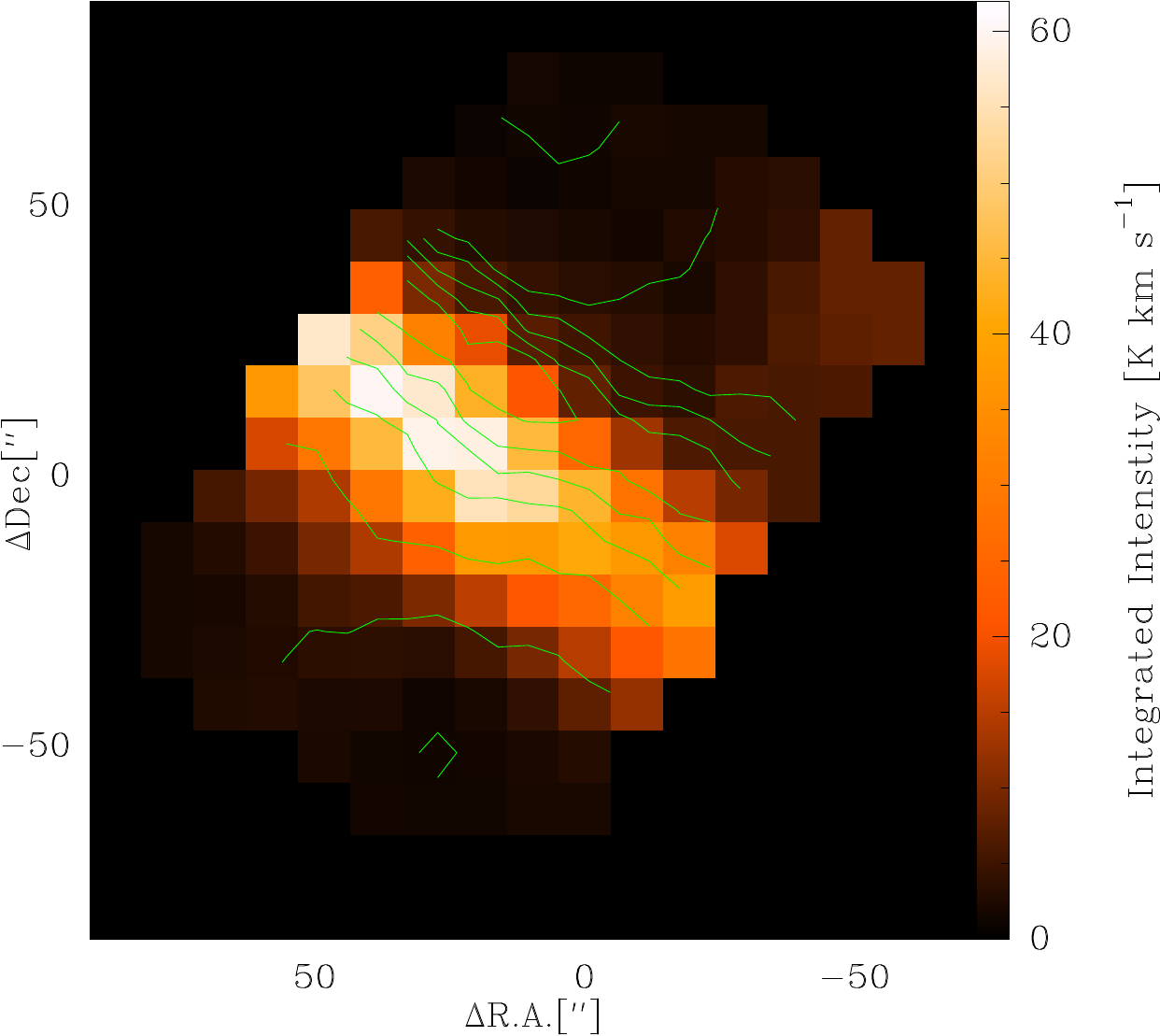}}
  \caption{The Orion Bar observed with HIFI/\emph{Herschel}. 
      The green contours show [C{\sc ii}] line intensities integrated between 7 and 13~km~s$^{-1}$. The contours range between 200 and
      800~K~km~s$^{-1}$ in steps of 100~K~km~s$^{-1}$. The colour scale gives the $^{13}$CO 10-9 line intensity, integrated between
      9 and 12~km~s$^{-1}$. The reference position is the ``CSO reference
      position'', (5$^h$35$^m$20.122$^s$, -5$^\circ$25'21.96'').}
  \label{Fig:CII}
\end{figure}

All presented HIFI/\emph{Herschel} observations are strips across the {\changed b}ar with a 
width of $1\arcmin$ or more, except for the CO~16-15 observations where a single 
cut has been observed. The observations have been taken in the on-the-fly (OTF) 
observing mode around the centre position 
($\alpha_{\rm J2000}$ = 5$^h$35$^m$20.81$^s$, $\delta_{\rm J2000}$ = -5$^\circ$25'17.1'') 
with a position angle perpendicular to the {\changed b}ar, i.e.~145$^\circ$ east of north, 
and an OFF position 6~arcminutes southeast of the map. The observations used the 
Wide-Band-Spectrometer (WBS) with a frequency resolution {\changed of 1.1~MHz which 
corresponds to 0.17\kms{} at the rest frequency of the [C{\sc ii}] line.} Both 
polarizations were averaged to improve the signal-to-noise ratio. Integration 
times varied between 4 and 30~s resulting in noise levels between a 0.02 and 0.3~K.
The high-frequency HIFI/\emph{Herschel} data, i.e.~the maps of [C{\sc ii}] and 
CO 16-15 have been reduced in HIPE as described by \citet{Ossenkopf_2013}. All other 
lines were analysed using the GILDAS software\footnote{\url{http://www.iram.fr/IRAMFR/GILDAS}}
for baseline subtraction and spatial {\changedTwo re-sampling}. An overlay of our data, 
[C{\sc ii}] overplotting $^{13}$CO~10-9, is shown in Fig.~\ref{Fig:CII}.

The line intensities (Table~\ref{Tab:observations}) are given on a 
$T_{\rm mb}$ scale. For the HIFI/\emph{Herschel} observations 
$T_{\rm mb}$ is a factor 1.26 to 1.5 higher than $T_A^*$, depending on 
the respective frequency \citep{Roelfsema_2012}. As discussed by 
\citet{Ossenkopf_2013}, the scaling from $T_A^*$ to $T_{\rm mb}$ is 
questionable for very extended emission (like [C{\sc ii}]) 
where the error beam of the telescope is likely to be filled with 
emission of approximately the same brightness as the main beam. Hence, 
for extended emission our intensities are upper limits.

Our data set is combined with ground-based observations of CO~$2-1$, CO~$3-2$, 
CO~$6-5$, $^{13}$CO~$3-2$, $^{13}$CO~$6-5$ and HCO$^+$~$3-2$ observed with the 
Caltech Submillimeter Observatory (CSO) (D. Lis, priv. comm.). The CSO 
observations are typically more extended but overlap with the HIFI/\emph{Herschel} 
maps. To facilitate the comparison between the maps, the reference positions
of all maps have been shifted to be equal to the CSO reference position
(5$^h$35$^m$20.122$^s$, -5$^\circ$25'21.96'').

\begin{figure}
  \resizebox{\hsize}{!}{\includegraphics{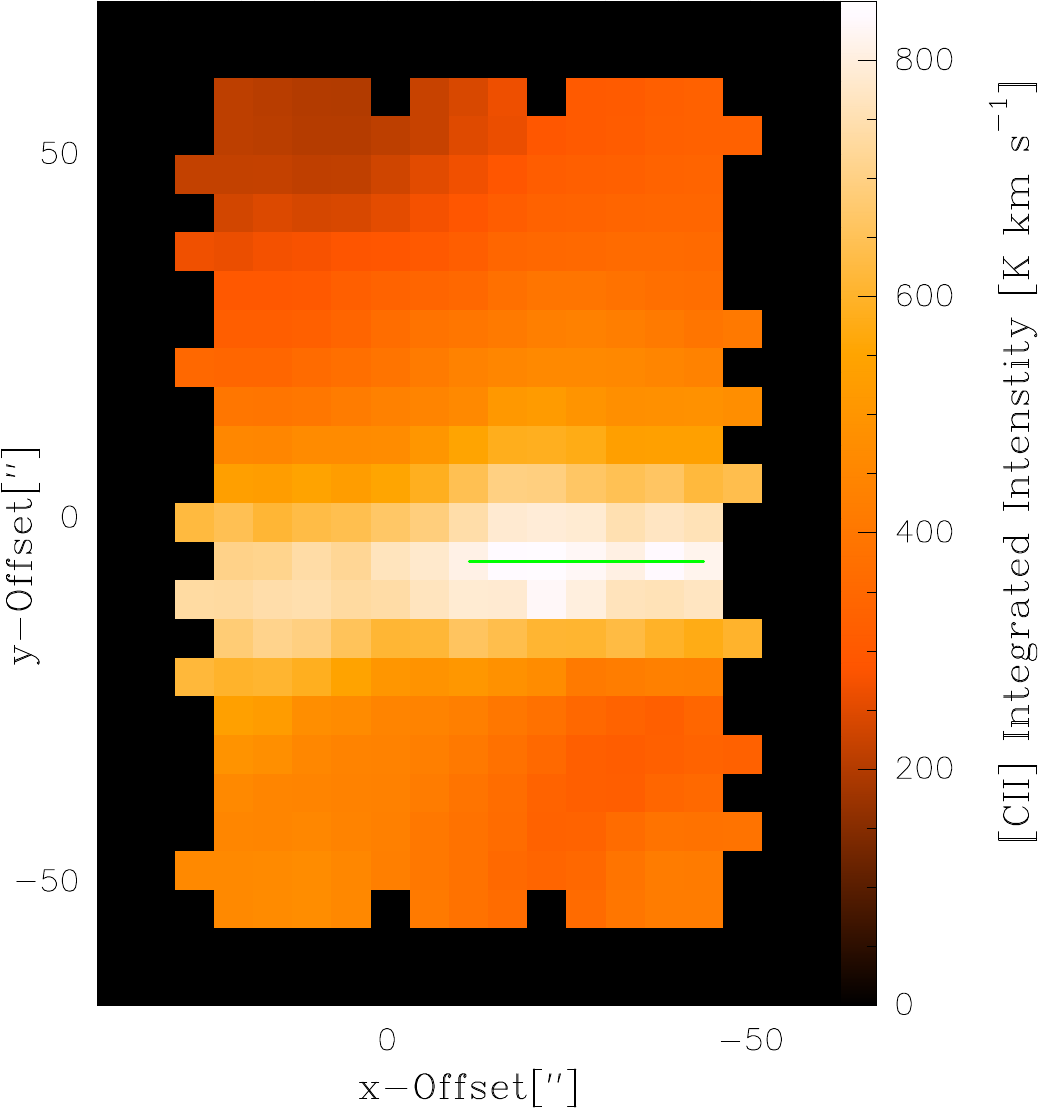}}
  \caption{[C{\sc ii}] integrated intensity (same as the contours in Fig.~\ref{Fig:CII}) 
  rotated by $-145^\circ$, i.e.~$\Theta^1$~Ori~C in the north-west of the {\changed b}ar is at the 
  bottom in this orientation. The green line marks the cut with the highest averaged line 
  integrated intensity, including all pixels with x-offsets between \mbox{-11.3'} and -43.5'. }
  \label{Fig:CII_rot}
\end{figure}

To simplify the analysis of the stratification profile, the maps have
been rotated around the CSO reference position by -145$^\circ$ 
(-145$^\circ$ clockwise), resulting in an orientation of the Orion Bar 
parallel to the ``$x$-axis'' ({\changed see} Figs.~\ref{Fig:CII} and 
\ref{Fig:CII_rot}). As we focus on the stratification of the chemical 
and excitation structure across the Orion Bar, the observed spectra have 
been averaged along rows of pixels\footnote{For the CO~16-15 cut, each 
``row'' only contains one pixel} parallel to the $x$-axis ensuring that 
we average over clumps and interclump medium. In the $x$-range between
$-11.3\arcsec$ and $-43.5\arcsec$ the Orion Bar has a very straight 
appearance in all of our maps and an average over $\sim30\arcsec$ 
guarantees that we are not affected by individual clumps, but consider
a clump-ensemble on the observational side as well{\changed. \citet{Lis_2003} 
observe the size of dense condensations in the Orion Bar and find sizes between 
$3.81\arcsec$ and $7.96\arcsec$ and \citet{YoungOwl_2000} discuss clumps of 
$9\arcsec$ size, supporting our approach.}

Gaussian profiles were fitted to the averaged spectra. We fit two 
Gaussian profiles, one profile fixed at a centre velocity of 
8~km~s$^{-1}$ to exclude the emission from the Orion Ridge 
\citep{vanderTak2013}. The other profile fits the main component at 
about 11~km~s$^{-1}$ originating from the Orion Bar. Integration of 
this component yields the line integrated intensity, averaged for the 
respective row ($y$-position). The peak position was determined 
by fitting a parabola to the row-averaged intensities at the 
different $y$-positions. As deviations between the fitting points and 
the fitted parabolas are very small, we assume the pointing error of 
the telescope as the main uncertainty in the determination of the 
peak position. The pointing errors are $2.4\arcsec$ for 
HIFI/\emph{Herschel} \citep{Pilbratt_2010} and $3\arcsec$ for CSO 
data\footnote{\url{http://cso.caltech.edu/wiki/cso/science/overview}}. The resulting peak intensities and $y$-offsets are 
summarised in Table~\ref{Tab:observations} for the different 
transitions. {\changedTwo Table~\ref{Tab:observations} indicates a 
peculiarity of the HCO$^+$~3--2 transition. It seems to peak in front 
of all the other molecular transitions, including HCO$^+$~6--5
that should tracer warmer gas, while the profiles of both lines are 
very similar. We have no evidence for a pointing problem in these 
data so that we stick to the formal errors, but as there is no 
physical scenario that would explain this peak offset we rather 
question the role of the HCO$^+$~3--2 peak position in the fit of the 
stratification pattern in the discussion
(Sect.~\ref{Sect:Discussion_inhomogeneous})}. 

\begin{table*}
\caption{Summary of averaged integrated intensities and spatial offsets of the observations.}
\label{Tab:observations}
\centering
\begin{tabular}{lccccccc} 
\hline\hline  
{\changed Transition} & Frequency\tablefootmark{a} & Observatory & Beamsize\tablefootmark{b} & Peak intensity\tablefootmark{c}
& $y$-offset\tablefootmark{d} & $y$-offset\tablefootmark{d} & $\Delta y_{\rm obs}$\tablefootmark{e}\\
  & [GHz]& & [arcsec]& [K km s$^{-1}$] & [pc] &[arcsec] & [pc] \\
  \hline
  [C{\sc ii}]                       & 1900.5369    & HIFI/Herschel & 11.2 & 1153   $\pm$ {\changed 115}  & -0.016 $\pm$ 0.005 & -7.8 $\pm$ 2.4 & 0\\
  CO~$2-1$                          & 230.5380000  & CSO           & 30.5 &  402   $\pm$ {\changed 32 }  &  0.029 $\pm$ 0.006 & 14.6 $\pm$ 3.0 & 0.045 $\pm$ 0.008 \\
  CO~$3-2$                          & 345.7959899  & CSO           & 21.9 &  406   $\pm$ {\changed 68 }  &  0.014 $\pm$ 0.006 &  7.2 $\pm$ 3.0 & 0.030 $\pm$ 0.008 \\
  CO~$6-5$                          & 691.4730763  & CSO           & 10.6 &  560   $\pm$ {\changed 244}  &  0.020 $\pm$ 0.006 &  9.8 $\pm$ 3.0 & 0.036 $\pm$ 0.008 \\
  CO~$10-9$                         & 1151.985452  & HIFI/Herschel & 18.4 &  374   $\pm$ {\changed 37 }  &  0.021 $\pm$ 0.005 & 10.5 $\pm$ 2.4 & 0.037 $\pm$ 0.007 \\
  CO~$16-15$                        & 1841.345506  & HIFI/Herschel & 11.5 &  128   $\pm$ {\changed 13 }  &  0.019 $\pm$ 0.005 &  9.4 $\pm$ 2.4 & 0.035 $\pm$ 0.007 \\
  $^{13}$CO~$3-2$                   & 330.5879653  & CSO           & 21.9 &  114   $\pm$ {\changed 18 }  &  0.026 $\pm$ 0.006 & 12.8 $\pm$ 3.0 & 0.042 $\pm$  0.008\\
  $^{13}$CO~$5-4$                   & 550.9262851  & HIFI/Herschel & 38.5 &  120   $\pm$ {\changed 12 }  &  0.042 $\pm$ 0.005 & 20.8 $\pm$ 2.4 & 0.058 $\pm$ 0.007 \\
  $^{13}$CO~$6-5$                   & 661.0672766  & CSO           & 10.6 &  157   $\pm$ {\changed 65 }  &  0.030 $\pm$ 0.006 & 14.8 $\pm$ 3.0 & 0.046 $\pm$ 0.008\\
  $^{13}$CO~$10-9$                  & 1101.3495971 & HIFI/Herschel & 19.3 &   92   $\pm$ {\changed 9.2}  &  0.018 $\pm$ 0.005 &  9.0 $\pm$ 2.4 & 0.034 $\pm$ 0.007\\
  HCO$^+$~$3-2$                     & 267.5576259  & CSO           & 30.5 &   46   $\pm$ {\changed 5  }  &  0.010 $\pm$ 0.006 &  5.0 $\pm$ 3.0 & 0.026 $\pm$ 0.008\\
  HCO$^+$~$6-5$                     & 535.0615810  & HIFI/Herschel & 39.6 &    8.7 $\pm$ {\changed 0.87} &  0.022 $\pm$ 0.005 & 11.2 $\pm$ 2.4 & 0.038 $\pm$ 0.007\\
\hline
\end{tabular}	
\tablefoot{
\tablefoottext{a}{Taken from ``The Cologne Database for Molecular Spectroscopy (CMDS)'' 
(\citealt{Mueller_2001, Mueller_2005}; \url{http://www.astro.uni-koeln.de/cdms/})}
\tablefoottext{b}{Calculated based on \citet{Roelfsema_2012} for HIFI/\emph{Herschel}. Taken from 
\url{http://www.submm.caltech.edu/cso/receivers/beams.html} ``calculated FWHM''
for CSO data. For non-circular beams an average has been used.}
\tablefoottext{c}{Line integrated intensity averaged along the {\changed b}ar at the position 
of the peak (see text), $T_{\rm mb}$ scale. 
{\changed For HIFI/\emph{Herschel} the error on $T_{\rm mb}$ is about 10\% \citep[see][]{Roelfsema_2012}.
The error given for the CSO data has been calculated (and extrapolated for frequencies $>345$~GHz)
based on the errors on $T_A^*$ and $\eta_{\rm mb}$ given in \citet{Mangum_1993}.}
}
\tablefoottext{d}{Measured spatial offset into the PDR (with position angle 145$^\circ$ east of north) relative to the CSO reference
position.}
\tablefoottext{e}{Measured spatial shift into the PDR (with position angle 145$^\circ$ east of north) relative to the 
[C{\sc ii}] peak 
position.}
}
\end{table*}

\section{3D Model of the Orion Bar PDR}
\label{Sect:OriModel}

We have composed a 3D model of the Orion Bar PDR from cubic voxels with an edge 
length of 0.01~pc, corresponding to $5.0\arcsec$ at the distance of 414~pc. 
The voxel size is small enough {\changed to trace physical changes in the PDR and} 
to analyse stratification effects, but large enough to ensure that the total 
number of voxels can be treated on a standard PC. {\changed Furthermore, for all 
observations that are fitted in this work, the resulting pixel size is at least a 
factor two smaller than the beamsize.} Our Cartesian coordinate system is chosen 
in such a way that the $x$-direction is parallel to the Orion Bar and the $z$-direction 
points towards the observer. As we are mainly interested in the stratification of 
the Orion Bar here, the current model ignores any variation of the density structure 
in $x$-direction. This reduces the number of free parameters, but excludes for the 
moment the simulation of additional structures like the Orion Ridge.
 
{\changed In this work we focus on geometries for the Orion Bar PDR that are 
based on the \citetalias{Hogerheijde_1995} series of papers, i.e.~on geometries 
that consist} of an almost edge-on cavity wall facing the illumination from 
$\Theta^1$~Ori~C (see Figs.~\ref{Fig:OriBarGeometry_FUV} and \ref{Fig:Ori_geometry}). 
{\changed Aiming for a fit of the observations presented in 
Sect.~\ref{Sect:Observations} different parameters have been varied in this 
model set-up. An overview over these parameters is provided in 
Sect.~\ref{Sect:parameters}. In Sect.~\ref{Sect:ModelAssessment} we 
discuss the measures that are used to evaluate our fits.} 
{\changed A} second geometry {\changed that has been discussed for the
Orion Bar is} the filament model proposed by \citet{Walmsley_2000}
and \citet{Arab_2012}. {\changed This model consists} of a
cylinder in the plane of the sky with the main symmetry axis along the 
bar (see Fig.~\ref{Fig:cyl_CII_intensity}). In
Appendix~\ref{appendix:cylindrical} we show preliminary tests of this
geometry, which indicate that a simultaneous reproduction of the
observed stratification pattern and the line integrated intensities
based on the cylindrical model is problematic. For this geometry, the 
short lines of sight through the compound close to $y=0$ enforce that 
the emission peaks appear deep in the cloud, where the FUV flux is low. 
This reduces the line integrated intensities and {\changedTwo increases 
the scatter between the $y$-offsets calculated for the different 
transitions.}

The main illuminating source $\Theta^1$~Ori~C is $111\arcsec$ away from 
the IF. This corresponds to a separation by 22.3 voxels in
$y$-direction between star and interface. In $x$-direction, the 
location of the star defines our zero point, i.e.~in voxel units 
$\Theta^1$~Ori~C is located at 
$\lbrack 0, y_{\rm IF}+22.3, z_{\rm star}\rbrack$ in the model. The $z$ 
position of the star ($z_{\rm star}$) is not exactly known (see 
Fig.~\ref{Fig:Ori_geometry}) and has become one of our fitting 
parameters.

{\changedTwo Based on C$^{18}$O~3-2 observations and assuming a 
conversion factor of 
$N_{\rm H_2}/N({\rm C}^{18}{\rm O})=5\times 10^6$, 
\citetalias{Hogerheijde_1995} derive} a total H$_2$ column density of 
$N_{{\rm H}_2}=6.5\times 10^{22}$~cm$^{-2}$ along the line of sight 
(peak) for a path length of 0.6~pc. For a uniform density along 
the line of sight this translates into 
$n_{{\rm H}_2}=N_{{\rm H}_2}/(0.6 {\rm pc})=3.5\times 10^4$~cm$^{-3}$.
Consequently, the total average mass in a voxel with volume 
(0.01~pc)$^3$ is:
\begin{eqnarray}
\label{Eq:Hogerheijde_mass}
{\rm M}_{\rm HJ}&=&3.5\times 10^4\, {\rm cm^{-3}} m_{\rm H_2}\left(\frac{3.086\times 10^{18} {\rm cm}}{\rm pc}\right)^3
\left(0.01\, {\rm pc}\right)^3 \nonumber\\
 &=& 0.00173\, {\rm M}_\sun\, ,
 \end{eqnarray}
which we use as a baseline for our simulations. 

The clump ensembles in the models contain clumps at the mass points 
$\lbrack 10^{-3}, 10^{-2}, 10^{-1}, 10^0\rbrack$~M$_\sun$, {\changedTwo implying that one voxel typically only contains fractions of clumps, i.e. $N_j<1$}. The upper 
mass limit matches the resolved clump masses in the range 
$0.5-1.5$~M$_\sun$ determined for the Orion Bar PDR by 
\citet{Lis_2003}. The lower limit of $10^{-3}$~M$_\sun$ is
used as the smallest mass contained in the available KOSMA-$\tau$ 
input grid because, to gain a good approximation of a fractal 
geometry, the inclusion of very small structures is desired. We 
discuss this choice and show simulation results based on models using 
different mass points in Sect.~\ref{Sect:MassPoints}. In the 
KOSMA-$\tau$ 3D code the pixels are scaled to superpixels (see 
Sect.~\ref{Sect:EnsembleStatistics}). In the simulations presented 
here, after the scaling process, each ``supervoxel'' usually contains 
one clump at mass point $10^{0}$~M$_\sun$ and consequently 
$\lbrace N_j \rbrace=\lbrack 331, 48, 7, 1\rbrack$ for the different 
mass points (see Eq.~\ref{Eq:Nbinned2}, results rounded to integer 
values).

The thin interclump medium is mimicked by a second clump ensemble 
with an averaged density that is about two orders of magnitude lower 
than the averaged density of the dense clumps. To approximate a 
relatively homogeneous interclump medium, we start our simulations 
using small clumps of $10^{-2}$~M$_\sun$. Furthermore, the VFF of the 
interclump medium should be equal to unity or smaller. Therefore, we 
add the condition
\begin{equation}
 M_{\rm inter,\, tot}\lbrack {\rm M}_\odot \rbrack \leq 0.01^3\, \frac{m_{\rm H}}{{\rm M}_\odot}\rho_{\rm inter}\lbrack {\rm pc}^{-3} \rbrack \, ,
\end{equation}
or equivalently
\begin{equation}
 \frac{M_{\rm inter,\, tot}\lbrack {\rm M_{\rm HJ}} \rbrack}{\rho_{\rm inter}\lbrack {\rm cm}^{-3} \rbrack }\leq 1.43\times 10^{-5}\, .
\label{Eq:inter_coupled}
\end{equation}
{\changedTwo For a discussion of the interclump parameter 
$M_{\rm inter,\, tot}$ and $\rho_{\rm inter}$ see 
Sect.~\ref{Sect:parameters}.}
We start our simulations with a VFF of unity in
Sect.~\ref{Sect:HomMassesDensities}. Different choices for the mass 
point and the VFF of the interclump medium are tested in 
Sect.~\ref{Sect:Inhomogeneous}.

In this work we do not fit the full line profiles. Therefore, the
velocity spread between the clumps in one ensemble 
($\sigma_{j,\, {\rm ens}}$, see Eq.~\ref{Eq:Nji}) has been fixed. We 
discuss our choice of the $\sigma_{j,\,{\rm ens}}$ and show examples 
of simulated line profiles in Sect.~\ref{Sect:line_profiles}. 

The KOSMA-$\tau$ 3D code allows for the simulation of (2D) maps. 
{\changed As an example, Figs.~\ref{Fig:CO3} and \ref{Fig:CO3_conv} show simulated 
maps of line integrated CO~$3-2$ intensities, before and after the convolution with a 
Gaussian beam of 21.9$\arcsec$} FWHM, matching the CSO beam used in the observations.
{\changed The maps are based on model 1m (see Table~\ref{Tab:models}).} 
We find a combination of the imprint of the {\changed sharp edge of the bar} and
a curvature stemming from the varying distance to the illuminating star.
{\changed The convolution blurs the edge and the emission peak}, but still
allows to recover the stratification of the emission.

For our systematic parameter study we have reduced the map size 
{\changed (and the beam convolution)} to a {\changed cut} of {\changed only
one pixel in $x$-direction across the Orion Bar.} {\changed Such a cut}
enables us to {\changed derive the line} integrated intensities and peak
offsets within {\changed a computing time of about six
hours\footnote{\changed The computing time strongly correlates with the
number of voxels used in a specific set-up. Six hours are needed for the computation on one core 
of a Intel\textsuperscript{\textregistered} Xeon E5620 2.4~GHz CPU with 64~GB
RAM.}. {\changed For the 2D maps 18~days of computing time are needed.
Typical simulated cuts are shown in Fig.~\ref{Fig:cuts_6j} based on model 6j
(see Sect.~\ref{Sect:Inhomogeneous} and Table~\ref{Tab:models})}.

\begin{figure}
  \resizebox{\hsize}{!}{\includegraphics{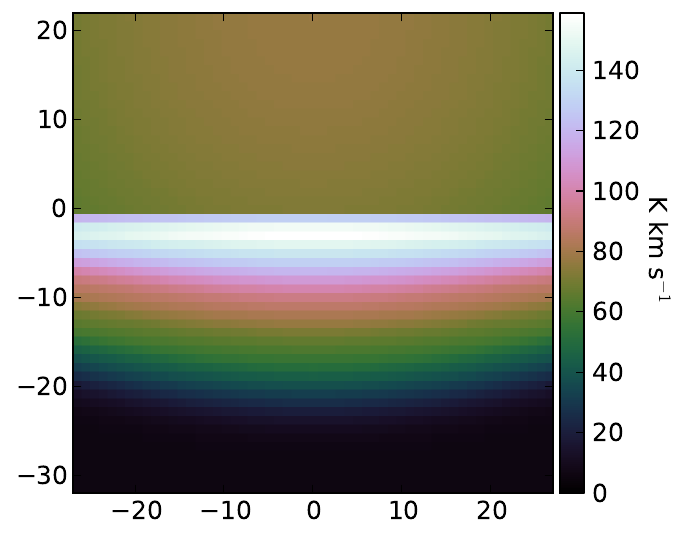}}
  \caption{Simulated map of line-integrated CO 3-2 emission of the Orion Bar PDR, 
  based on model {\changed 1m (see Table~\ref{Tab:models})}. The coordinates 
  are given in {\changed units of} pixels{\changed ,} one pixel corresponds to 0.01~pc 
  or $5\arcsec$ on the sky. The illuminating star $\Theta^1$~Ori~C is located 
  {\changed at $x=0$ and $y=22.3$ on top} of the map.
  }
  \label{Fig:CO3}
\end{figure}

\begin{figure}
  \resizebox{\hsize}{!}{\includegraphics{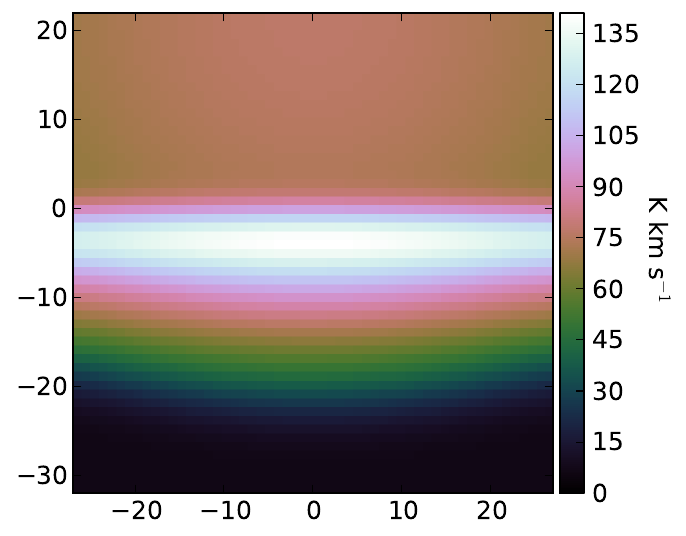}}
  \caption{Same as Fig.~\ref{Fig:CO3} but after convolution with a Gaussian beam
  of {\changed 21.9}$\arcsec$ or 4.4 pixels FWHM (see Table.~\ref{Tab:observations}).}
  \label{Fig:CO3_conv}
\end{figure}

\begin{figure*}
\centering
   \includegraphics[width=17cm]{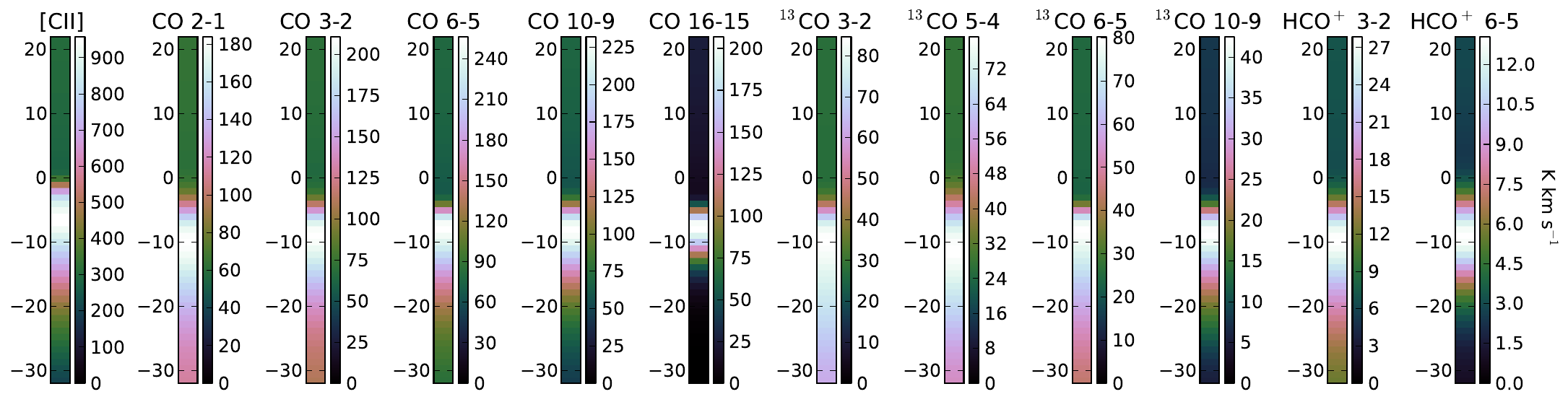}
     \caption{Simulated cuts perpendicular to the Orion Bar, based on model {\changed 6j,
     (see Table~\ref{Tab:models})}. {\changed Each} colour scales give{\changed s} the line 
     integrated intensity {\changed of} the transitions indicated above the respective {\changed cut}.
     }
     \label{Fig:cuts_6j}
\end{figure*}

\subsection{\changed C$^{18}$O: upper limit for the total column of molecular gas}
\label{Sect:C18O}

\begin{table}
\caption{\changed C$^{18}$O emission adopted from
\citetalias{Hogerheijde_1995} and simulated based on 
model{\changedTwo s} 2b {\changedTwo and 6j}.}
\label{Tab:C18O}
\centering                        
\begin{tabular}{l c c c c c}       
\hline\hline                
  Transition & $\theta$\tablefootmark{a} & \multicolumn{4}{c}{$\int T_{\rm mb}\, {\rm dv}$~[\kms{}]}  \\
 \hline
            &       &     \citetalias{Hogerheijde_1995} & 2b & 2b\_ext & 6j \\   
\hline                        
  C$^{18}$O~$2-1$ & 13$''$ & 16.1 & 38.7 & 39.4 & 35.5 \\    
  C$^{18}$O~$3-2$ & 21$''$ & 30.2 & 41.6 & 42.2 & 37.4 \\
\hline                                 
\end{tabular}
\tablefoot{
\tablefoottext{a}{Telescope HPBW.}
}
\end{table}

\begin{figure}
  \resizebox{\hsize}{!}{\includegraphics{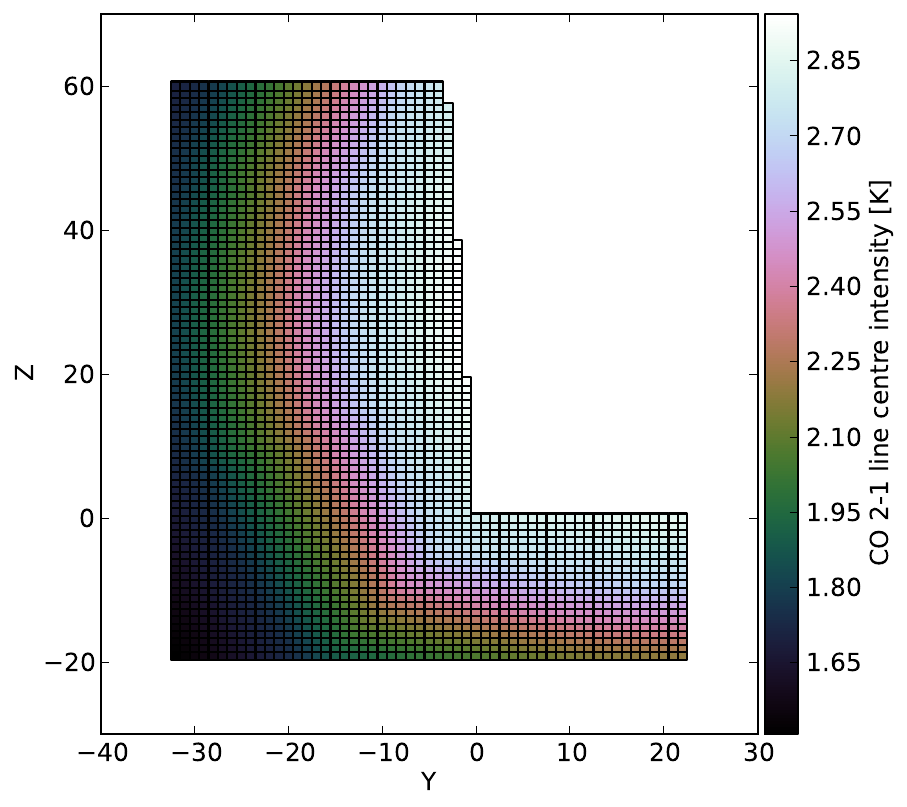}}
  \caption{\changed Cut through the Orion Bar model 2b. For each voxel the colour 
  scale gives the CO~$2-1$ line intensity of dense clumps and 
  interclump medium, at the line centre (at 11.3~\kms{}). The illuminating star 
  $\Theta^1$~Ori~C is located at $\lbrack 0, 22.3, 30\rbrack$.
  }
  \label{Fig:CO2-1_intensity}
\end{figure}

\begin{figure}
  \resizebox{\hsize}{!}{\includegraphics{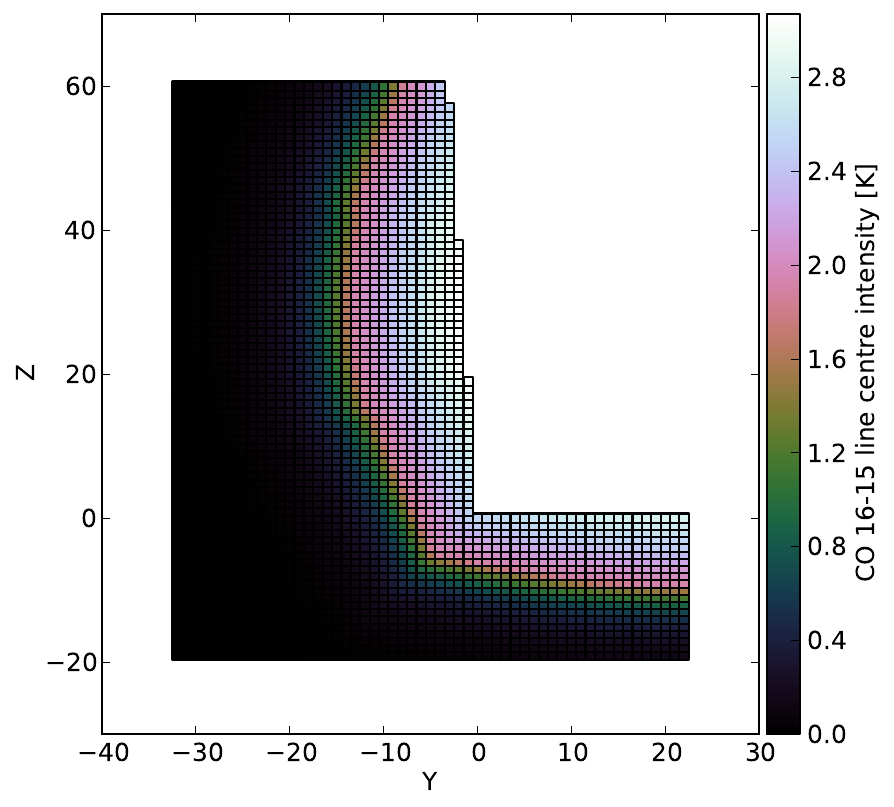}}
  \caption{\changed Same as Fig.~\ref{Fig:CO2-1_intensity} but with the CO~$16-15$ line 
  intensity given on the colour scale.
  }
  \label{Fig:CO16-15_intensity}
\end{figure}

{\changed 
The maps of the Orion Bar can include radiation from the background molecular
cloud (see Fig.~\ref{Fig:Ori_geometry}). Therefore, we should in principle
extend our model into the negative $z$-direction until we have reached a
depth were non of the investigated tracers is excited anymore. However, to
reduce computing time, the background molecular cloud is cut off at $z=-20$ 
in our systematic parameter study. At $z=-20$ the FUV flux has usually
dropped below one Draine field (see Fig.~\ref{Fig:OriBarGeometry_FUV}). To
investigate possible contributions to the final maps/cuts from the background
molecular cloud we have re-run the simulation of model 2b (which provides one
of the best fits of the line integrated intensities; see
Sect.~\ref{Sect:AlphaIUV} and Table~\ref{Tab:models}), but with the compound 
extended to $z=-100$.

Figures~\ref{Fig:CO2-1_intensity} and \ref{Fig:CO16-15_intensity} show simulated 
cuts through the Orion Bar model 2b, before the extension. The colour scales in 
these plots give the line centre intensity emitted by each voxel, in 
Fig.~\ref{Fig:CO2-1_intensity} for CO~$2-1$ and in Fig.~\ref{Fig:CO16-15_intensity} 
for CO~$16-15$. The figures show that CO~$16-15$ is only excited close to the PDR surface, 
where the FUV flux is relatively high. Hence, the background molecular cloud will 
not be visible in the final line integrated maps. For CO~$2-1$ the situation is 
different: the excitation only depends weakly on the FUV flux and hence, the voxels 
still emit at $z=-20$. However, the effect of adding the background cloud to the 
simulation is still small due to the high optical depth of the CO~$2-1$ line. 
Overall, we find that adding the background molecular cloud slightly changes
the quality of the fit of the line integrated intensities (see 
Table~\ref{Tab:models}), but it does not change the outcome of our systematic 
parameter study.

{\changedTwo The total column density of the Orion Bar can be constrained from
optically thin tracers that are only weakly sensitive to the PDR conditions.}
\citetalias{Hogerheijde_1995} provide line integrated intensities of the
C$^{18}$O~$2-1$ and $3-2$ transitions at the emission peak of the Orion 
Bar PDR. Due to the low optical depths of these transitions compared to the
other CO isotopologues, {\changedTwo they provide} an upper limit for the total
(volume-averaged) column density of the dense clumps, including the
background cloud. Table~\ref{Tab:C18O} {\changedTwo compares the intensities from 
\citetalias{Hogerheijde_1995} to the simulated line
integrated intensities based on models 2b, 6j, and 2b\_ext, having 
cut-offs at $z=-20$ and at $z=-100$. 
In contrast to all PDR simulations, \citetalias{Hogerheijde_1995} 
observed a C$^{18}$O~$2-1$ line that is significantly weaker than 
the $3-2$ line. This} could be explained by a cold foreground layer. 
However, as we have not included foreground material into our 
models, a detailed fit of that line is beyond the scope of this 
work. {\changedTwo Therefore, we concentrate on the C$^{18}$O~$3-2$ 
line for the column density estimate like 
\citetalias{Hogerheijde_1995}.}

We find that the contribution from the interclump medium to the 
C$^{18}$O~$2-1$ and $3-2$ line emission is negligible. Furthermore, the increase 
of the C$^{18}$O line integrated intensities due to the background extension 
is low. {\changedTwo Using the \citetalias{Hogerheijde_1995} column density 
of $N_{{\rm H}_2}=6.5\times 10^{22}$~cm$^{-2}$ leads to line intensities
that are too low by more than a factor of two. Models 2b and 6j
contain a mass per voxel of 2~M$_{\rm HJ}$ combined with a total depth of 
0.8~pc (cut-off at $z=-20$, parameters $M_{\rm cl,\, tot}$, $d_{\rm cavity}$, 
see Sect.~\ref{Sect:parameters}).}
The total (volume-averaged) column of the ensembles of dense clumps is 
$N_{{\rm H}_2}\approx 1.7\times 10^{23}$~cm$^2$, a factor 2.7 higher than the 
value that was found by \citetalias{Hogerheijde_1995}. {\changedTwo The two 
models provide intensities that are too high by 40\,\% and 25\,\%  compared
to the observations so that we consider the column density of 
$1.7\times 10^{23}$~cm$^2$ as the upper limit.} Consequently, we exclude 
models with higher column densities from our simulation runs.
Lower columns are always allowed in our models, as they could be compensated 
by a deeper background cloud that is invisible in all the PDR tracer discussed
here.
}

\section{Parameter scans}
\label{Sect:ParameterScans}

\subsection{Parameters}
\label{Sect:parameters}
 
In the following we summarise the parameters that are varied within 
our simulation runs. If available we also give values taken from 
\citetalias{Hogerheijde_1995} which will be used as an initial guess 
for our simulations. 
\begin{description} 
    \item[$M_{\rm cl,\, tot}$:] The mass contained in dense clumps per 
    voxel. Based on \citetalias{Hogerheijde_1995} we have estimated the 
    \emph{total} mass per voxel, M$_{\rm HJ}$, in 
    Eq.~\ref{Eq:Hogerheijde_mass}. Furthermore, 
    \citetalias{Hogerheijde_1995} state that about 10\% 
    (i.e.~0.1~M$_{\rm HJ}$) of the molecular material\footnote{Atomic 
    hydrogen is only contained in a thin surface layer 
    ($A_{\rm V}\lesssim 0.1$) of a PDR 
    \citep[see for example][]{Roellig_2007}. Hence, in the comparison
    between simulated (column) densities and the results stated in
    \citetalias{Hogerheijde_1995}, we assume that the contribution of 
    atomic hydrogen is negligible, i.e.~we use 
    $N\approx 2\ N_{\rm H_2}$ and $\rho\approx 2\ n_{\rm H_2}$ when 
    comparing to the molecular densities from 
    \citetalias{Hogerheijde_1995}.} is contained in clumps.
\item[$M_{\rm inter,\, tot}$:] The mass contained in the interclump
    medium per voxel. Following \citetalias{Hogerheijde_1995} the 
    interclump medium accounts for 90\% of the total molecular 
    column density. {\changedTwo Using their values, 
    i.e.~$\rho_{\rm inter}=2\ n_{\rm H_2}=2\times 3\times 10^{4}$~cm$^{-3}$ and a total interclump mass of 0.9~M$_{\rm HJ}$ in 
    one voxel with a volume of (0.01~pc)$^3$, the VFF is about 1.05 
    (see Sect.~\ref{Sect:OriModel}; per voxel this corresponds 
    (statistically) to 0.156 clumps with a mass of $10^{-2}$~M$_\sun$ 
    and a volume of $6.74\times 10^{-6}$~pc$^3$).} In some of the 
    presented simulations {\changedTwo (see 
    Sect.~\ref{Sect:HomMassesDensities})} 
    $M_{\rm inter,\, tot}$ is no independent parameter, but coupled 
    to $\rho_{\rm inter}$ to ensure an interclump-VFF of 
    unity. {\changedTwo Therefore, in our first simulation runs where 
    $M_{\rm inter,\, tot}$ (and $\rho_{\rm inter}$) are varied,
    $\rho_{\rm inter}=6\times 10^4$~cm$^{-3}$ corresponds to 
    $M_{\rm inter,\, tot}=0.848$~M$_{\rm HJ}$ (instead of 
    0.9~M$_{\rm HJ}$).} 
 \item[$\rho_{\rm cl}$:] The ensemble-averaged hydrogen nucleus density of the dense clumps. 
       \citetalias{Hogerheijde_1995} derive $n_{{\rm H}_2}\approx 1^{+3.0}_{-0.7}10^6$~cm$^{-3}$
       (i.e.~$\rho_{\rm cl}\approx 2\times  10^6$~cm$^{-3}$)
       using only one type of dense clumps .
 \item[$\rho_{\rm inter}$:] The ensemble-averaged hydrogen nucleus density of the interclump medium. 
      \citetalias{Hogerheijde_1995} derive $n_{{\rm H}_2}\approx 3^{+2.0}_{-2.2}10^4$~cm$^{-3}$ 
      (i.e.~$\rho_{\rm inter}\approx 2\times 3\times 10^4$~cm$^{-3}$) for 
      a homogeneous interclump medium. In some simulations $\rho_{\rm inter}$ is coupled to 
      $M_{\rm inter,\, tot}$ (see above).
 \item[$\alpha'$:] The inclination angle of the bar (see Fig.~\ref{Fig:Ori_geometry}). 
       \citet{Jansen_1995} discuss that an inclination angle $\alpha' < 3^\circ$
       is probable.
 \item[$z_{\rm star}$:] The z-position of the illuminating source, which is not discussed 
       in \citetalias{Hogerheijde_1995}. Here, we start our simulations using $z_{\rm star}=0.3$~pc,
       i.e.~with the illuminating source located at half height of the cavity.
 \item[$I_{\rm UV}$:] The FUV flux from the illuminating source, 
    $\Theta^1$~Ori~C, at the position of the IF.  
    \citetalias{Hogerheijde_1995} state $4.4\times 10^4\,\chi_0$. We   
    provide an estimate of $I_{\rm UV}$ in
    Appendix~\ref{appendix:FUVatIF}. {\changedTwo The flux 
    $I({\mathbf r})$ at a position ${\mathbf r}$ of a voxel at the 
    cloud surface (i.e.~not affected by FUV absorptions) is given by}
    \begin{equation}
    I({\mathbf r})=I_{\rm UV}\frac{\left(0.223~{\rm pc}/\Delta s\right)^2}{|{\mathbf r}_{star}-{\mathbf r}|^2}\, ,
    \end{equation}
    {\changedTwo where 0.223~pc is the observed distance in 
    $y$-direction between $\Theta^1$ Ori C and the Orion Bar 
    (see Sect.~\ref{Sect:geometry}), ${\mathbf r}_{\rm star}$ is the 
    position of the illuminating source in our model and $\Delta s$ is 
    the edge-length of one voxel in pc.
    }
 \item[$d_{\rm cavity}$:] The depth of the cavity (see 
       Fig.~\ref{Fig:Ori_geometry}). \citetalias{Hogerheijde_1995} 
       assume 0.6~pc. 
 \item[$d_{\rm clumps}$:] The depth into the cloud at which dense clumps
       appear. With this parameter we want to test whether the same 
       amount of clump and interclump mass is contained in each voxel, 
       independent from the depth into the PDR, or if there is there a 
       process that removes dense clumps from the surface. The models 
       discussed in \citetalias{Hogerheijde_1995} do not account for 
       such an effect, but it is proposed in the text.
 \item[$m_{\rm l, cl}$:] The lowest mass point of the ensemble of dense
       clumps. \citetalias{Hogerheijde_1995} find that a single-density 
       model cannot explain the observed line ratios and assume that a 
       ``range of densities'' appears in the beam. They construct their 
       model with two density components (clump and interclump medium) 
       as a ``first order approximation''. Here, we start our 
       simulations with dense clumps down to $10^{-3}$~M$_\sun$.
 \item[$m_{\rm inter}$:] Mass point used for the interclump medium, 
 i.e.~the interclump medium is represented by identical clumps of mass 
 $m_{\rm inter}$. Here, we start with $m_{\rm inter}=10^{-2}$~M$_\sun$.
\end{description}

\begin{table}
\caption{Overview over fitting parameters.}
\label{Tab:Fit_parameter}
\centering                          
\begin{tabular}{l c c }        
\hline\hline                 
Parameter & Initial value & Best fit \\    
\hline                        
   $M_{\rm cl,\, tot}$~[M$_{\rm HJ}$]    & 0.1                    & $\sim 2$ \\      
   $M_{\rm inter,\, tot}$~[M$_{\rm HJ}$] & 0.858\tablefootmark{a} & 0.1...0.4   \\
   $\rho_{\rm cl}$~[cm$^{-3}$]           & $2\times 10^6$         & $\geq 4\times 10^6$    \\
   $\rho_{\rm inter}$~[cm$^{-3}$]        & $6\times 10^4$         & $\geq 1\times 10^5$   \\
   $\alpha'$~[$\degr$]                   & $3$                    & 0...3  \\ 
   $z_{\rm star}$~[pc]                   & $0.3$                  & $0.3$ \\  
   $I_{\rm UV}$                          & $4.4\times 10^4$       & $\geq 6.6\times 10^4$\\    
   $d_{\rm cavity}$~[pc]                 & $0.6$                  & $0.6$ \\ 
   $d_{\rm clumps}$~[pc]                 & 0                      & 0.02...0.04\\  
   $m_{\rm l, cl}$~[M$_\sun$]            & $10^{-3}$              & $\leq 10^{-3}$ \\  
   $m_{\rm inter}$~[M$_\sun$]            & $10^{-2}$              & $10^{-2}$\\ 
\hline                                
\end{tabular}
\tablefoot{
\tablefoottext{a}{\citetalias{Hogerheijde_1995} find 
$M_{\rm inter,\, tot}=0.9$~M$_{\rm HJ}$, however, we start with 0.858~M$_{\rm HJ}$ for $\rho_{\rm inter}=6\times 10^4$~cm$^{-3}$, 
enforcing a{\changedTwo n interclump-}VFF of unity in our initial models.}
}
\end{table}

\subsection{Model assessment}
\label{Sect:ModelAssessment}

To evaluate the goodness of the fit, accounting for the simultaneous 
reproduction of the integrated line intensities and the Orion Bar 
stratification structure, we determine the $y$-position and the 
integrated intensity of the pixel with the highest integrated intensity for each transition from our simulated cuts.

The stratification is measured in terms of the $y$-offset of the 
intensity peak relative to the [C{\sc ii}] peak position: 
$\Delta y_i = y_{\rm [CII]}-y_i$ where the index $i$ refers to different 
transitions (a negative $\Delta y_i$ indicates a shift towards 
$\Theta^1$~Ori~C). Simulations and observations are compared by deriving 
the difference between the respective offsets,
\begin{equation}
\label{Eq:yrel}
y_{{\rm diff},i}=\Delta y_i - \Delta y_{{\rm obs},i}\, ,
\end{equation}
where $\Delta y_{{\rm obs},i}$ refers to the observations (by definition, 
$y_{\rm diff, CII}=0$ for [C{\sc ii}], our reference coordinate).
The relative differences between simulated and observed peak integrated 
intensities are given by 
$I_{{\rm rel},i}=I_{{\rm fit},i}/{I_{{\rm obs},i}}$. We summarise the $y_{{\rm diff},i}$ and the $I_{{\rm rel},i}$ of different 
models and transitions in ``scatter plots'' (see for example 
Figs.~\ref{Fig:scatter_intensity_1_2_3} and 
\ref{Fig:scatter_offset_1_2_3}), which yield a clear way to compare the 
models and to identify systematic behaviour. In addition, we define 
measures to evaluate the goodness of our fits. For the $y$-offsets we use 
a chi-square test, namely
\begin{equation}
\chi_{\rm off}^2 = \sum_i \left[ \frac{\Delta y_i - \Delta y_{{\rm obs}, i}}{{\rm Err}(\Delta y_{{\rm obs}, i})}\right]^2\, ,
\label{Eq:chi_off}
\end{equation}
where ${\rm Err}(\Delta y_{{\rm obs}, i})$ denotes the error of the 
offsets derived from the observations, as stated in 
Table~\ref{Tab:observations}, and the sum runs over all simulated 
transitions. A typical chi-square test, as used in 
Eq.~\ref{Eq:chi_off}, evaluates the model in terms of absolute 
(squared) differences between the observed and the fitted values. For 
the line integrated intensities we want a figure of merit which 
evaluates our models in terms of factors, for instance {\changedTwo
a simulation result that deviates from the observations} by a factor 
two {\changedTwo has the same quality as a results that is} wrong by 
a factor 1/2. Therefore, we define 
\begin{equation}
\chi_{\rm I}^2 = \sum_i \left[ \frac{{\rm Log}_{10}(I_{{\rm fit},i}) - {\rm Log}_{10}(I_{{\rm obs},i})}{0.434\ {\rm Err}(I_{{\rm obs},i})/I_{{\rm obs},i}} \right]^2\, ,
\label{Eq:chi_inten}
\end{equation}
where the errors ${\rm Err}(I_{{\rm obs},i})$ are stated in Table~\ref{Tab:observations} and 
the denominator has been derived by error propagation, i.e.
\begin{equation}
{\rm Err}({\rm Log}_{10}(I_{{\rm obs},i}))=\left|\frac{\partial {\rm Log}_{10}(I_{{\rm obs},i})}{\partial I_{{\rm obs},i}}\right|{\rm Err}( I_{{\rm obs},i})=\frac{{\rm Err}( I_{{\rm obs},i})}{{\rm ln}(10)\  I_{{\rm obs},i}}\, .
\end{equation}
When parameters are varied during the fitting process the line intensities and the $y$-offsets are usually 
affected in a very different manner. To make the effects of parameter variations
visible we state $\chi_{\rm off}^2$ and $\chi_{\rm I}^2$ separately for each simulation run. 
However, to evaluate how well the stratification pattern and the line integrated intensities are 
matched by a specific model we use the sum
\begin{equation}
\chi_{\rm tot}^2 = \chi_{\rm I}^2 + \chi_{\rm off}^2\, .
\label{Eq:chi_tot}
\end{equation}
Furthermore, the quality of a fit in the statistical sense is given by the ``reduced chi-square''
\begin{equation}
\tilde{\chi}_{\rm tot}^2 = \frac{\chi_{\rm tot}^2}{f}\, ,
\label{Eq:reduced_chi_tot}
\end{equation}
where $f=N-M$ are the degrees of freedom \citep[see][]{Press_1992}. Here, 
$N = 23$ is the number of quantities that are fitted, namely the line 
integrated intensities of 12 different transitions plus the 11 
$y$-offsets relative the [C{\sc ii}] peak position, and $M$ is the number 
parameters that can be adjusted. In this work $M$ varies between 
different series of simulations. 

The best method to derive a good fit would be to systematically explore 
the parameter space in all directions. Unfortunately, due to the large 
number of free parameters, this cannot be done within an acceptable 
amount of computing time. Hence, we choose the following approach: the 
values taken from \citetalias{Hogerheijde_1995}, as summarised in 
Sect.~\ref{Sect:parameters}, are used as an initial guess for our 
simulations. Successively ``series of models'' are run, where in each 
series at least two parameters are varied at a time. Based on the 
$\tilde{\chi}_{\rm tot}^2$~-~test the best model from each series is 
selected and the parameters are kept for the next series. If there are 
interdependences between parameters we try to vary these parameters at 
the same time.

\subsection{Simulation runs}
\label{Sect:SimulationRuns}

\begin{sidewaystable*}
\caption{\changedTwo Overview of different model set-ups.}
\label{Tab:models} 
\centering
\begin{tabular}{l c c c c c c c c c c c c c c c c c} 
\hline\hline             
Model & $M_{\rm cl, tot}$ & $\rho_{\rm cl}$ & $M_{\rm inter, tot}$ & $\rho_{\rm inter}$ & VFF\tablefootmark{a} & $\alpha'$ & $I_{\rm UV}$ & $d_{\rm cavity}$ & $z_{\rm star}$ & $m_{\rm l, cl}$ & $m_{\rm inter}$ & $d_{\rm clumps}$ & $\chi_{\rm I}^2$& $\chi_{\rm off}^2$ & $\chi_{\rm tot}^2$ & $\tilde{\chi}_{\rm tot}^2$ & $f$\tablefootmark{b} \\ 
\hline
 & [M$_{\rm HJ}$] & [cm$^{-3}$] & [M$_{\rm HJ}$] & [cm$^{-3}$] & & [$\degr$] & [$\chi_0$] & [pc] & [pc] & [M$_\sun$] & [M$_\sun$] & [pc] \\ 
\hline
   1c        & 0.1 & $1\times 10^6$ & 0.429  & $3\times 10^4$ & 1     & 3  & $4.4\times 10^4$  & 0.6 & 0.3 & $10^{-3}$ & $10^{-2}$ & 0 & 8903  & 1169  & 10072 & 504 & 20 \\ 
   1m        & 2.0 & $1\times 10^6$ & 0.429  & $3\times 10^4$ & 1     & 3  & $4.4\times 10^4$  & 0.6 & 0.3 & $10^{-3}$ & $10^{-2}$ & 0 & 1148  & 214   & 1362  & 68  & 20 \\ \hline  
   1d        & 0.1 & $2\times 10^6$ & 0.858  & $6\times 10^4$ & 1     & 3  & $4.4\times 10^4$  & 0.6 & 0.3 & $10^{-3}$ & $10^{-2}$ & 0 & 8315  & 557   & 8872  & 444 & 20 \\ 
   1i        & 0.5 & $2\times 10^6$ & 0.858  & $6\times 10^4$ & 1     & 3  & $4.4\times 10^4$  & 0.6 & 0.3 & $10^{-3}$ & $10^{-2}$ & 0 & 3642  & 356   & 3998  & 200 & 20 \\  
   1n        & 2.0 & $2\times 10^6$ & 0.858  & $6\times 10^4$ & 1     & 3  & $4.4\times 10^4$  & 0.6 & 0.3 & $10^{-3}$ & $10^{-2}$ & 0 & 1194  & 158   & 1352  & 68  & 20 \\ \hline
   1j        & 0.5 & $2\times 10^6$ & 1.43   & $10^5$         & 1     & 3  & $4.4\times 10^4$  & 0.6 & 0.3 & $10^{-3}$ & $10^{-2}$ & 0 & 3863  & 170   & 4033  & 202 & 20 \\ \hline
   1z        & 2.0 & $4\times 10^6$ & 0.0858 & $6\times 10^3$ & 1     & 3  & $4.4\times 10^4$  & 0.6 & 0.3 & $10^{-3}$ & $10^{-2}$ & 0 & 422   & 178   & 600   & 30  & 20 \\ 
   1A        & 2.0 & $4\times 10^6$ & 0.143  & $10^4$         & 1     & 3  & $4.4\times 10^4$  & 0.6 & 0.3 & $10^{-3}$ & $10^{-2}$ & 0 & 463   & 134   & 597   & 30  & 20 \\
   1B        & 2.0 & $4\times 10^6$ & 0.429  & $3\times 10^4$ & 1     & 3  & $4.4\times 10^4$  & 0.6 & 0.3 & $10^{-3}$ & $10^{-2}$ & 0 & 711   & 214   & 925   & 46  & 20 \\
   1C        & 2.0 & $4\times 10^6$ & 0.858  & $6\times 10^4$ & 1     & 3  & $4.4\times 10^4$  & 0.6 & 0.3 & $10^{-3}$ & $10^{-2}$ & 0 & 1009  & 242   & 1251  & 63  & 20 \\ 
   1D        & 2.0 & $4\times 10^6$ & 1.43   & $10^5$         & 1     & 3  & $4.4\times 10^4$  & 0.6 & 0.3 & $10^{-3}$ & $10^{-2}$ & 0 & 1248  & 151   & 1399  & 70  & 20 \\ \hline   
   1E        & 2.0 & $4\times 10^6$ & 0      & -              & -     & 3  & $4.4\times 10^4$  & 0.6 & 0.3 & $10^{-3}$ & $10^{-2}$ & 0 & 323.3 & 125.7 & 449.0   & 22.5  & 20 \\ 
   1bb & 2.0 & $4\times 10^6$ & 0.05   & $10^5$         & 0.035 & 3  & $4.4\times 10^4$  & 0.6 & 0.3 & $10^{-3}$ & $10^{-2}$ & 0 & 291.6 & 131.4 & 423.0   & 22.3  & 19\\ 
   1P  & 2.0 & $4\times 10^6$ & 0.1    & $10^5$         & 0.07  & 3  & $4.4\times 10^4$  & 0.6 & 0.3 & $10^{-3}$ & $10^{-2}$ & 0 & 279.3 & 142.3 & 421.6   & 22.2  & 19\\  
   1aa & 2.0 & $4\times 10^6$ & 0.2    & $10^5$         & 0.14  & 3  & $4.4\times 10^4$  & 0.6 & 0.3 & $10^{-3}$ & $10^{-2}$ & 0 & 280   & 174   & 454   & 24  & 19\\   \hline
   2b(\_ext)\tablefootmark{c} & 2.0 & $4\times 10^6$ & 0.1    & $10^5$         & 0.07  & 3  & $6.6\times 10^4$  & 0.6 & 0.3 & $10^{-3}$ & $10^{-2}$ & 0 & 271(266) & 131(136) & 402(402) & 24(24) & 17 \\    
   2d  & 2.0 & $4\times 10^6$ & 0.1    & $10^5$         & 0.07  & 0  & $5.5\times 10^4$  & 0.6 & 0.3 & $10^{-3}$ & $10^{-2}$ & 0 & 274.3 & 60.5  & 334.8  & 19.7 & 17 \\     
   2e  & 2.0 & $4\times 10^6$ & 0.1    & $10^5$         & 0.07  & 0  & $6.6\times 10^4$  & 0.6 & 0.3 & $10^{-3}$ & $10^{-2}$ & 0 & 272.7 & 60.5  & 333.2  & 19.6 & 17 \\    
   2h  & 2.0 & $4\times 10^6$ & 0.1    & $10^5$         & 0.07  & 7  & $6.6\times 10^4$  & 0.6 & 0.3 & $10^{-3}$ & $10^{-2}$ & 0 & 272   & 130   & 402  & 24 & 17 \\ 
   2i  & 2.0 & $4\times 10^6$ & 0.1    & $10^5$         & 0.07  & 15 & $6.6\times 10^4$  & 0.6 & 0.3 & $10^{-3}$ & $10^{-2}$ & 0 & 303   & 318   & 621  & 37 & 17 \\ \hline   
   3a  & 2.0 & $4\times 10^6$ & 0.1    & $10^5$         & 0.07  & 0  & $6.6\times 10^4$  & 0.1 & 0.1 & $10^{-3}$ & $10^{-2}$ & 0 & 856   & 1702  & 2558 & 171 & 15  \\      
   3c  & 2.0 & $4\times 10^6$ & 0.1    & $10^5$         & 0.07  & 0  & $6.6\times 10^4$  & 0.3 & 0.2 & $10^{-3}$ & $10^{-2}$ & 0 & 390   & 61    & 451  & 30 & 15\\
   3n  & 4.0 & $4\times 10^6$ & 0.1    & $10^5$         & 0.07  & 0  & $6.6\times 10^4$  & 0.3 & 0.3 & $10^{-3}$ & $10^{-2}$ & 0 & 250   & 111   & 361  & 24 & 15 \\ 
   3r  & 3.0 & $4\times 10^6$ & 0.1    & $10^5$         & 0.07  & 0  & $6.6\times 10^4$  & 0.4 & 0.4 & $10^{-3}$ & $10^{-2}$ & 0 & 252   & 93    & 345  & 23 & 15 \\ \hline    
   4a  & 2.0 & $4\times 10^6$ & 0.1    & $10^5$         & 0.07  & 0  & $6.6\times 10^4$  & 0.6 & 0.3 & $10^{-3}$ & $10^{-3}$ & 0 & 277   & 78    & 355  & 27 & 13 \\    
   4b  & 2.0 & $4\times 10^6$ & 0.1    & $10^5$         & 0.07  & 0  & $6.6\times 10^4$  & 0.6 & 0.3 & $10^{-2}$ & $10^{-2}$ & 0 & 325   & 131   & 456  & 35 & 13 \\    
   4d  & 2.0 & $4\times 10^6$ & 0.1    & $10^5$         & 0.07  & 0  & $6.6\times 10^4$  & 0.6 & 0.3 & $10^{0}$  & $10^{-2}$ & 0 & 606   & 142   & 748  & 58 & 13 \\ 
   4f  & 2.0 & $4\times 10^6$ & 0.1    & $10^5$         & 0.07  & 0  & $6.6\times 10^4$  & 0.6 & 0.3 & $10^{-3}$ & $10^{-1}$ & 0 & 280   & 126    & 406  & 31 & 13 \\ \hline
   6b  & 2.0 & $4\times 10^6$ & 0.2    & $1\times 10^5$ & 0.14  & 3  & $6.6\times 10^4$  & 0.6 & 0.3 & $10^{-3}$ & $10^{-2}$ & 0.02 & 272   & 83    & 355  & 30 & 12\\     
   5h  & 2.0 & $4\times 10^6$ & 0.2    & $1\times 10^5$ & 0.14  & 0  & $6.6\times 10^4$  & 0.6 & 0.3 & $10^{-3}$ & $10^{-2}$ & 0.02 & 272   & 78    & 350  & 29 & 12\\    
   6a  & 2.0 & $4\times 10^6$ & 0.1    & $1\times 10^5$ & 0.07  & 3  & $6.6\times 10^4$  & 0.6 & 0.3 & $10^{-3}$ & $10^{-2}$ & 0.02 & 273   & 60    & 333  & 28 & 12\\  
   6j  & 2.0 & $4\times 10^6$ & 0.2    & $1\times 10^5$ & 0.14  & 3  & $6.6\times 10^4$  & 0.6 & 0.3 & $10^{-3}$ & $10^{-2}$ & 0.04 & 285   & 31    & 316  & 26 & 12\\    
   6k  & 2.0 & $4\times 10^6$ & 0.3    & $1\times 10^5$ & 0.21  & 3  & $6.6\times 10^4$  & 0.6 & 0.3 & $10^{-3}$ & $10^{-2}$ & 0.04 & 336   & 17    & 353  & 29 & 12\\  
   6l  & 2.0 & $4\times 10^6$ & 0.4    & $1\times 10^5$ & 0.28  & 3  & $6.6\times 10^4$  & 0.6 & 0.3 & $10^{-3}$ & $10^{-2}$ & 0.04 & 429   & 15    & 444  & 37 & 12\\      
\hline
\end{tabular}
\tablefoot{
\tablefoottext{a}{\changed Volume filling factor of the interclump medium, derived from $M_{\rm inter, tot}$ and $\rho_{\rm inter}$ (see Eq.~\ref{Eq:inter_coupled}).}
\tablefoottext{b}{\changed Degrees of freedom used for the calculation of $\tilde{\chi}_{\rm tot}^2$ (see Sect.~\ref{Sect:ModelAssessment}).}
\tablefoottext{c}{\changed Numbers in brackets refer to the model extended to $z=-100$ (see Sect.~\ref{Sect:C18O}).}
}
\end{sidewaystable*}

\subsubsection{\changed Ensemble-averaged densities and masses per voxel}
\label{Sect:HomMassesDensities}

\begin{figure*}
\centering
   \includegraphics[width=17cm]{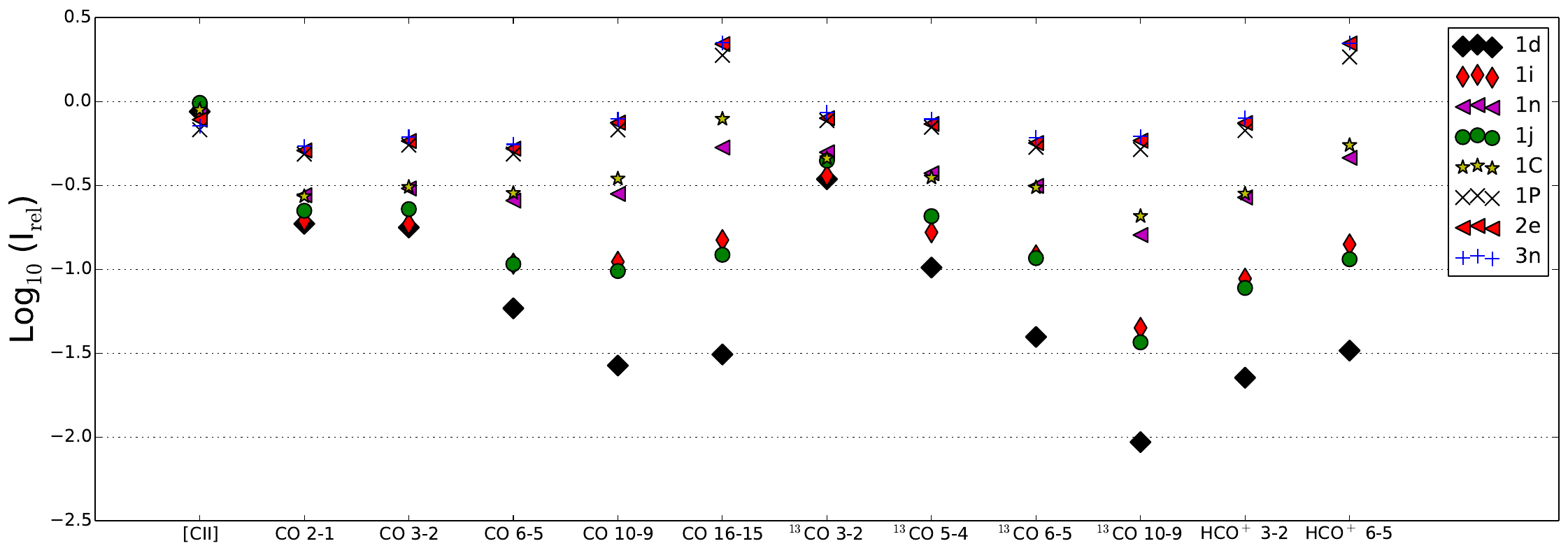}
     \caption{
     Scatter plot of the line integrated intensities for selected models from 
     Sects.~\ref{Sect:HomMassesDensities}, \ref{Sect:AlphaIUV} and \ref{Sect:DcavityZstar}. For each 
     transition the ratio between simulated and observed line integrated intensity at the respective 
     peak position, $I_{{\rm rel},i}=I_{{\rm fit},i}/I_{{\rm obs},i}$, is plotted on a logarithmic 
     scale. The different transition are indicated on the abscissa and different symbols mark the 
     different models.
     }
   \label{Fig:scatter_intensity_1_2_3}
\end{figure*}

\begin{figure*}
\centering
   \includegraphics[width=17cm]{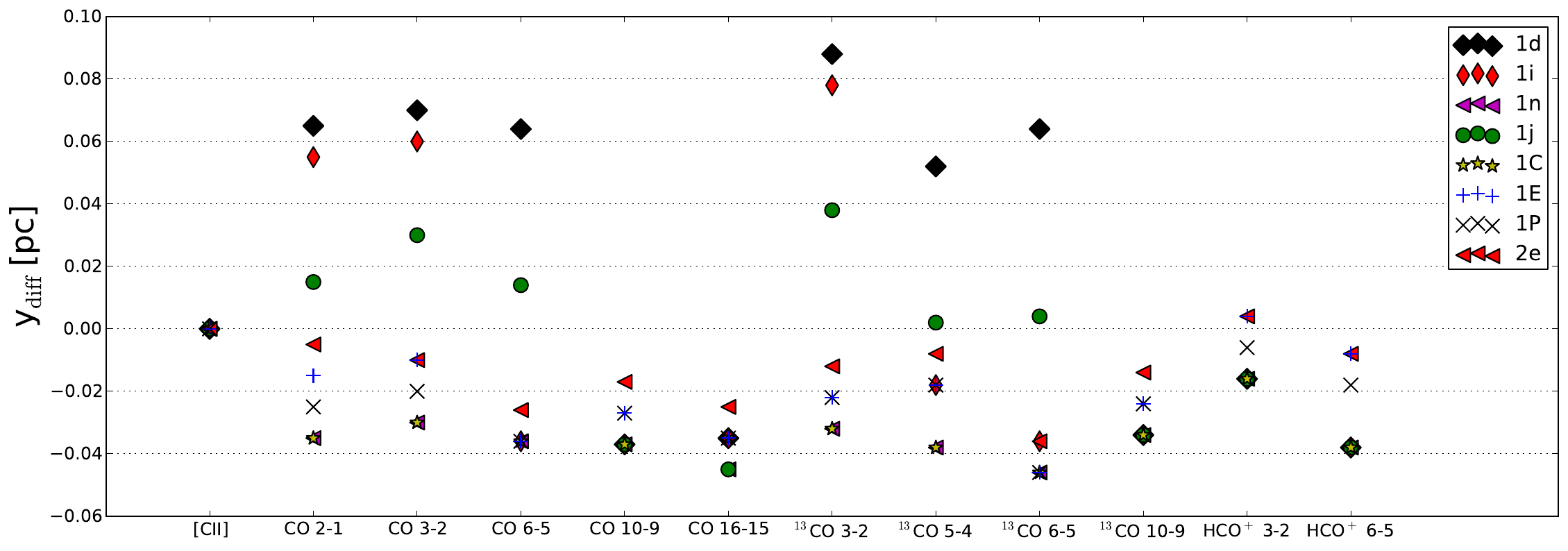}
     \caption{
     Scatter plot of the $y$-offsets of selected models from Sects.~\ref{Sect:HomMassesDensities}, 
     \ref{Sect:AlphaIUV} and \ref{Sect:DcavityZstar}. For each transitions 
     the difference between the offsets of the simulated and the observed peak position, 
     as defined in Eq.~\ref{Eq:yrel}, is plotted. All offsets are relative to the [C{\sc ii}] 
     peak position, hence, for the reference [C{\sc ii}] transition $y_{\rm diff}$ is always zero.
     A negative offset indicates that the simulated emission peak is shifted too far into the direction 
     of $\Theta^1$~Ori~C, a positive offset indicates that the emission appears too deep in the cloud.
     The different transitions are indicated on the abscissa and different symbols mark different models.
     }
     \label{Fig:scatter_offset_1_2_3}
\end{figure*}

Each voxel within a compound is filled by two ensembles, one 
representing the dense clumps and one representing the interclump 
medium. We start our simulation runs investigating models where 
these ensembles are the same within each voxel 
(i.e.~$d_{\rm clumps}=0$, see Sect.~\ref{Sect:parameters}) and refer 
to these models as ``homogeneous''. Examples of inhomogeneous models 
are given in Sect.~\ref{Sect:Inhomogeneous}.

The total mass of the clump and interclump medium contained in each 
voxel, as well as the related ensemble-averaged densities, have a 
strong influence on the simulation outcome. Especially in the 
homogeneous models, these parameters are interdependent: both 
components can contribute\footnote{In model 1c the dense clumps do 
only account for 1.4\% of the total FUV attenuation. In model 1m, 
where the total mass of the ensemble of dense clumps has been 
increased, 29\% of the total FUV attenuation is due to this 
ensemble.} to the FUV attenuation and hence control the line 
intensities emitted by both components. Therefore, we vary these 
parameters ($M_{\rm cl, tot}$, $\rho_{\rm cl}$, $M_{\rm inter, tot}$ 
and $\rho_{\rm inter}$) first. {\changedTwo For all other 
parameters we use our initial guess (Table~\ref{Tab:Fit_parameter}).
An overview of different model set-ups} is given in 
Table~\ref{Tab:models}. 

{\changedTwo In our first series of models} (represented by models 
1c to 1D in Table~\ref{Tab:models}) we couple $\rho_{\rm inter}$ to 
$M_{\rm inter, tot}$, enforcing a VFF of unity for the interclump 
medium. In these runs we have tested ensemble-averaged densities of $10^6$, $2\times 10^6$ and 
$4\times 10^6$~cm$^{-3}$ for the dense clumps\footnote{An ensemble-averaged 
density of $4\times 10^6$~cm$^{-3}$ combined with the four 
mass points implies that the smallest clumps have densities of about $10^7$~cm$^{-3}$ 
(see Eq.~\ref{Eq:rho_cl}). Higher densities are not possible with the current 
input grid (see Table~\ref{Tab:Grid_parameter}).} combined with total ensemble masses 
between 0.1 and 2~M$_{\rm HJ}$. As discussed in Sect.~\ref{Sect:C18O} models with 
$M_{\rm cl, tot}>2$~M$_{\rm HJ}$ (combined with $d_{\rm cavity}=0.6$~pc) have been 
excluded. For the interclump medium we have tested densities between $6\times 10^3$ 
and $10^5$~cm$^{-3}$, which implies total masses per voxel between 0.0858 and 1.43~M$_{\rm HJ}$. 
For these models the degree of freedom (see Sect.~\ref{Sect:ModelAssessment}) is
$f=23-3=20$ if a VFF of unity is enforced and $f=19$ if $M_{\rm inter, tot}$ and 
$\rho_{\rm inter}$ are treated independently.

In models 1c and 1m or similarly in models 1d, 1i and 1n, the total 
mass of the dense clumps has been increased from 0.1 to 2~M$_{\rm HJ}$ 
while all other parameters remain unchanged. From all tested models the 
parameters of model 1d are closest to the values from 
\citetalias{Hogerheijde_1995}. Figure~\ref{Fig:scatter_intensity_1_2_3} 
gives an overview over the ratios between simulated and observed line 
integrated intensities for selected transitions. We find that the 
``\citetalias{Hogerheijde_1995}-model'' 1d does not reproduce any of 
the fitted line integrated intensities, except for [C{\sc ii}]. 
However, all other intensities
are too low, often by orders of magnitude. Model 1i uses the same parameters, 
but with M$_{\rm cl, tot}$ increased to 0.5~M$_{\rm HJ}$, which increases the line 
intensities of the species that are (dominantly) emitted by the dense clumps,
namely the high-$J$ CO isotopologues and the HCO$^+$ transitions. Still, the 
resulting line intensities are too low, except for [C{\sc ii}]. In model 1n,
which uses M$_{\rm cl, tot}=2$~M$_{\rm HJ}$, the line integrated intensities 
of most transitions are still too low, but the fit does significantly improve 
compared to the models discussed above.

Model 1j uses the same parameters as model {\changedTwo 1i} except for 
the density (and therefore also the total mass) of the interclump 
medium, which has been increased. In
Fig.~\ref{Fig:scatter_intensity_1_2_3} one can see how the line 
integrated intensities of the transitions which are (at least 
partially) emitted by the interclump medium, namely [C{\sc ii}] and the 
low-$J$ CO isotopologues, increase. The line intensities of the other 
transitions \emph{decrease} due to the stronger FUV attenuation in the 
cloud. 

Overall, we find that increasing M$_{\rm cl, tot}$ improves the 
quality of our fit (lower $\tilde{\chi}_{\rm tot}^2$). Furthermore, 
from comparing for example model 1m with 1B (or 1n and 1C; see 
Table~\ref{Tab:models} {\changedTwo or 
Fig.~\ref{Fig:scatter_intensity_1_2_3}}) we find that a higher 
ensemble-averaged density of $\rho_{\rm cl}=4\times 10^6$~cm$^{-3}$ 
provides lower $\chi_{\rm I}^2$ and, {\changedTwo although 
$\chi_{\rm off}^2$ can increase, lower} 
$\tilde{\chi}_{\rm tot}^2$. 

The spatial offsets of the different peak positions do mainly depend 
on the FUV attenuation in the cloud and on the peak position of the
[C{\sc ii}] line which is the reference for all other transitions. 
Figure~\ref{Fig:scatter_offset_1_2_3} gives an overview over the 
$y_{{\rm diff},i}$ (see Eq.~\ref{Eq:yrel}) of the models that have 
already been included in Fig.~\ref{Fig:scatter_intensity_1_2_3}.
For models 1d and 1i we find that the emission peaks of the CO~$2-1$ 
and $^{12/13}$CO~$3-2$ transitions (for model 1d also of the 
$^{13}$CO~$5-4$ and $^{12/13}$CO~$6-5$ transitions) are shifted too 
far into the cloud by about five to nine pixels (0.05 to 0.09~pc). 
For model 1c (not shown, but note the high $\chi^2_{\rm off}$ for 
this model) the CO~$2-1$ and the $^{12/13}$CO~$3-2$ emission peaks 
are shifted even further into the cloud, they appear about 20~pixel 
behind the [C{\sc ii}] emission peak. Based on these transitions and
for the current set-up, we have to conclude that the FUV attenuation 
in the cloud is significantly too weak. However, for most other 
transitions, the offsets are found to be too small and hence a deeper 
FUV penetration would be necessary to increase their $y$-offsets. 
Model 1j with M$_{\rm cl, tot}=0.5$ and M$_{\rm inter, tot}=1.43$ 
shows a similar, but less pronounced behaviour compared to models 1d 
and 1i (see Fig.~\ref{Fig:scatter_offset_1_2_3}). We conclude that a 
fit of the stratification pattern based on the set-ups presented in 
this section and an interclump-VFF=1 is not possible.

\subsubsection{Reduction of the interclump medium}

For models 1E to 1aa the constraint that the 
{\changedTwo interclump-}VFF needs to be unity 
has been dropped. Models 1z to 1aa in Table~\ref{Tab:models} use 
M$_{\rm cl, tot}=2$ and $\rho_{\rm cl}=4\times 10^6$~cm$^{-3}$.
Models 1z to 1D show how the composition of the interclump medium 
affects the fit in terms of the chi-square tests. We find that the 
fit improves in terms of $\chi_{\rm I}^2$ and 
$\tilde{\chi}_{\rm tot}^2$ if the amount and VFF of the interclump 
medium is reduced, allowing for a deeper FUV penetration into the 
cloud. The $\chi^2_{\rm off}$ are somewhat more random, which is 
probably due to changing $y$-offsets of the [C{\sc ii}] reference 
position (see below). {\changedTwo The models with a  
${\rm interclump-VFF}\ll 1$, i.e.~models 1E, 1bb and 1P, provide 
similar 
$\tilde{\chi}^2$-values and we cannot discriminate between these 
models based on our set-up. However, as we need to decide with which 
model we want to continue our simulations in the next section, we 
provide the $\tilde{\chi}_{\rm tot}^2$ with a higher precision than 
integer values in Table~\ref{Tab:models}. Based on these results, 
model 1P with $\tilde{\chi}_{\rm tot}^2=22.2$} provides the best fit 
of all models discussed so far. {\changedTwo It} uses 
$\rho_{\rm inter}=10^5$~cm$^{-3}$ and 
$M_{\rm inter, tot}=0.1$~M$_{\rm HJ}$, resulting in an VFF of 0.07 
for the interclump medium. We note that for our best fitting models 
the intensities of the CO~$16-15$ and HCO$^+$~$6-5$ transitions are 
now somewhat too high compared to the observations (see model 1P in 
Fig.~\ref{Fig:scatter_intensity_1_2_3}). 

In all models the simulated $y$-offsets of the line peak positions 
of (nearly) all transitions tend to be too 
low. The best fit in terms of the stratification pattern 
($\chi_{\rm off}^2=126$) is obtained by model 1E (see 
Table~\ref{Tab:models}). However, for this model the simulated 
$y$-offsets relative to the [C{\sc ii}] emission peak are too small for 
all simulated tracers (except HCO$^+$~$3-2$, see 
Fig.~\ref{Fig:scatter_offset_1_2_3}) and some transitions 
($^{12/13}$CO~$6-5$ and CO~$16-15$) appear at the same offset as 
[C{\sc ii}]. Furthermore, in model 1E the interclump medium has been 
completely removed, while the existence of some interclump medium 
has been deduced from several observations. Overall, model 1P provides 
the best combined fit of intensity and stratification, but when we 
actually look at the stratification pattern it becomes clear that the 
fit of the pattern is not satisfactory. Good fits of the 
stratification pattern are only found if inhomogeneous models are used (see 
Sect.~\ref{Sect:Inhomogeneous}).

\subsubsection{$\alpha'$ and $I_{\rm UV}$}
\label{Sect:AlphaIUV}

{\changed
In a second series of simulations the effects of varying the
inclination angle $\alpha'$ and the FUV flux at the cloud surface have 
been investigated. We have tested inclination angles of 
$\alpha'=0\degr$, $3\degr$, $7\degr$ and $15\degr$, and have 
simultaneously increased the FUV flux by 25\% and by 50\% compared to 
the original \citetalias{Hogerheijde_1995} value. All other parameters 
in this second series have been adopted from the best model from the 
previous section, namely model 1P. As we are optimising two additional 
parameters in this section we use $f=23-6=17$ degrees of freedom for 
the calculation of the $\tilde{\chi}_{\rm tot}^2$. 

For all tested $\alpha'$ {\changedTwo and in the tested parameter 
range}, the fit of the line integrated intensities 
{\changedTwo depends only weakly on $I_{\rm UV}$ (see for example 
the comparison between models 1P and 2b, where the FUV flux at the 
PDR surface has been increased by 50\%, in Table~\ref{Tab:models}).}  
{\changedTwo Based on the $\chi_{\rm I}^2$,} the best fit of the line 
integrated intensities is provided by models 2b, 2e and 2h 
($\chi_{\rm I}^2=271$, 273 and 272 respectively), which use 
$I_{\rm UV}=6.6\times 10^4$~$\chi_0$ combined with $\alpha'=3\degr$, 
$0\degr$ or $7\degr$. These models are found to fit the line integrated 
intensities of all transition within a factor of about two or better, 
for model 2h we find $I_{\rm rel, CO~6-5}=0.49$ and 
$I_{\rm rel, HCO^+~6-5}=2.04$ with the $I_{\rm rel}$ of all other transitions lying between these values. For models 2b and 2e the 
$I_{\rm rel}$ are slightly higher (see also model 2e in 
Fig.~\ref{Fig:scatter_intensity_1_2_3}). 
{\changedTwo As} increasing $I_{\rm UV}$ mainly increases the line 
intensities of the "outlier transitions" (see 
Sect.~\ref{Sect:Discussion_intensities}), we have not tested models with 
$I_{\rm UV}$ higher than $6.6\times 10^4$~$\chi_0$.

Changing the inclination angle has two effects. In general, 
choosing a small 
$\alpha'$ provides more excited column along one line of sight through the 
compound and hence increases the line integrated intensities of the optically 
thin transitions. However, increasing $\alpha'$ broadens the emission peak,
which can also lead to an increase of line integrated intensities after the beam 
convolution. Overall, we find that changing $\alpha'$ from $3\degr$ to $0\degr$
or $7\degr$ has an negligible effect on the line integrated intensities, however, 
for $\alpha'=15\degr$ the fit becomes worse (see for example model 2i in 
Table~\ref{Tab:models}).

A more significant effect is that the fit of the stratification pattern 
is found to be best for $\alpha'=0\degr$, independent from the choice 
of $I_{\rm UV}$. For larger inclination angles, especially for 
$\alpha'=15\degr$, $\chi_{\rm off}^2$ is found to increase. The lowest 
$\chi_{\rm off}^2=61$ is provided by model 2e, which fits the 
$y$-offsets of two transitions (CO~$2-1$ and HCO$^+$~$3-2$) within the 
observational uncertainty, however, for all other transitions the 
simulated $y$-offsets relative to the [C{\sc ii}] peak are too small. 
Model 2e is also included in the scatter plots, 
Figs.~\ref{Fig:scatter_intensity_1_2_3} and 
\ref{Fig:scatter_offset_1_2_3}.

\subsubsection{$d_{\rm cavity}$ and $z_{\rm star}$}
\label{Sect:DcavityZstar}

In a third simulation run two geometrical parameters have been varied, 
the depth of the cavity that defines the length of the line of sight 
through the bar ($d_{\rm cavity}$) and the $z$-position of the 
illuminating source ($z_{\rm star}$). These parameters are partly 
interdependent: the column of material that is excited in the bar 
depends on the position of the star relative to the cavity wall. We 
vary both parameters, $d_{\rm cavity}$ and $z_{\rm star}$, between 0.1 
and 0.6~pc, which includes the range of values found in literature (see 
Fig.~\ref{Fig:Ori_geometry}). In addition, we have investigated two 
models where the star is lying outside of the cavity
($d_{\rm cavity}=0.3$~pc and $z_{\rm star}=0.4$~pc, and 
$d_{\rm cavity}=0.6$~pc and $z_{\rm star}=0.7$~pc). In 
Sect.~\ref{Sect:C18O} we have derived an upper limit for the total 
molecular column density along a line of sight through a model 
compound, which was found to correspond to 
$M_{\rm cl, tot}=2$~M$_{\rm HJ}$ for $d_{\rm cavity}=0.6$~pc. Hence, 
for the models where $d_{\rm cavity}$ has been reduced, we also test 
models where $M_{\rm cl, tot}$ has been increased to provide the same 
upper column density limit (see for example models 3r and 3n in 
Table~\ref{Tab:models}}). All other parameters used in this section 
have been adopted from model 2e. For the degree of freedom we use 
$f=23-8=15$ as eight different parameters are adjusted by now.

Briefly, the result of this simulation run is that model 2e, i.e.~the deep 
cavity ($d_{\rm cavity}=0.6$~pc) with the star at half-height 
($z_{\rm star}=0.3$~pc) is already the best configuration. Model 3a (see 
Table~\ref{Tab:models}) with the very shallow cavity ($d_{\rm cavity}=0.1$ 
and $z_{\rm star}=0.1$~pc) can be excluded, because in this model the emission 
of some transitions from the back of the cavity (``below'' the illuminating 
source) is about as strong as the emission from the bar itself. Most 
extremely for HCO$^+$~$6-5$ the emission at $y\gtrsim 20$ of 
the simulated cut is significantly stronger than the emission from the bar, 
causing the high $\chi_{\rm off}^2$ of this model. 

For all models where the total column density of the ensembles of dense 
clumps is reduced compared to model 2e, the simulated line intensities 
decrease for all transitions, increasing the $\chi_{\rm I}^2$ and hence 
decreasing the quality of the fit (see for example model 3c). On the 
other side, for models 3n ($d_{\rm cavity}=0.3$~pc, $z_{\rm star}=0.3$~pc 
and $M_{\rm cl, tot}=4$~M$_{\rm HJ}$) and 3r ($d_{\rm cavity}=0.4$~pc, 
$z_{\rm star}=0.4$~pc and $M_{\rm cl, tot}=3$~M$_{\rm HJ}$), which have 
the same column density of the dense clumps as model 2e, the fit of 
the line integrated intensities improves slightly compared to model 2e. 
This improvement stems from two effects. Placing the illuminating source 
close to the outer edge of the cavity puts the hottest material in the 
PDR closest to the observer. Therefore, after the radiative transfer, the 
line integrated intensities tend to be increased. Second, if the same 
amount of material is comprised in a shorter bar, the material is -- on 
average -- closer to the illuminating source, which increases the line 
integrated intensities (see Sect.~\ref{Sect:AlphaIUV}). Therefore, the 
$\chi_{\rm I}^2$ of a model with the same column density as models 3n and 
3r, but with $d_{\rm cavity}=0.6$~pc and $z_{\rm star}=0.6$, is slightly 
higher. However, as model 2e provides a better fit of the stratification 
pattern, is still has the lowest $\chi_{\rm tot}^2$ and 
$\tilde{\chi}_{\rm tot}^2$.

\subsubsection{$m_{\rm l, cl}$ \changed{and} $m_{\rm inter}$}
\label{Sect:MassPoints}

\begin{figure*}
\centering
   \includegraphics[width=17cm]{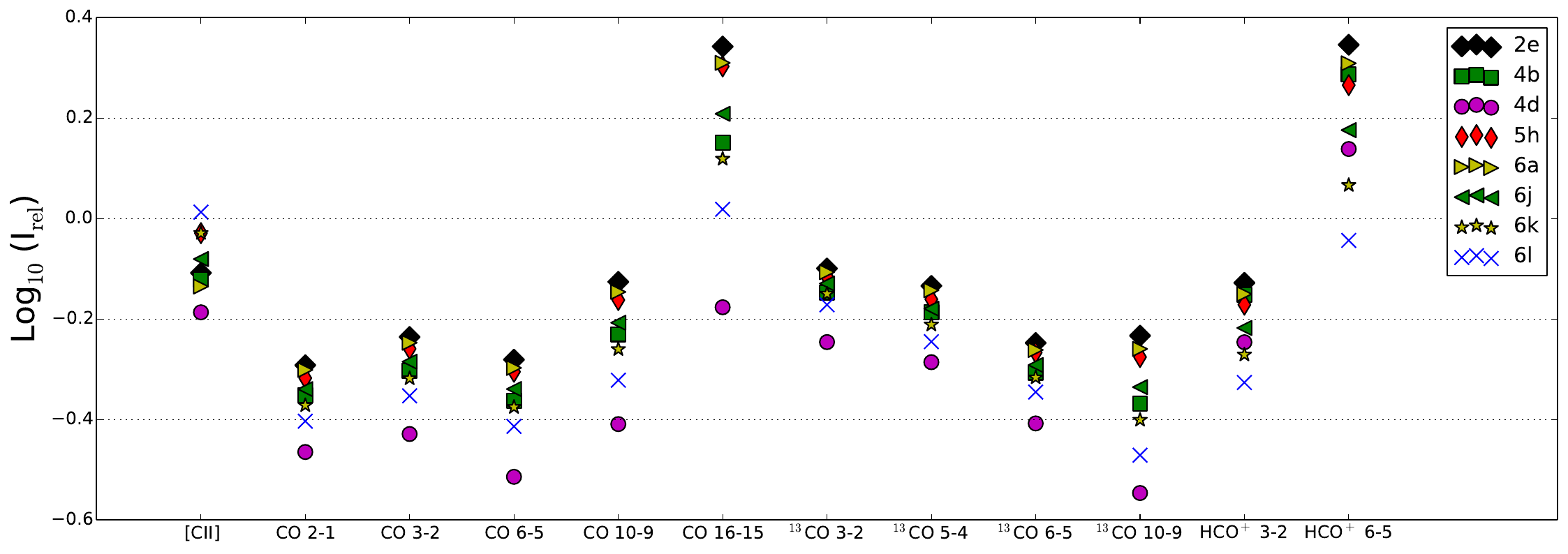}
     \caption{\changed 
     Same as Fig.~\ref{Fig:scatter_intensity_1_2_3}, plotted for selected models from 
     Sects.~\ref{Sect:MassPoints} and \ref{Sect:Inhomogeneous}. In addition, model 2e 
     from Sect.~\ref{Sect:AlphaIUV} is given for comparison.
     }
   \label{Fig:scatter_intensity_4_5_6}
\end{figure*}

\begin{figure*}
\centering
   \includegraphics[width=17cm]{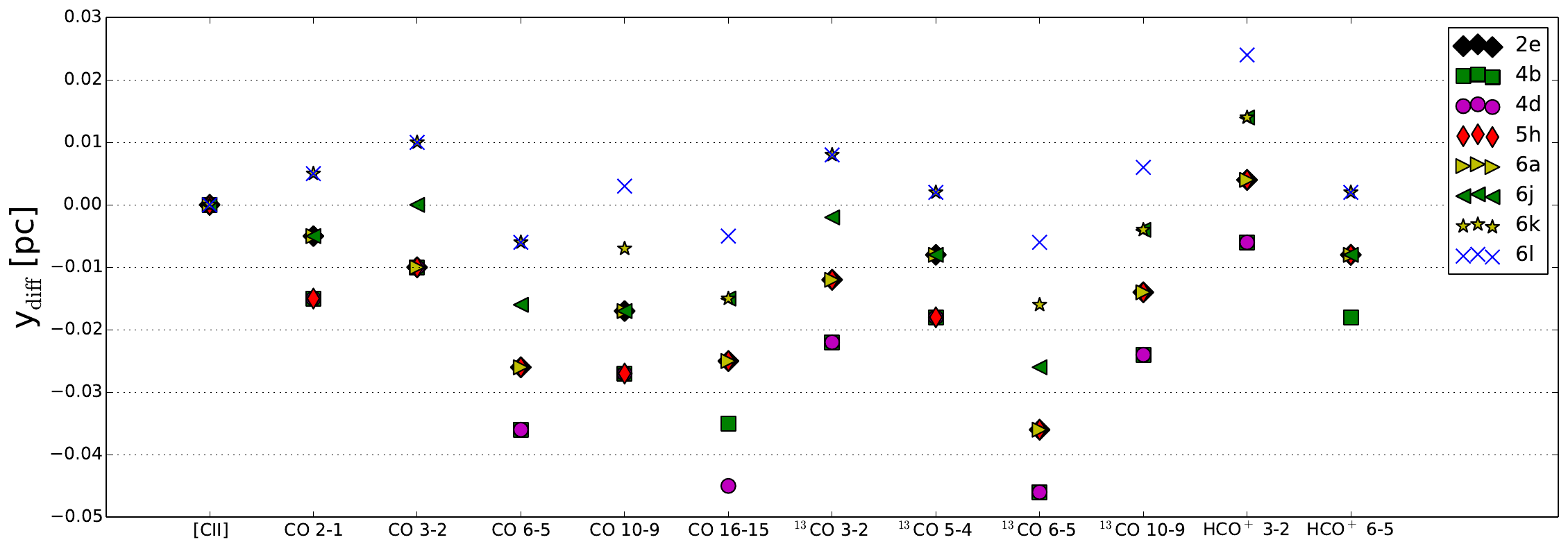}
     \caption{\changed
     Same as Fig.~\ref{Fig:scatter_offset_1_2_3}, plotted for selected models from 
     Sects.~\ref{Sect:MassPoints} and \ref{Sect:Inhomogeneous}. In addition, model 2e 
     from Sect.~\ref{Sect:AlphaIUV} is given for comparison.
     }
     \label{Fig:scatter_offset_4_5_6}
\end{figure*}

As discussed in Sect.~\ref{Sect:OriModel}, all models presented so far
have used an ensemble with four mass points, $\lbrack 10^{-3}, 10^{-2},
10^{-1}, 10^0\rbrack$~M$_\sun$ (i.e.~$m_{\rm l, cl}=10^{-3}$~M$_\sun$), 
for the ensemble of dense clumps. Only the upper limit of one~M$_\sun$ 
can be inferred from observations. Furthermore, the interclump medium was
represented by identical clumps of $10^{-2}$~M$_\sun$. In this section,
we test the impact of these choices. For the lower cut-off mass of the 
ensemble of dense clumps we test $m_{\rm l, cl}=10^{-2}$~M$_\sun$ and 
$m_{\rm l, cl}=10^0$~M$_\sun$ in addition to the initial value, where the 
second choice implies that the ensemble only contains clumps with one
M$_\sun$. Each of these choices is combined with $m_{\rm inter}=10^{-3}$,
$10^{-2}$, $10^{-1}$ or $10^{0}$~M$_\sun$ for the clumps representing
the interclump medium. For the degrees of freedom we use $f=23-10=13$ as
ten different parameters have been adjusted by now.

Independent of the choice of $m_{\rm inter}$, we find that the models
that use $m_{\rm l, cl}=10^{-3}$~M$_\sun$ provide the best fits. For
example models 2e, 4b and 4d (see Table~\ref{Tab:models}) all use 
$m_{\rm inter}=10^{-2}$~M$_\sun$, but different $m_{\rm l, cl}$. 
The $I_{\rm rel}$ of the different transitions of these models are shown 
in the scatter plot Fig.~\ref{Fig:scatter_intensity_4_5_6}. We can see 
how the line integrated intensities of the different transitions
systematically decrease when $m_{\rm l, cl}$ increases, especially 
CO~$16-15$ is affected. As the line integrated intensities of all 
transitions except for CO~$16-15$ and HCO$^+$~$6-5$ tend to be too low, 
$\chi_{\rm I}^2$ increases with increasing $m_{\rm l, cl}$. 

Varying the mass of the individual clumps of the interclump medium
(while keeping the total mass $M_{\rm inter, tot}$ fixed) changes the FUV 
attenuation and hence has an impact on the line integrated intensities and 
on the stratification pattern. For example for $\rho_{\rm inter}=10^5$~cm$^{-3}$ 
and $M_{\rm inter}=0.1$~M$_{\rm HJ}$ used in this series, decreasing $m_{\rm inter}$ 
from $10^{-2}$ to $10^{-3}$~M$_\sun$ increases the FUV attenuation per voxel by 
a factor 1.7, due to the more even distribution of the ISM. 
The line emission of the interclump medium also depends on the size of the 
individual clumps. Model 4f shows that, if we we increase the size of 
the clumps in the interclump medium, the lower emission from the
interclump gas shifts the peak of the [C{\sc ii}] line, strongly
reducing the observable stratification. 
The optimal choice of $m_{\rm inter}$ 
depends on $m_{\rm l, cl}$, however, for the models that use 
$m_{\rm l, cl}=10^{-3}$~M$_\sun$ and hence provide the best fits, 
the further result clearly favour  
$m_{\rm inter}=10^{-2}$~M$_\sun$. Overall, we find that model 2e, with 
$m_{\rm l, cl}=10^{-3}$~M$_\sun$ and $m_{\rm inter}=10^{-2}$~M$_\sun$ 
provides the best fit in terms of stratification and line integrated 
intensities.

\subsubsection{\changed Inhomogeneous models}
\label{Sect:Inhomogeneous}

{\changed 
Based on the homogeneous models used so far, the line integrated 
intensities can be fitted within a factor of about two for all 
simulated transitions. However, the homogeneous models fail in 
reproducing the observed stratification pattern (see discussion of 
models 2h/2e in Sect.~\ref{Sect:AlphaIUV}). Furthermore, different 
authors (\citealt{Parmar1991}; \citetalias{Hogerheijde_1995}; 
\citealt{YoungOwl_2000}) have proposed that the ISM in the Orion Bar is 
not uniformly distributed. Instead they present observations and models, 
which suggest that the inset of dense material, embedded in the thinner 
interclump medium, is only found at about $10\arcsec-20\arcsec$ into 
the cloud. 

Consequently, we have tested series of models incorporating such a 
step. We limit ourselves to the idealised density profile that is 
produced if the parameter $d_{\rm clumps}$ (see 
Sect.~\ref{Sect:parameters}) is chosen to be larger than zero. Note, 
that for these inhomogeneous models the FUV attenuation close to the PDR 
surface (as defined by $d_{\rm clumps}$) only depends on the 
composition of the interclump medium. The dense clumps only contribute 
deep in the cloud, hence the impact of varying $\rho_{\rm cl}$ and 
$M_{\rm cl, tot}$ becomes smaller.

We have {\changedTwo varied} the parameter $d_{\rm clumps}$ to be 
equal to 0.02, 0.03 and 0.04~pc, as suggested by the observations. As 
the FUV attenuation within the cloud does significantly change 
compared to the 
homogeneous models when the dense clumps are removed, we have 
simultaneously varied the composition of the interclump medium a 
second time. Therefore, we have tried values of $\rho_{\rm inter}$ 
between $6\times 10^3$ and $10^5$~cm$^{-3}$ and $M_{\rm inter}$ 
between 0.05 and 0.5~M$_{\rm HJ}$. Furthermore, previous simulation 
runs have shown that for some inhomogeneous models $\alpha'=3\degr$ 
provides better results than $\alpha'=0\degr$ and hence {\changedTwo
we have repeated the simulation runs of our most promising models 
with $\alpha'=3\degr$ instead of $\alpha'=0\degr$. In 
Table~\ref{Tab:models} the names of the inhomogeneous 
$\alpha'=0\degr$-models start with a "5" (only model 5h is listed), while 
the names of the inhomogeneous $\alpha'=3\degr$-models start with a "6".} 
All other parameters have been adopted from model 2e. As, compared to the 
models from the previous section, we have fitted one additional parameter 
($d_{\rm clumps}$), we have used $f=23-11=12$ for the calculation 
of the related $\tilde{\chi}_{\rm tot}^2$.

The best-fitting models of these simulation runs are (sorted for 
increasing $\chi_{\rm I}^2$ and decreasing $\chi_{\rm off}^2$ in 
Table~\ref{Tab:models}) models 6b, 5h, 6a, 6j, 6k and 6l. We find 
that all of these models use $M_{\rm inter}$ between 0.1 and 
0.4~M$_{\rm HJ}$ combined with $\rho_{\rm inter}=10^5$~cm$^{-3}$ and 
hence have VFFs of the interclump medium between 0.07 and 0.28. 
{\changedTwo For these models and based on our set-up we cannot 
clearly discriminate between $\alpha'=0\degr$ and $\alpha'=3\degr$.
Models 6a, 6j, 6k and 6l provide lower $\chi^2_{\rm tot}$-values than 
the corresponding $\alpha'=0\degr$-models (not shown) while model 
5h is slightly better than model 6b.} Furthermore, $d_{\rm clumps}$ is 
0.02 or 0.04~pc for these models. The observed stratification pattern is 
not sensitive to the exact choice.

Models 6b, 5h and 6a provide the best fit of the line integrated 
intensities ($\chi_{\rm I}^2=272$ or 273) of all models that have been 
investigated in this section, but they do not provide any improvements 
compared to model 2e from Sect.~\ref{Sect:AlphaIUV}. In terms of the
fit of the line intensities, the inhomogeneous model does not provide
any improvement, but it also does not deteriorate the fit. 
The situation is very different when considering the stratification 
pattern.
Model 6l provides $\chi_{\rm off}^2=15$, which has not been reached by 
any other model. In Fig.~\ref{Fig:scatter_offset_4_5_6} we can see how 
for model 6l the relative $y$-offsets lie in the $(0\pm 0.01)$~pc
interval for all transitions, except for the HCO$^+$~3-2 transition. 
The $I_{\rm rel}$ simulated for model 6l lie between 0.33 
($^{13}$CO~$10-9$) and 1.04 (CO~$16-15$; see also 
Fig.~\ref{Fig:scatter_intensity_4_5_6}).

Models 6j and 6k (see Table~\ref{Tab:models} and 
Figs.~\ref{Fig:scatter_intensity_4_5_6} and 
\ref{Fig:scatter_offset_4_5_6}) are examples for compromises between 
models 2e/6b/5h/6a and model 6l. Model 6k fits the $y$-offsets of 
\emph{all} transitions within 0.016~pc and has $I_{\rm rel}$ between 
0.4 and 1.3. For model 6j the $y$-offset of $^{13}$CO~$10-9$ is too 
small (-0.026~pc), for all other transitions the $y$-offsets are 
fitted within 0.017~pc. Furthermore, the $I_{\rm rel}$ of model 6j 
lie between 0.46 and 1.6. Overall, model 6j provides the lowest 
$\chi_{\rm tot}^2$ of all models tested in the scope of this work. 
The simulated cuts of this model are shown in Fig.~\ref{Fig:cuts_6j}. 
In terms of a simple best fitting model, 2e is very good 
(2e provides the lowest $\tilde{\chi}_{\rm tot}^2$), but when 
explicitly asking for a reproduction of the stratification, only
inhomogeneous models work. Then models 6j and 6k are much better.
}

\subsubsection{\changed Line Profiles}
\label{Sect:line_profiles}

\begin{figure}
  \resizebox{\hsize}{!}{\includegraphics{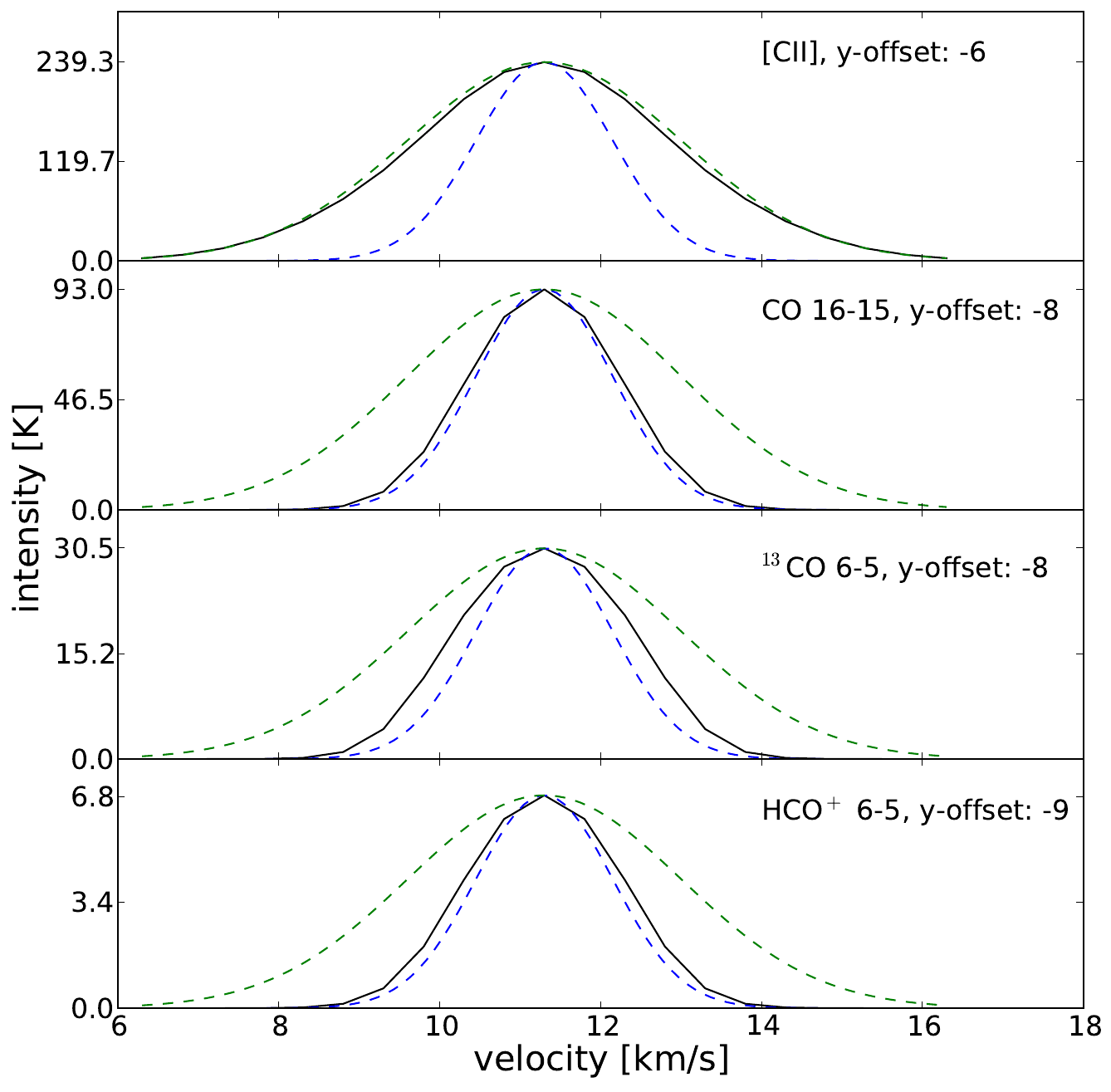}}
  \caption{\changed Line profiles of selected transitions of model 6j. 
   The full lines show the simulated profiles, the dashed lines are Gaussian 
   profiles with the same peak intensity as the respective profile, but 
   with linewidths of ${\rm FWHM}=2$~km~s$^{-1}$ and ${\rm FWHM}=4$~km~s$^{-1}$,
   matching the observed total velocity dispersions. 
   For each transition the $y$-offset corresponds to the position with the highest 
   line integrated intensity. 
  }
  \label{Fig:specs6j}
\end{figure}
 
In addition to the maps and cuts presented in Sect.~\ref{Sect:OriModel} 
the KOSMA-$\tau$ 3D code is capable of simulating line profiles 
{\changedTwo for each individual pixel of a simulation. The beam 
convolution, which is performed by KOSMA-$\tau$ 3D for the line 
integrated intensity maps, is not applied to the line profiles.} The 
simulation {\changedTwo of the line profiles} is based on the velocity 
dependent ensemble averaged line intensities and optical depths,
which have been discussed in Sect.~\ref{Sect:EmissivityOpacity}, and on 
the velocity dependent radiative transfer as discussed in 
Sect.~\ref{Sect:radTrans}. The line width of single clumps 
($\sigma_{j,\,{\rm line}}${\changedTwo ,} which may depend on the mass point $j$ of the individual clump, see Eqs.~\ref{Eq:Ixi} and 
\ref{Eq:tauxi}) in the KOSMA-$\tau$ input grid, is 
$\sigma_{j,\,{\rm line}}\approx 0.71$~km~s$^{-1}$ (i.e.~${\rm FWHM}=1.67$~km~s$^{-1}$ 
or a Doppler broadening parameter of $b=1$~km~s$^{-1}$). The ensemble velocity 
dispersion ($\sigma_{j,\,{\rm ens}}$, see Eq.~\ref{Eq:Nji}) is an additional 
input/fit parameter. Both, $\sigma_{j,\,{\rm line}}$ and $\sigma_{j,\,{\rm ens}}$, 
can be different for clump and interclump medium.

Here, we are not aiming for a fit of the line profiles observed in the Orion 
Bar PDR, and use a fixed velocity dispersion based on some observed line 
widths $\sigma_{{\rm tot},j}$. For the total velocity dispersion of the 
dense clumps we use 
$\sigma_{{\rm tot},j}\approx 0.85$~\kms{} (${\rm FWHM}=2$~\kms{}) 
and for the interclump medium we use 
$\sigma_{{\rm tot},j}\approx 1.70$~\kms{} (${\rm FWHM}=4$~\kms{}) 
as typical values  for the Orion Bar PDR (see \citealt{Nagy_2013} 
and references therein). The clump and interclump ensemble velocity 
dispersion then follows from 
\begin{equation}
 \sigma_{{\rm ens},j} = \left(\sigma_{{\rm tot},j}^2 - \sigma_{{\rm line},j}^2\right)^{1/2}\, .
\end{equation}
Consequently, our input parameters are 
$\sigma_{{\rm ens},j}\approx 0.47$~\kms{} for the ensemble of 
dense clumps and $\sigma_{{\rm ens},j}\approx 1.54$~\kms{} for the ensemble representing 
the interclump medium. The systematic velocity of all voxel (see parameter ${\rm v_{sys}}$ 
in Eq.~\ref{Eq:Nji}) has been set to ${\rm v}_{\rm sys}=11.3$~\kms{}. Sampling
the spectra at 21 different velocities around ${\rm v_{sys}}$, with a spacing of $0.5$~km s$^{-1}$ provides
sufficiently smooth profiles (see velocities ${\rm v}_i$ and ${\rm v}_{\rm obs}$ in 
Sect.~\ref{Sect:EmissivityOpacity}).

The full lines in Fig.~\ref{Fig:specs6j} show selected line profiles from the model 
with the lowest $\chi^2_{\rm tot}$, namely model 6j (see Sect.~\ref{Sect:Inhomogeneous}). 
The dashed lines in the figure are Gaussian line profiles with the same peak intensity 
as the respective profile, but with linewidths of ${\rm FWHM}=2$~km~s$^{-1}$ and 
${\rm FWHM}=4$~km~s$^{-1}$, corresponding to the input linewidths. By comparing the 
simulated profiles with the Gaussian profiles we can see that a large fraction of 
the [C{\sc ii}] emission is emitted by the interclump medium, while the transitions 
of the CO isotopologues and of HCO$^+$ are dominantly emitted by the dense clumps.
Furthermore, some profiles appear broadened as suggested by observations 
(Nagy et al., in prep), however, fit and quantitative comparison are left for future work.

\subsection{Discussion}
\label{Sect:discussion}

We find that a geometry of the Orion Bar region similar to the geometry 
derived in \citetalias{Hogerheijde_1995} is well suited to reproduce the 
observed stratification and line intensities. Significant conclusions 
can be drawn, however, from our {\changedTwo unsuccessful} attempts to 
simultaneously fit the line intensities and the peak positions measuring 
the geometrical stratification of the Orion Bar PDR. Initial simulations 
based on a cylindrical geometry (see 
Appendix~\ref{appendix:cylindrical}) are less promising.

\subsubsection{The stratification pattern}
\label{Sect:Discussion_inhomogeneous}

The definition of $\tilde{\chi}_{\rm tot}^2$ based on the measurement
and modelling accuracies leads to a fit that is typically dominated by 
the contribution of the intensity mismatches. It seems much more difficult 
to tweak the model towards a fit of all line intensities within the
measured accuracy than towards a fit of the observed stratification pattern.
Consequently, we aim for two goals that are not easily unified: on the
one hand we want to provide the best fit in the statistical sense, i.e.
with the lowest value of $\tilde{\chi}_{\rm tot}^2$; on the other hand
we want to reproduce the observed stratification pattern as good as 
possible.

The lowest $\tilde{\chi}_{\rm tot}^2$, namely 
$\tilde{\chi}_{\rm tot}^2=20$, and hence the best fit in the 
statistical sense, is provided by models 2d and 2e, which fit the 
line integrated intensities of all transitions within a 
factor\footnote{The $I_{\rm rel}$ of the different transitions lie 
between 0.50 and 2.1 for model 1d and between 0.51 and 2.2 for model 
2e.} 2.2. However, these models {\changedTwo cannot reproduce the 
stratification pattern observed for the Orion Bar PDR}. All models 
from Sects.~\ref{Sect:HomMassesDensities} {\changedTwo to} 
\ref{Sect:MassPoints} indicate that a reproduction of the 
stratification pattern based on a homogeneous {\changedTwo set-up} (i.e.~models 
that contain the same ensembles of clumps in each voxel; 
$d_{\rm clumps}=0$) is impossible.

The observable stratification in the line integrated intensity maps must
stem from a combination of spatially varying excitation conditions, 
column densities, and line widths from clump and interclump medium (see
Fig.~\ref{Fig:specs6j} and also Fig.~\ref{Fig:cyl_CII_intensity})
along the different lines of sight through the inclined 
cavity wall, modified by the beam convolution. As the homogeneous models
fail to reproduce the stratification, we had to switch to models where
dense clumps only exist at some depth into the cloud; $d_{\rm 
clumps}>0$) in Sect.~\ref{Sect:Inhomogeneous}. 

This is in agreement with previous models and observations. 
\citetalias{Werf_1996} {\changedTwo identify an elongated clump 
with a thickness of about $10\arcsec$ at $\sim$20$\arcsec$ from the IF 
into the cloud, and deduce a density 
increase in this region based on CS~5~-~4 and C$^{34}$S~3~-~2 
observations and a Large Velocity Gradient (LVG) model.} \citet{Parmar1991} proposed already a 
clumpy picture of the PDR with an increasing size and number of clumps 
from the IF into the molecular cloud. This was later adopted by 
\citetalias{Hogerheijde_1995}. A detailed investigation was performed 
by \citet{YoungOwl_2000}. They found that the Orion Bar is best 
modelled when incorporating a ridge of dense clumps into a thinner 
interclump medium at a depth of $20\arcsec$. This is in agreement with 
our model fit where dense clumps have to be added to the interclump 
medium at a depth of $10\arcsec$ to $20\arcsec$, to reproduce the 
observed chemical stratification. 

The best fit of the stratification pattern is 
provided by model 6l, which has $\chi_{\rm off}^2=15$, fitting the relative $y$-offsets 
within $(0\pm 0.01)$~pc, i.e. in about the accuracy of the simulations, 
except for HCO$^+$~3-2. As the HCO$^+$~3-2 peak position provides a major
outlier in the observational data (see Sect.~\ref{Sect:Observations}), we
rather speculate that the unsuccessful fit of this position might be
caused by {\changedTwo a problem in the observational data}.
For further improvements in the fit of the stratification pattern, based 
on our current method, higher resolution observations and smaller voxel 
sizes are necessary. {\changedTwo We have also found models 
that provide g}ood compromises between the fit of the stratification 
pattern and the line integrated intensities, overall, model 6j provides 
the lowest $\chi_{\rm tot}^2$ (see Sect.~\ref{Sect:Inhomogeneous}).

The parameters of the ``best-fitting'' models are summarized in 
Table~\ref{Tab:Fit_parameter}. Our simulations show that these models can 
reproduce the stratification pattern, however a discrimination between 
the tested values for the depth at which the inset begins, 
$d_{\rm clumps}=0.02$, 0.03 or 0.04~pc, is not possible. The best fits of 
the line integrated intensities come from homogeneous models, or, if 
these are excluded, from models with $d_{\rm clumps}=0.02$~pc. The best fit 
of the stratification pattern comes from a model with 
$d_{\rm clumps}=0.04$~pc. 

\subsubsection{Line intensities}
\label{Sect:Discussion_intensities}

When using the original parameters from \citetalias{Hogerheijde_1995}
(see Table~\ref{Tab:Fit_parameter}) we find that almost all
line intensities are too low compared to the observed line integrated intensities.
Only the [C{\sc ii}] intensity, which 
originates predominantly from the interclump medium, can be roughly reproduced by the 
original \citetalias{Hogerheijde_1995}-model. Even when considering the
optically thin C$^{18}$O lines that were used to derive the column
densities we need at least a factor 1.5 more mass per voxel than suggested 
by \citetalias{Hogerheijde_1995} (see Sect.~\ref{Sect:C18O}).
To enable a fit of the other line integrated 
intensities within a factor between two or three, the ``best-fitting'' models (see
Sect.~\ref{Sect:Discussion_inhomogeneous}) use 
$M_{\rm cl, tot}=2$~M$_{\rm HJ}$\footnote{\changed Or correspondingly 
$M_{\rm cl, tot}=4$~M$_{\rm HJ}$ for $d_{\rm cavity}=0.3$~pc and so on.} for a depth 
of the cavity of 0.6~pc. Consequently, our total mass per voxel is a factor 
2.1 to 2.4 higher (for $M_{\rm inter, tot}=0.1$ to 0.4~M$_{\rm HJ}$; see 
Sect.~\ref{Sect:Discussion_interclump}) than the value inferred from
\citetalias{Hogerheijde_1995}.

The difference can be explained from the nature of the different models.
\citetalias{Hogerheijde_1995} use a two component (clump and interclump medium) 
model with a uniform kinetic temperature of $T_{\rm kin}=(85\pm 30)$K (for both 
components) to fit the observations. In 
the KOSMA-$\tau$ PDR code, the full gas temperature distribution is calculated as a function 
of the clump radius. For a clump of 1~M$_\sun$ with a hydrogen surface density of 
$n_{\rm s}=10^6$~cm$^{-3}$ and an FUV flux at the clump surface of $10^4$~$\chi_0$,
the KOSMA-$\tau$ PDR code computes a \emph{clump-averaged} temperature of 67K,
relatively close to the value by \citetalias{Hogerheijde_1995}. However, in 
the KOSMA-$\tau$ simulations there is a significant change in temperature between 
surface and core. If we compute the average temperature ``felt'' by a particular
molecule, i.e. obtained when weighing the temperature profile by the abundance 
of the different species, we obtain very different temperatures. C$^+$, being 
abundant in a hot surface layer, ``feels'' an average temperature of about 1600K
while CO and $^{13}$CO ``feel'' temperatures of 38K, and HCO$^+$ of about 32K. 

The emission of all optically thin molecular species is therefore significantly
weaker in our model than in the 85~K model from \citetalias{Hogerheijde_1995}.
As the line intensities are roughly proportional to the source function, which
is itself determined by the excitation temperature, we find that a change of 
the gas temperature from 85~K to 38~K reduces the line intensities by a factor
of about 2.5 for the lower-$J$ CO lines. This explains the new upper limit on 
the molecular column density, deduced from the same C$^{18}$O observations in
Sect.~\ref{Sect:C18O}, compared to the model of \citetalias{Hogerheijde_1995}. 
{\changedTwo By using a 2.7 times higher column density than 
\citetalias{Hogerheijde_1995}, our best fitting model (6j) overestimates 
the C$^{18}$O~3~-~2 line integrated intensity by about 25\,\% (see 
Sect.~\ref{Sect:C18O}). This is better than the deviation that we find 
between our fit and observations for the other CO isotopologues. 
Taking the constraints from the C$^{18}$O 
observations we cannot further increase} the total column densities. 
However, all our model fits show the tendency that in particular the low- 
and mid-$J$ lines from CO, $^{13}$CO, and HCO$^+$ are somewhat too weak.

The line integrated intensities of our best-fitting models follow 
a general trend, {\changedTwo which is visible in the scatter plots, 
e.g.~Fig.~\ref{Fig:scatter_intensity_4_5_6}.} While for many models 
the fit of the [C{\sc ii}] line integrated intensity is satisfactory, 
it tends to be too low for other transitions, especially for CO~$2-1$, 
$3-2$ and $6-5$ and for $^{13}$CO~$6-5$ and $10-9$. Even with model 
3n, which provides the best fit of the line integrated intensities, 
only a factor 0.54 of the observed CO~$2-1$ line integrated intensity 
is reproduced (see Fig.~\ref{Fig:scatter_intensity_1_2_3}). On the 
other side, the line integrated intensities of the CO~$16-15$ and 
HCO$^+$~$6-5$ transitions are predicted too high, in our best-fitting 
models by a factor of about two.

The numerical experiments show that the intensity of the CO~$16-15$ 
and HCO$^+$~$6-5$ transitions depend strongly on the FUV flux that is 
available for the excitation of the gas in the dense clumps. Other 
transitions (e.g.~low-$J$ transitions of the CO isotopologues) 
are less affected. The FUV flux is governed mainly by the composition 
of the interclump medium. {\changedTwo Due to the} numerous low-$J$ 
line observations {\changedTwo used within our fitting process, the 
models are forced into a parameter range where the interclump gas
is strongly reduced, resulting in a high FUV flux within the cloud and 
therefore overall increased} line integrated intensities. {\changedTwo 
The effect on [C{\sc ii}], which stems from both dense clumps and 
interclump gas, is minor}, but for the sensitive high-$J$ 
rotational lines, we obtain {\changedTwo the} ``outliers'' with too 
high predictions, {\changedTwo that have been described above.} The 
pattern becomes even stronger when we add very small and dense clumps, 
i.e.~when we {\changedTwo reduce} $m_{\rm l, cl}$. {\changedTwo In the 
opposite case, by increasing the amount of interclump medium and hence 
reducing the FUV flux in the cloud, we can adjust} all relative 
intensities to similar values, i.e.~remove the characteristic pattern, 
but then all intensities (except for [C{\sc ii}]) are too low. This 
{\changedTwo issue} asks for a new mechanism, which systematically 
increases the line integrated intensities of all, in particular the 
low-$J$, molecular transitions.

Adding the background molecular cloud (see Sect.~\ref{Sect:C18O} and
model 2b\_ext in Table~\ref{Tab:models}) increases the line 
integrated intensities of the species that can be excited by very low
FUV fluxes, i.e.~of the low-$J$ transitions of the CO isotopologues
and of HCO$^+$~$3-2$. Consequently, the extension of the compound
improves the fit of line integrated intensities (while the influence
on the stratification pattern is small) and slightly ``flattens'' the
scatter plot of the $I_{\rm rel}$. The amount of background material
is, however, constrained by the C$^{18}$O limit as discussed in 
Sect.~\ref{Sect:C18O}, so that no significant change to our results
is possible from this side.

An enhancement of the CO line intensities on the modelling site 
might be possible when including non-equilibrium effects due
{\changedTwo to} an advancing IF: \citet{Stoerzer_1998} find that in 
such models low-$J$ CO lines can be enhanced by a factor 
{\changedTwo two} compared to equilibrium models. The mid-$J$ CO lines can also be affected. 

\subsubsection{Composition of the dense clumps}
\label{Sect:Discussion_clumps}

The characteristic intensity pattern obtained based on the original
model parameters from \citetalias{Hogerheijde_1995} (Model 1d) shows a
growing discrepancy between the predicted line intensities and the
observed values with the rotational level for our linear molecules.
Moreover, it does not reproduce the observed stratification pattern
at all. This indicates that in this {\changedTwo set-up} the molecular material is
too cold and too evenly distributed. As a consequence, we had
to concentrate more mass in dense clumps and reduce the fraction of 
interclump mass to explain the observed intensities. Apart from the
need for a higher total voxel mass discussed above, a reasonable fit
of the integrated intensities is only possible when increasing the fraction 
of material that is contained in the dense clumps from the 10\% in
\citetalias{Hogerheijde_1995} to 83\% -- 95\%. 

Clump densities found in the literature cover $n_{{\rm H}_2}\approx 1^{+3.0}_{-0.7}10^6$~cm$^{-3}$ 
\citepalias[][corresponding to $\rho_{\rm cl}=2\times 10^6$~cm$^{-3}$]{Hogerheijde_1995}, 
$3\times 10^6$~cm$^{-3}$ \citep{YoungOwl_2000} or between $3\times 10^6$ 
and $1.2\times 10^7$~cm$^{-3}$ (assuming that the clumps are virialised, see \citealt{Lis_2003}).

When comparing ensemble-averaged densities of $\rho_{\rm cl}=10^6$~cm$^{-3}$ and 
$4\times 10^6$~cm$^{-3}$ for the  dense clumps we find that the higher density 
slightly improve the fit of the line integrated intensities. Unfortunately,
we could not test higher ensemble-averaged densities than $4\times 10^6$~cm$^{-3}$
due to the boundaries of the KOSMA-$\tau$ parameter grid. It only provided
densities up to $10^7$~cm$^{-3}$, while a full clump spectrum with an average
density of  $10^7$~cm$^{-3}$ would contain some clumps with significantly higher 
density.

In Sect.~\ref{Sect:MassPoints}, we have tested models for which
the lower cut-off mass of the ensemble of dense clumps has been 
increased relative to the {\changedTwo initial} value of $10^{-3}$~M$_\sun$. We 
notice that the small (and dense) clumps contribute to the molecular 
emission. Removing them, while keeping the ensemble-averaged density 
and total mass fixed, systematically decreases the line integrated 
intensities of all transitions (except [C{\sc ii}]), especially for 
the high-$J$ transitions of the CO isotopologues. As the simulated 
line integrated intensities tend to be too low in our models, the 
fit of the line integrated intensities improves if small and dense 
clumps are included. Consequently, our simulations do not 
support the "turn-over" in the clump-mass function that has been
suggested by some authors (see Sect.~\ref{Sect:FractalISM}).

{\changedTwo For model 6j the radii of the clumps at the different mass 
points are $\lbrace R_j\rbrace_{j=1...4}=\lbrace 0.0008, 0.0022, 0.0059, 
0.0159\rbrace$~pc for the ensemble of dense clumps and 0.010~pc for the 
interclump medium. At the distance of 414~pc to earth the size of the 
largest clumps ($2\times 0.0159$~pc) correspond to 15.8\arcsec, which is
about a factor two lager than the sizes that have been observed/derived 
by \citet{Lis_2003} and \citet{YoungOwl_2000} (see 
Sect.~\ref{Sect:Observations}).
} 

\subsubsection{Composition of the interclump medium}
\label{Sect:Discussion_interclump}

The interclump medium has to account for a significant fraction of the 
[C{\sc ii}] emission and partly for the emission of the low-$J$ 
HCO$^+$, CO and $^{13}$CO lines. Its main effect is, however, the 
attenuation of the FUV field, effectively lowering the gas temperature 
in the dense clumps. Ensembles with a more homogeneous distribution of 
the ISM within the voxels, which is for example achieved by reducing 
the ensemble-averaged density while keeping the total mass constant, 
cause stronger FUV attenuation.

The original \citetalias{Hogerheijde_1995}-parameters correspond to a 
VFF of the interclump medium of (nearly) unity. In a homogeneous {\changedTwo set-up} 
the combination of dense clumps (with $M_{\rm cl, tot}=2$~M$_{\rm HJ}$) 
and interclump medium provides too much FUV attenuation to allow for 
sufficient excitation of the (especially high-$J$) transitions that emit 
at some depth into the cloud.
Based on simulations of \emph{inhomogeneous} models with an
interclump-VFF of unity (not shown) we conclude that if we limit ourselves to an 
interclump-VFF of one, we can fine-tune the composition of the interclump medium 
to optimise the fit of the st{\changedTwo r}atification pattern \emph{or} of the line integrated 
intensities, but we do not find a set of parameters providing both.

In the homogeneous {\changedTwo set-up} the fits converges towards models that use a 
large amount of mass in the dense clumps, but no interclump
medium (see models 1z and 1E). However, such models clearly 
{\changedTwo contradict} the observed stratification pattern and 
direct observations of the interclump medium 
\citep[e.g.][]{Stutzki_1988, Ossenkopf_2013}. If we optimise 
the inhomogeneous models after dropping the constraint of the fixed 
interclump-VFF we find the ``best-fitting'' models with an 
interclump-VFF between 0.07 and 0.28, discussed in 
Sect.~\ref{Sect:Discussion_inhomogeneous}. 

In our {\changedTwo initial} model we used clumps of $10^{-2}$~M$_\sun$ to represent 
the interclump medium. In Sect.~\ref{Sect:MassPoints}, we varied this
assumption to use clumps of $10^{-3}$~M$_\sun$, $10^{-2}$~M$_\sun$, and
$1$~M$_\sun$ instead.
Changing the mass points has two major effects. First, the mass points 
yield a possibility to fine-tune the FUV attenuation in the cloud. 
Representing the interclump medium by clumps with $10^{-3}$~M$_\sun$, 
which allows for a more homogeneous distribution of the ISM, increases 
the FUV attenuation in the cloud, degrading the quality of the fit.
Second, the small clumps also contribute to the molecular emission. 
If we increase the size of the clumps in the interclump medium, 
the reduced emission from the interclump gas leads to
stratification offsets that {\changedTwo are} too small compared to the observed 
structure. Hence, our initial guess turned out to be a good choice.

For the interclump medium densities in the range $n_{{\rm H}_2}\approx 3^{+2.0}_{-2.2}10^4$~cm$^{-3}$ 
\citepalias[][corresponding to $\rho_{\rm inter}=6\times 10^4$~cm$^{-3}$]{Hogerheijde_1995}, 
$5\times 10^4$~cm$^{-3}$ \citep{YoungOwl_2000} or $2\times 10^5$~cm$^{-3}$ \citep{Simon_1997} 
have been proposed. Our best-fitting models have a hydrogen nucleus density of 
$10^5$~cm$^{-3}$. 
{\changedTwo If we correct this number by the} VFFs between 0.07 and
0.28 found for our best models, we obtain an average density of the 
interclump medium \emph{over the volume of the voxel} of $7\times 10^3$ 
to $2.8\times 10^4$~cm$^{-3}$, which is somewhat lower.

{\changedTwo The \emph{Herschel} observation (Fig.~\ref{Fig:CII})
shows that C[{\sc ii}] has a rather smooth emission profile across the 
bar. Fig.~\ref{Fig:BruteForceCII_inter_80voxel} shows C[{\sc ii}] 
emission of the interclump medium of model 6j. In this figure we have 
used a number surface density corresponding to 80 voxels along one line of 
sight, which is typical for the models discussed in this work. The 
figure shows that although the VFF of the interclumps medium
of model 6j is only 0.04, the emission becomes rather smooth, 
especially if we consider that it is observed with an instrument 
resolution of about 0.02~pc.}

\subsubsection{The FUV illumination}
\label{Sect:Discussion_FUV}

{\changedTwo In this work we have assumed that the FUV radiation field
incident on a clump surface is isotropic. However, we have also 
applied the KOSMA-$\tau$ 3D code to the Orion Bar PDR, where the FUV 
irradiation comes from one dominating source, i.e.~from a defined 
direction. Therefore, one could argue that our (isotropic) approach 
increases the hot PDR surface area inside the cloud leading to an 
overestimation of the species originating form the hot gas, for 
instance the high-$J$ CO lines. However, at some depth into 
the cloud, the FUV photons are efficiently scattered 
\citep{Stoerzer_1996} and the assumption of an isotropic radiation 
field becomes reasonable. For the description of the \emph{surface} 
of the Orion Bar a model using beamed illumination would be more 
precise.} 

We have started our simulations with an FUV flux at the IF of 
$4.4\cdot 10^4\,\chi_0$, adopted from \citetalias{Hogerheijde_1995}. 
Furthermore, in Sect.~\ref{Sect:AlphaIUV} we have shown how 
increasing this values by 25\% and 50\% {\changedTwo slightly} improves the fit of 
the line integrated intensities, but the effect is small compared to 
variations of other parameters. Changes of the FUV flux due to FUV 
attenuation inside the cloud, which is governed by the composition of 
the ensembles, has a significantly stronger impact on the simulation 
outcome. Hence, our models cannot be used to determine the FUV flux 
that is incident on the cloud surface.

In Sect.~\ref{Sect:DcavityZstar} we have presented models with
inclination angels of the bar of $0\degr$, $3\degr$, $7\degr$ or
$15\degr$. From investigations of the line intensities emitted by the 
different voxels of the model compound we find that some species emit 
rather locally (see for example Fig.~\ref{Fig:CO16-15_intensity}). 
For these species increasing the inclination angle can reduce the 
excited column density along a line of sight through the compound, 
into the direction of the observer. Still, an increase of the line 
integrated intensity in the final map is possible, due to the beam 
convolution. In the simulations we have to compromise between these 
effects. For other tracers the excitation is less dependent on the 
FUV flux and the emission comes from a more extended region (see for 
example Fig.~\ref{Fig:CO2-1_intensity}). For these transitions the 
impact of the inclination angle is smaller. The stratification 
pattern is produced by the rather complicated combination between FUV 
attenuation, inclined geometry of the bar and beam convolution. It is 
difficult to disenta{\changedTwo n}gle these effects, however, in 
general the pattern becomes more {\changedTwo random}, and hence 
harder to fine-tune, for large inclination angles. {\changedTwo
Our best-fitting inhomogeneous models use $\alpha=3\degr$ or less.}
Such a small inclination angle was proposed by \citet{Melnick_2012}. 



\subsubsection{Further model improvements}

As a solution to the overall insufficient fit of the line 
intensities of the combination of all tracers considered here, it 
might be useful to play with the isotopic abundances. A lower 
$^{18}$O abundance {\changedTwo could} potentially fit the 
observed high intensities of the CO and $^{13}$CO lines under the 
column density constraint from the C$^{18}$O observations.
Non-equilibrium models for the PDR chemistry as proposed by 
\citet{Stoerzer_1998} {\changedTwo could} also lead to an 
enhancement of the CO line intensities.

To deal with the relatively high mismatch of the modelled line
intensities compared to the mismatch of the stratification 
pattern, it may be useful to create a new chi-square criterion 
with a higher weight of the $y$-offsets that allows for a better 
combined fit of both model aspects. Another possibility to obtain 
further constraints would be to fit the whole intensity profile 
along the cuts and not only the pixels that provide the peak 
intensities. A model with a more complex density structure than 
our idealised sharp transition might further improve our 
fit. Furthermore, a systematic fit of the line profiles, as 
discussed in Sect.~\ref{Sect:line_profiles}, would provide further 
constraints. Important information could come e.g. from the fit to 
more complicated line profiles as, in the context of the Orion 
Bar, seen for different transitions of CH$^+$ and SH$^+$ 
\citep{Nagy_2013}. 

A further improvement of the fit based on the chi-square tests 
could come from a more deliberate selection of the fitted tracers 
(if available). For example, with the current selection, only 
[C{\sc ii}] is sensitive to the thin and hot interclump gas at the 
IF. An inclusion of other transitions that are expected to peak at 
the IF, as for instance {\changedTwo CH$^+$ and SH$^+$ 
\citep{Nagy_2013} or OH$^+$ \citep{vanderTak2013}} could increase 
the weight of a correct fit of the interclump medium.

{\changedTwo In an ideal approach we should have repeated the 
two-dimensional scans using different sets of initial parameters 
and different successions for the parameter variations, to 
guarantee to find the globally best fit. This was, however, 
practically impossible due to the relatively large computational 
effort for each model run. In this way our fit may not be the 
best possible one, but our approach guarantees that we understand 
the effect of every individual parameter.
}

\section{Summary}
\label{Sect:summary}

The observation of high-density, high temperature tracers combined with
a layer{\changed ed} structure of the Orion Bar spanning over more than $15\arcsec$
rules out any description of the PDR in terms of a simple plane-parallel
model. PDRs, like molecular clouds, are clumpy and filamentary.

We propose a numerical model that is based on the representation of any PDR 
by an ensemble of clumps (possibly immersed in a thin interclump medium). 
Our new {\changed model ''KOSMA-$\tau$ 3D``} builds on the KOSMA-$\tau$ 
PDR code. It enables us to simulate {\changed the emission of star forming 
regions with arbitrary 3D geometries}. The region is modelled using cubic 
voxels where each voxel is filled with {\changed clumpy structures, following 
a discrete mass distribution.} The ensemble properties can vary between different 
voxels. Using a probabilistic algorithm for the calculation of {\changed the 
ensemble-averaged FUV extinction the {\changedTwo incident} FUV flux is derived for each 
voxel. A velocity dependent version of the probabilistic algorithm is used 
to calculate ensemble-averaged line intensities and optical depths and allows 
us to simulate the radiative transfer through the compound. The output of our 
code includes line integrated intensity maps and full line profiles.}

As a first test of the new model we performed simulations of the Orion Bar PDR.
{\changed In these simulations we tried to} simultaneously reproduce the 
{\changed line} integrated intensi{\changed ties} and the spatial 
offset{\changed s} of the emission peak{\changed s,} for different transitions 
of CO isotop{\changed ologues}, HCO$^+$ and [C{\sc ii}]. Different geometries 
and parameter combinations have been tried out. We find that
\begin{itemize}
  \item a line fit is only possible if we invoke a process that removes dense 
  clumps close to the PDR surface. The detailed stratification profile cannot 
  be reproduced by models with a {\changed spatially} constant ratio of clump 
  to interclump gas.
  \item{} the {\changed composition (i.e.~the ensemble-averaged density and 
  the total mass of the ensemble) of the interclump medium as well as the 
  total molecular column density provided by the dense clumps} are the most 
  critical parameters for the simultaneous fit of the line intensities and the 
  stratification structure of the Orion Bar. The interclump medium governs the 
  FUV attenuation and hence the spatial layering while the clumps produce most 
  of the emission from the molecular tracers. Our fit requires {\changed an 
  ensemble-averaged density of $4\times 10^6$~cm$^{-3}$ for 
  the ensemble of dense clumps and an ensemble-averaged 
  density of $10^5$~cm$^{-3}$ for ensemble 
  representing the interclump medium. Furthermore, for our 
  best models we need a VFF of the interclump medium between 
  0.07 and 0.28, which indicates that 
  also the interclump medium is not homogeneous, but breaks 
  up into sub-structure.}
  \item to reproduce the observed line integrated 
  intensities, a large ratio 
  (between 5 and 20) of total clump to total interclump mass is needed. This 
  contradicts the earlier estimates by \citetalias{Hogerheijde_1995}.
 \item{} The depth of the cavity and the position of the 
 illuminating source are of minor importance, as long as the total 
 molecular column density is fixed. Only positions of the illuminating source
 close ($\sim 0.1$~pc) to the {\changedTwo background} molecular cloud can be excluded.
 Furthermore, the observations are best reproduced if a small inclination angle
 of the bar, in the order of $3\degr$, is used in the model.
\end{itemize}

{\changed The focus of this work was on testing the new 3D PDR model and 
on fitting line integrated intensity maps of the Orion Bar PDR.} A systematic 
comparison of the {\changed simulated line} profiles to observations 
{\changed is left for future work.}

\begin{acknowledgements}
We thank D. Lis for providing the CSO data and helpful comments.
S.~Andree-Labsch thanks the Deutsche Telekom Stiftung and the Bonn-Cologne Graduate School of Physics 
and Astronomy for support by means of stipends. Modelling of irradiated molecular clouds is carried out 
within the Collaborative Research Center 956, sub-project C1, funded by the Deutsche Forschungsgemeinschaft 
(DFG). {\changed Furthermore, we wish to thank the anonymous referee for 
his/her suggestive and detailed comments.} 
\end{acknowledgements}
\bibliographystyle{aa} 
\bibliography{references}

\appendix
\section{Testing the probabilistic approach}
\label{appendix:bruteForce}

In Sect.~\ref{Sect:EmissivityOpacity} ensemble-averaged line 
intensities and optical depths have been derived based on binomial 
distributions ``probabilistic approach''. This formalism was verified 
using a direct approach that calculates the ensemble-averaged 
quantities for a specific realisation of an ensemble. Here, the 
direct approach is presented for the [C{\sc ii}], CO~$3-2$ and 
CO~$16-15$ lines. As it requires a relatively large amount of 
computing time\footnote{The calculation of the ensemble-averaged line 
intensity and optical depth of one ``small'' map 
(i.e.~$1184\times 1184$~pixel, for instance Fig.~\ref{Fig:BruteForceCO3_tau}) takes about two hours.} 
it is not used in the simulations of the KOSMA-$\tau$ 3D 
code, where ensemble-averaged quantities need to be calculated for about $10^5$ 
different ensembles. However, we used it to calculate the ensemble-averaged 
quantities of selected ensembles with known parameters to be compared to 
the results of the probabilistic approach. This serves three purposes: 
\begin{itemize}
\item to verify that the ensemble-averaged quantities calculated with
    the probabilistic approach match the averaged quantities of real 
    random realisations (random positions for each clump) of the 
    ensemble.
\item to verify that averaging over the clump projected area (see    
    Sect.~\ref{Sect:KOSMA}) can be used (the method presented here   
    accounts for the full (non-averaged) ``profiles'' 
    $I_{\rm line}(p)$ and $\tau_{\rm line}(p)$ with $p$ being the 
    \emph{impact parameter}, or radial distance from the centre point 
    of the clump).
\item to understand how much the ensemble-averaged quantities of the 
    same ensemble can vary depending on the random positions of the 
    individual clumps.
\end{itemize}

As discussed in Sect.~\ref{Sect:EnsembleStatistics} we consider 
randomly distributed clumps of masses $M_j$ with a fixed number 
surface density $N_{j}/ (\Delta s)^2$. The projected surfaces of the 
clumps may overlap. We focus on the discussion of one ensemble 
representing dense clumps and in addition, for the 1.9 THz 
[C{\sc ii}] transition, we show results of one typical ensemble 
representing the interclump medium (see below). The 
``dense-clump'' ensemble tested in this appendix contains clumps with 
masses $\lbrace M_{\it j} \rbrace_{j=1...n_M}=\lbrace 10^{-3}, 
10^{-2}, 10^{-1}, 10^{0}\rbrace$~M$_{\sun}$ and a total mass of 
0.00346~$M_{\sun}$ per $(0.01\, {\rm pc})^2$ projected surface area
(which translates into $N_{j}/ (\Delta s)^2$ using 
Eq.~\ref{Eq:Nbinned2}). The ensemble averaged density has been fixed 
to be $4\times 10^6$~cm$^{-3}$, the averaged densities of individual 
clumps have been calculated using Eq.~\ref{Eq:rho_cl}.

As discussed in Sect.~\ref{Sect:EnsembleStatistics} a clumpy ensemble 
can be scaled to arbitrary numbers of clumps as long as the number 
surface density for each mass point, $N^{'}_{j}/ (\Delta s')^2$, is 
kept constant. For the direct approach we create squared maps of a 
size $(\Delta s')^2$ in which the (centre points of the) $N^{'}_{j}$ 
clumps are randomly distributed. Two different map sizes are used in 
this appendix, ``small'' maps contain one and ``large'' maps contain 
two clumps at mass point 1~M$_{\sun}$. {\changedTwo In the maps we 
ignore an area of the thickness $R_{\rm cl,max}/2$ around the edges 
because that region could be affected by clumps with centres outside 
of the map, that are no taken into account here.}

Maps\footnote{For all presented maps an FUV flux of $10^4\,\chi_0$ has 
been used.} of random representations of the ensemble of dense clumps 
are shown in Figs.~\ref{Fig:BruteForceCII_Tb}, 
\ref{Fig:BruteForceCII_tau}, \ref{Fig:BruteForceCO3_tau} and \ref{Fig:BruteForceCO16_Tb}.
For instance, the colour scale of Fig.~\ref{Fig:BruteForceCII_Tb} shows the 
line centre intensity of the [C{\sc ii}] line, summed up along lines of sight perpendicular 
to the printed surface. [C{\sc ii}] emission is strongest in the outer layers of the
clumps. It is affected by limb brightening. The resulting ``intensity profile'' $I_{\rm line}(p)$ 
(see Eq.~\ref{Eq:IClAv}), where the [C{\sc ii}] line {\changed intensity} is highest at the edge of 
each (sufficiently large) projected clump, is clearly visible in Fig.~\ref{Fig:BruteForceCII_Tb}.
Analogously to Fig.~\ref{Fig:BruteForceCII_Tb} ``optical depth maps'' have been simulated. For 
example Fig.~\ref{Fig:BruteForceCII_tau} shows the same map as 
Fig.~\ref{Fig:BruteForceCII_Tb} but with the colour scale giving the optical depth of the 
[C{\sc ii}] transition.

\begin{figure}
  \resizebox{\hsize}{!}{\includegraphics{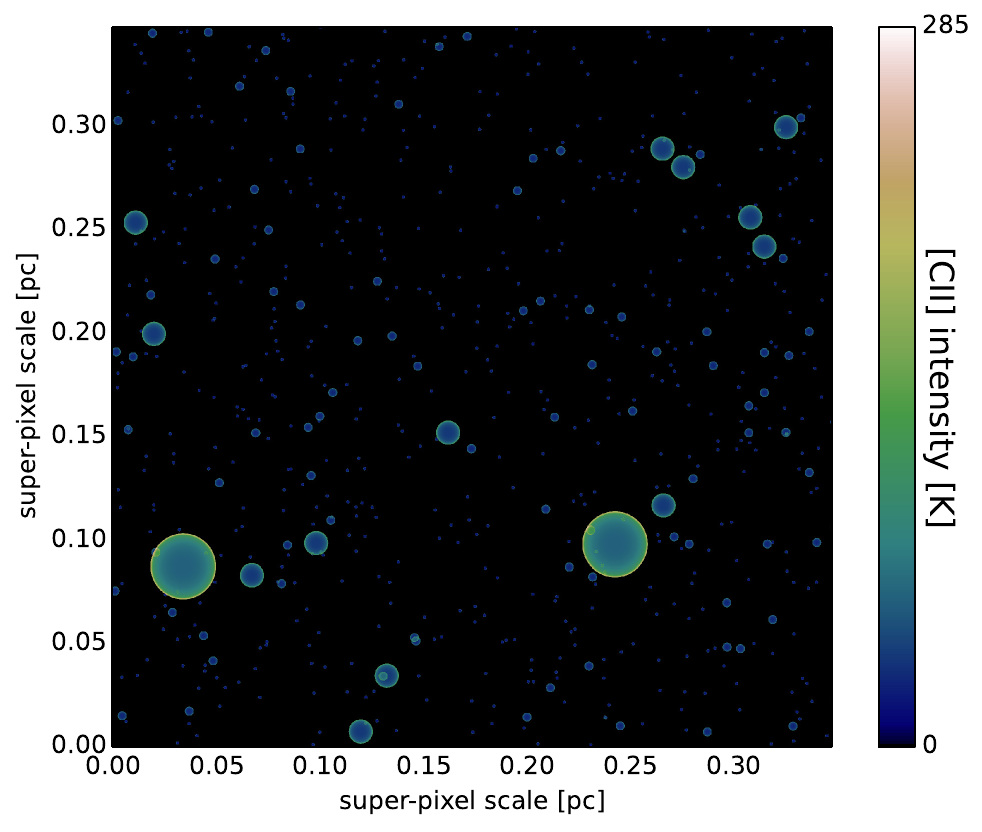}}
  \caption{\changed A ``large'' map showing one representation of the 
  test ensemble, consisting of randomly distributed clumps.
  The colour scale gives the line centre intensity of the 1.9~THz 
  [C{\sc ii}] transition  where the line intensity is highest in the 
  outer layer of each clump. This maps shows the 22th representation 
  from Fig.~\ref{Fig:BruteForce_statistics_CII_Tb}. {\changedTwo The 
  ensemble is shown in a ``superpixel'', i.e. with the same area
  filling factor of a normal 0.01~pc pixel, but scaled in the lateral
  directions to contain enough clumps of each size.  }}
  \label{Fig:BruteForceCII_Tb}
\end{figure}

\begin{figure}
  \resizebox{\hsize}{!}{\includegraphics{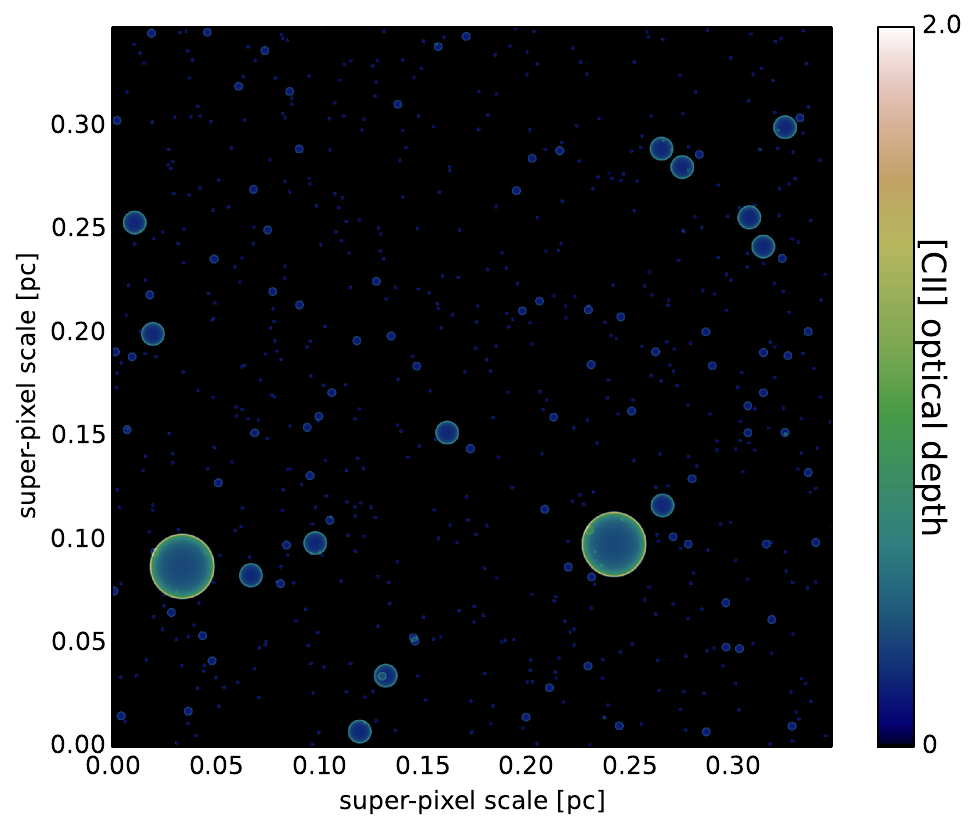}}
  \caption{\changed Same as Fig.~\ref{Fig:BruteForceCII_Tb} but with the colour scale giving 
  the optical depth of the 1.9~THz [C{\sc ii}] transition.}
  \label{Fig:BruteForceCII_tau}
\end{figure}

\begin{figure}
  \resizebox{\hsize}{!}{\includegraphics{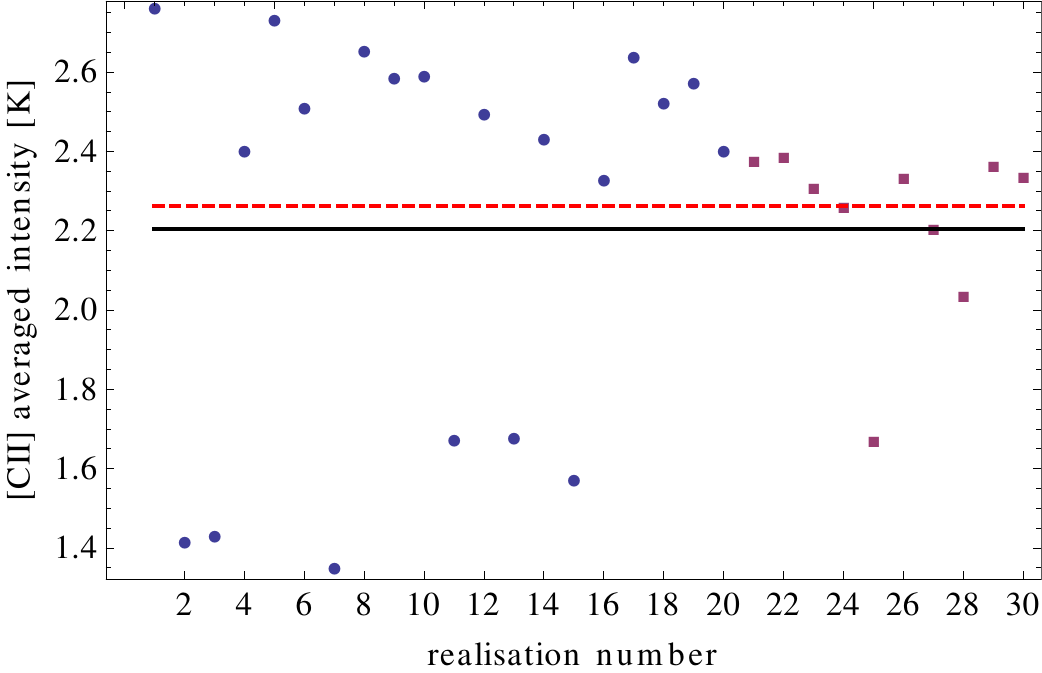}}
  \caption{Ensemble-averaged {\changed [C{\sc ii}] line centre intensities}. The 
  {\changed circles} show the ensemble-averaged {\changed intensities} of 20 different 
  {\changed small intensity maps} and the squares give the values for ten different 
  {\changed large intensity maps as presented in Fig.~\ref{Fig:BruteForce_statistics_CII_Tb}.}
  The black line gives the mean value of the 30 maps, weighted by the respective number of 
  grid points. {\changed For the 20 small maps the average is (2.24~$\pm$~0.50)~K and for 
  the ten large maps it is (2.17~$\pm$~0.33)~K. The stated error is the standard deviation, 
  which, as expected, is larger for the smaller maps.} The red, dashed line gives the value 
  derived using the {\changed probabilistic} approach for the same ensemble. The difference 
  between the two approaches is about 2.6\%.}
  \label{Fig:BruteForce_statistics_CII_Tb}
\end{figure}

\begin{figure}
  \resizebox{\hsize}{!}{\includegraphics{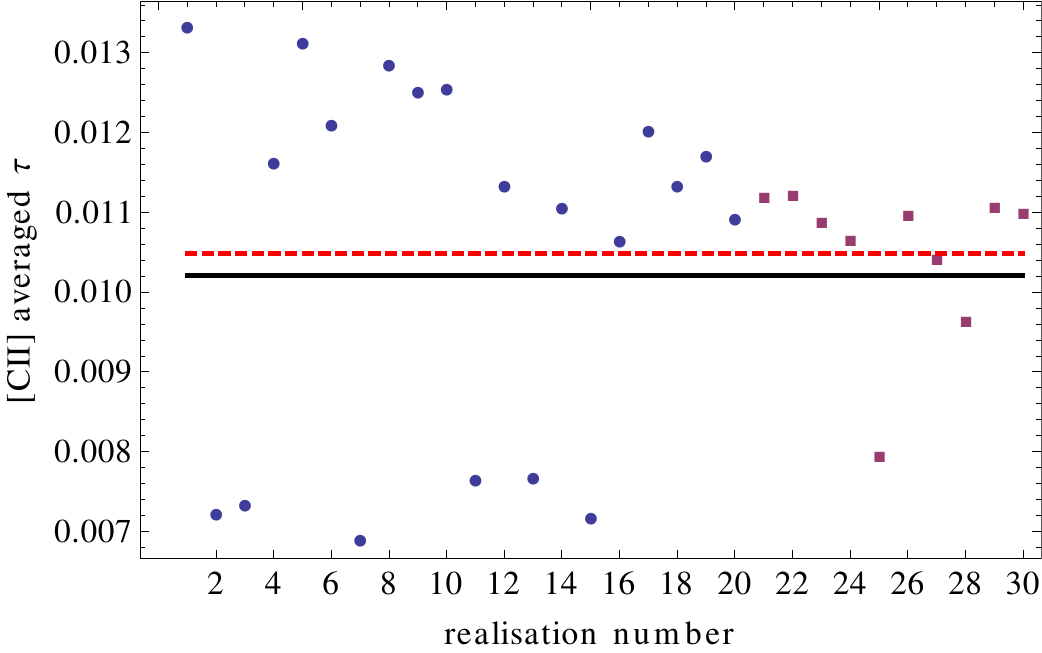}}
  \caption{Same as Fig.~\ref{Fig:BruteForce_statistics_CII_Tb} but 
  plotted for the ensemble-averaged optical depths, $\tau$, of the 	
  [C{\sc ii}] line. For the [C{\sc ii}] line the difference between 
  the probabilistic approach and the average over the 30 results 
  derived with the direct approach is about 2.6\% for line intensities 
  and optical depths.}
  \label{Fig:BruteForce_statistics_CII_tau}
\end{figure}

\begin{figure}
  \resizebox{\hsize}{!}{\includegraphics{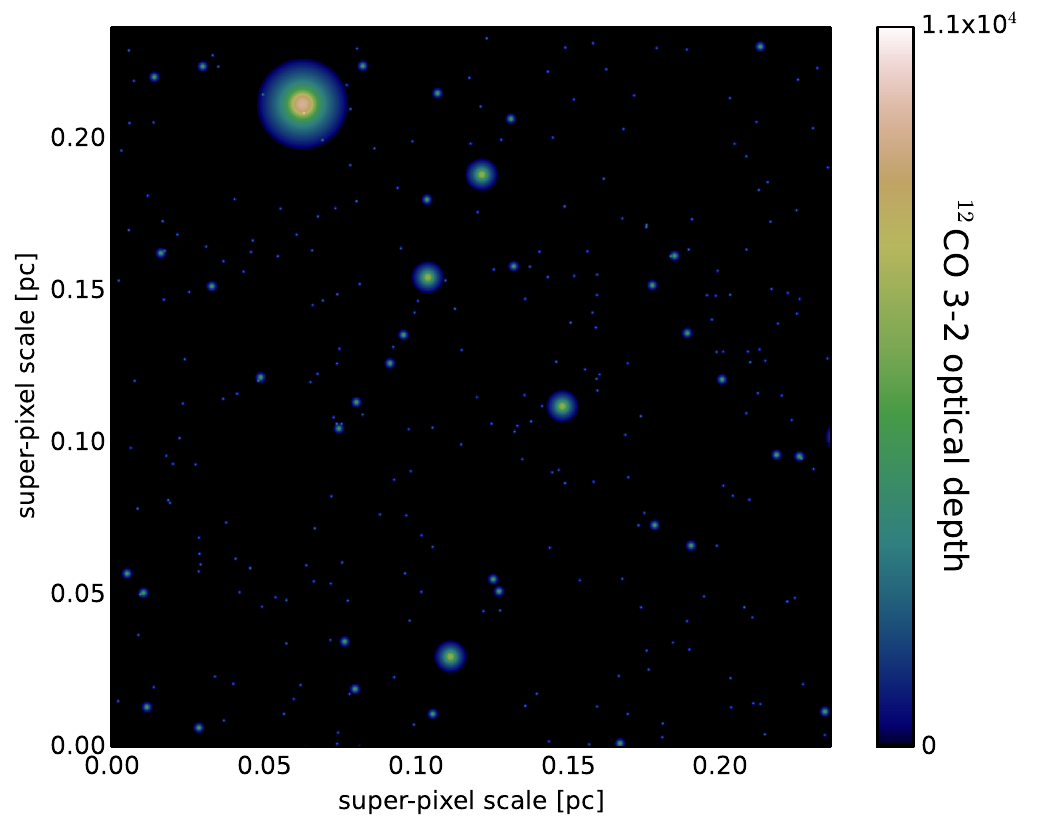}}
  \caption{A small map of one representation of the test ensemble.
  The colour scale gives the optical depth of the CO~3-2 line (line 
  centre), which is highest for lines of sight through the dense cloud 
  cores. This map shows the first realisation from 
  Fig.~\ref{Fig:BruteForce_statistics_CO3_tau}. {\changedTwo The 
  increasing optical depth towards the clump centres provide the 
  visual impression that the clumps are smaller than their actual 
  size.}}
  \label{Fig:BruteForceCO3_tau}
\end{figure}

\begin{figure}
  \resizebox{\hsize}{!}{\includegraphics{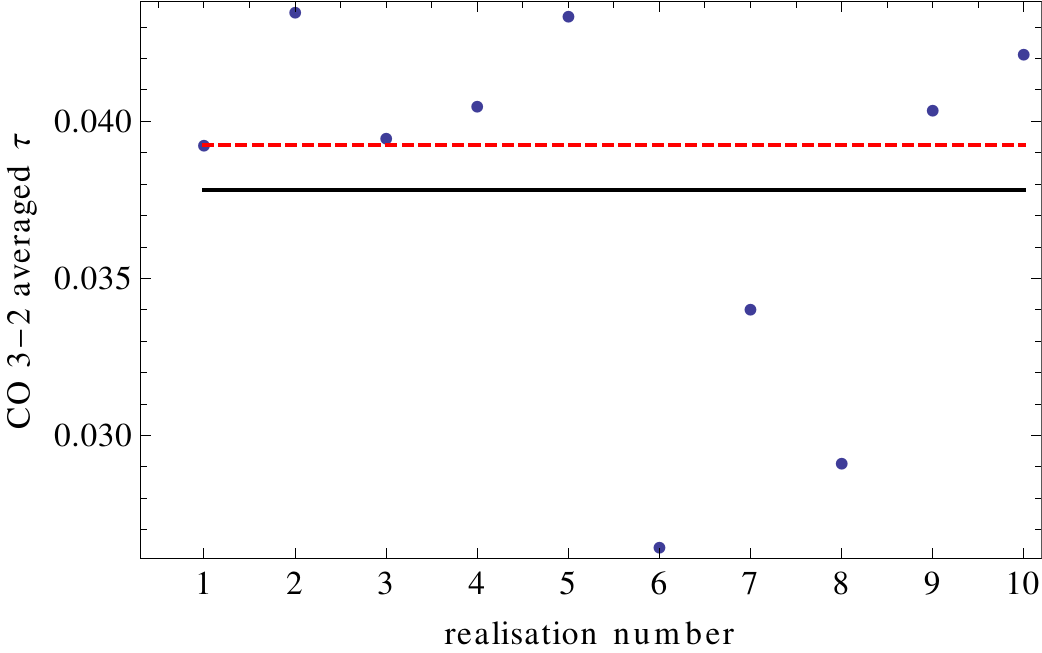}}
  \caption{Ensemble-averaged optical depths of the CO~$3-2$ line. The 
  circles show the results for ten different realisations of the 
  ensemble as presented in Fig.~\ref{Fig:BruteForceCO3_tau}. The black 
  line gives the mean value of the ten maps, namely 
  (0.038~$\pm$~0.006) where 0.006 is the standard deviation. The red, 
  dashed line gives the value derived with the probabilistic approach 
  for the same ensemble. The difference between the two approaches is 
  about 3.7\% (and about 3.5\% for the line intensities which are not 
  shown).}
  \label{Fig:BruteForce_statistics_CO3_tau}
\end{figure}

\begin{figure}
  \resizebox{\hsize}{!}{\includegraphics{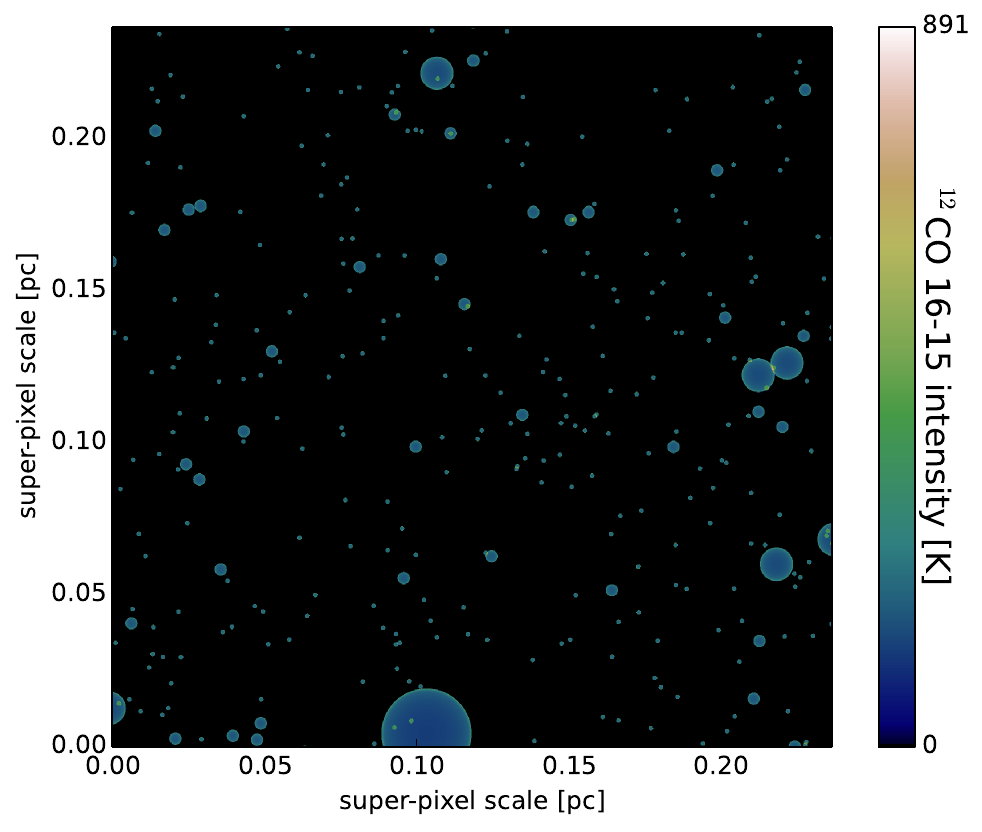}}
  \caption{A small map showing the line centre intensity of the 
  CO~16~-~15 line (colour scale). This map corresponds to the 5th 
  representation from Fig.~\ref{Fig:BruteForce_statistics_CO16_I} 
  where the large clump is sitting at the lower edge, causing a small 
  ensemble-averaged intensity and optical depth.}
  \label{Fig:BruteForceCO16_Tb}
\end{figure}

\begin{figure}
  \resizebox{\hsize}{!}{\includegraphics{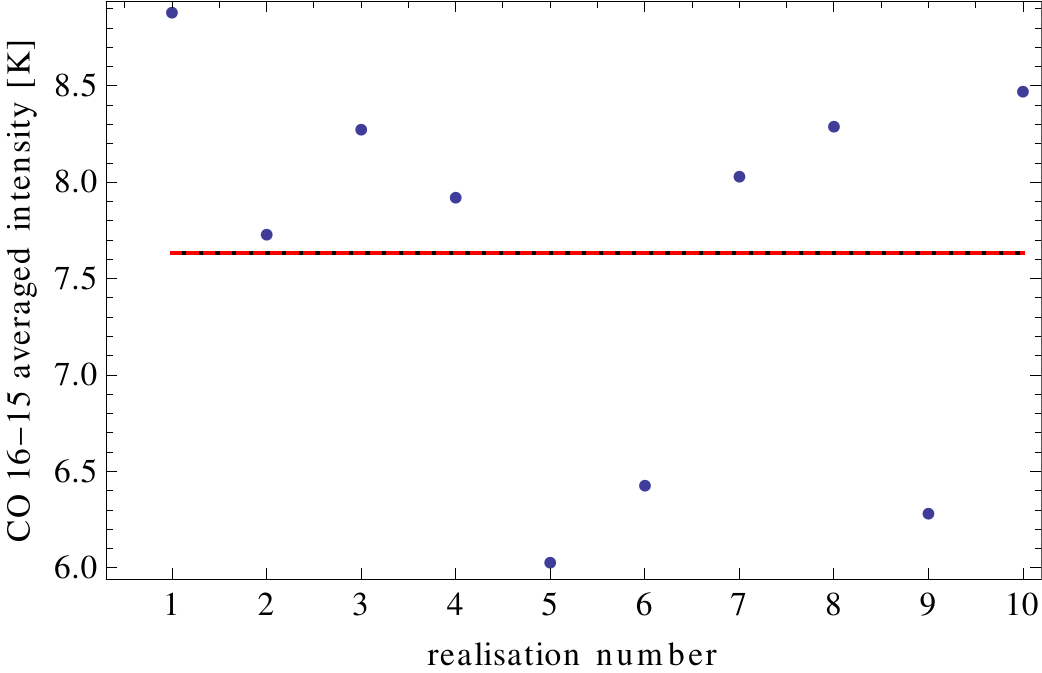}}
  \caption{Ensemble-averaged intensities of the CO~16~-~15 line (line 
  centre). The circles show the results for ten different realisations 
  of the ensemble as presented in Fig.~\ref{Fig:BruteForceCO16_Tb}. 
  The black line gives the mean value of the ten maps, namely 
  (7.6~$\pm$~1.0)~K where 1.0 is the standard deviation. The red, 
  dashed line gives the value derived by the probabilistic approach 
  for the same ensemble. Here, the two lines are lying on top of each 
  other (the difference between the two approaches is about 0.004\% 
  for the line intensities and about 0.8\% for the optical depths 
  which are not shown here).}
  \label{Fig:BruteForce_statistics_CO16_I}
\end{figure}

\begin{figure}
  \resizebox{\hsize}{!}{\includegraphics{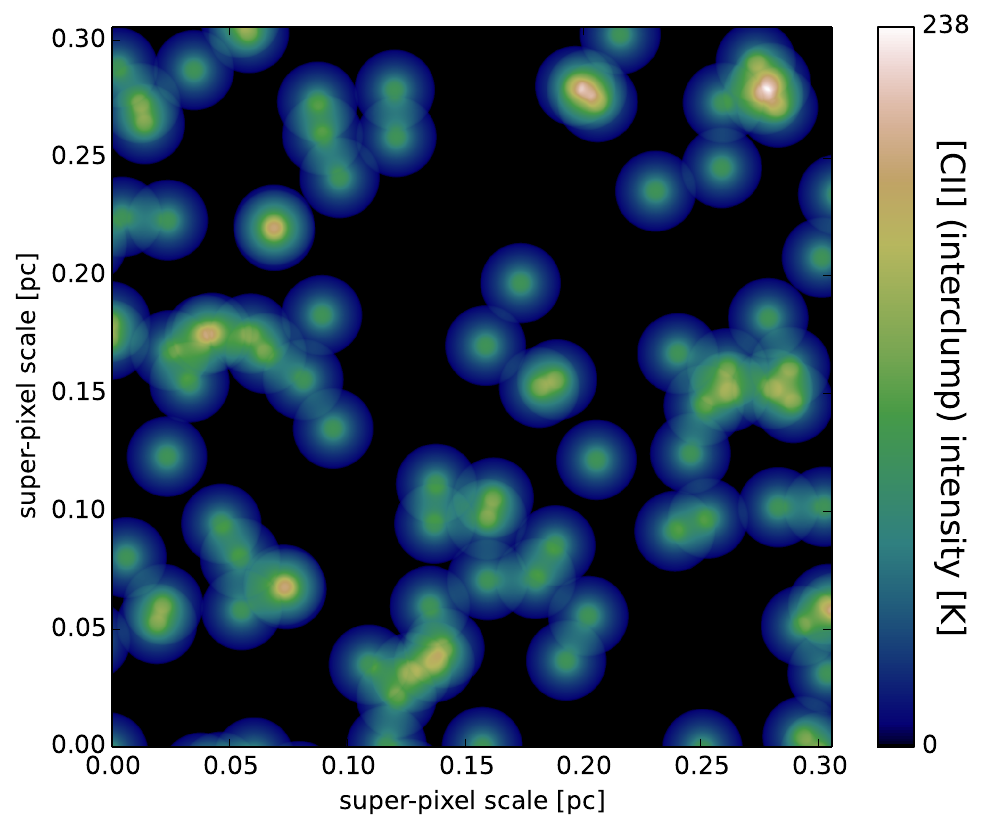}}
  \caption{A map showing a representation of the interclump medium with 
  initially 100 identical clumps (a few clumps have been cut away 
  during the removal of the edges of the map). The colour scale gives 
  the line centre intensities of the [C{\sc ii}] line.}
  \label{Fig:BruteForceCII_inter}
\end{figure}

\begin{figure}
  \resizebox{\hsize}{!}{\includegraphics{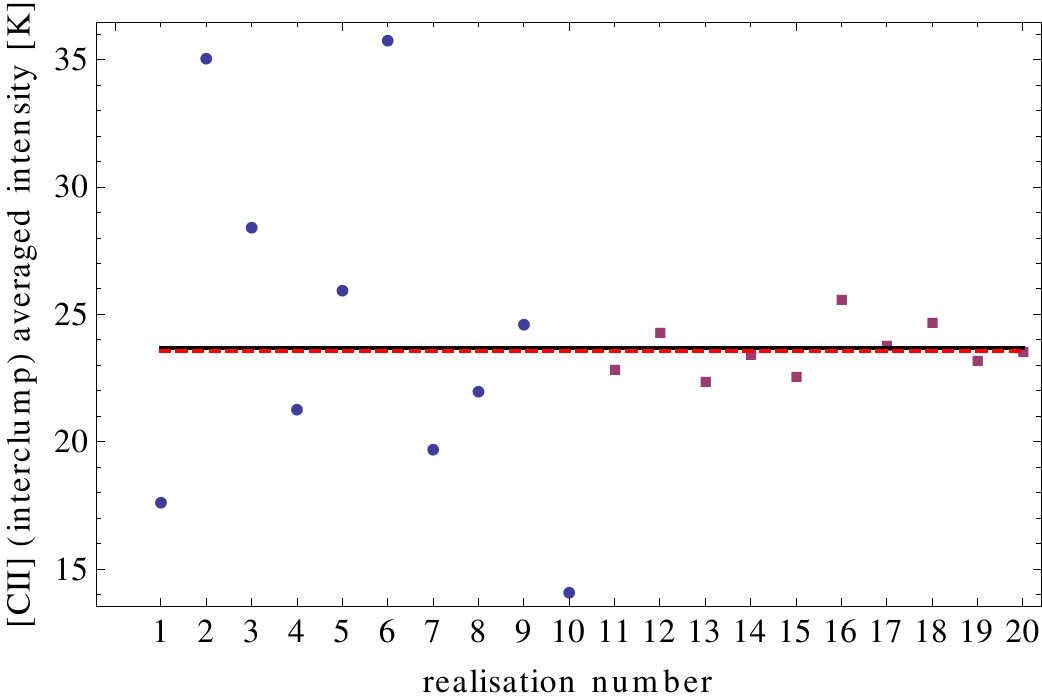}}
  \caption{Ensemble-averaged [C{\sc ii}] line centre intensities for 
  the ``interclump'' ensemble. The circles show the ensemble-averaged 
  intensities of ten different maps containing ten clumps and the 
  squares give the values for ten different maps with 100 clumps as 
  presented in Fig.~\ref{Fig:BruteForceCII_inter}. The black line 
  gives the mean value of the 20 maps, weighted by the respective 
  number of grid points. For the maps with ten clumps the average is 
  (24.4~$\pm$~7.1)~K and for the maps with 100 clumps it is
  (23.6~$\pm$~1.0)~K, where the stated error is the standard deviation.
  The red, dashed line gives the value derived by the probabilistic 
  approach (for $N_{n_M=1}=100$ clumps) for the same ensemble. The two 
  results agree within 0.4\% (intensities) and 0.8\% (optical depths, 
  not shown).}
  \label{Fig:BruteForce_statistics_CII_inter}
\end{figure}

\begin{figure}
  \resizebox{\hsize}{!}{\includegraphics{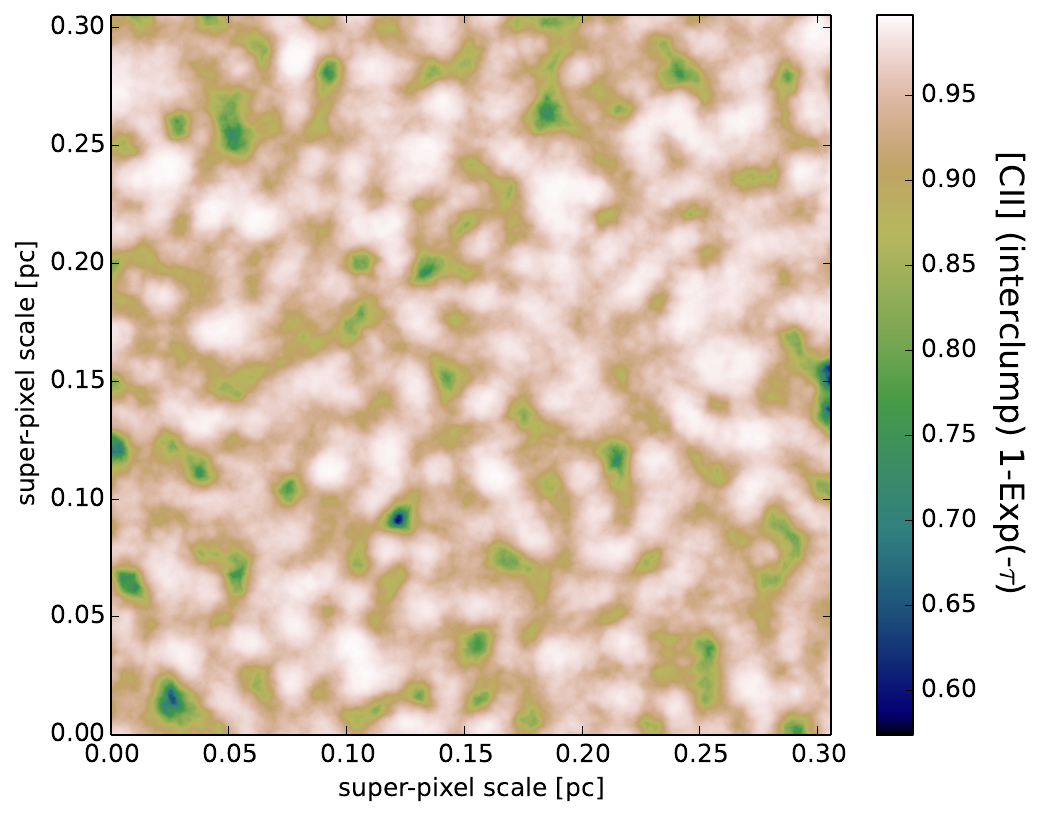}}
  \caption{\changedTwo A map showing the interclump medium of model 
  6j, with a surface density corresponding to the full line of sight
  of the model provided by 80 voxels. The colour scale shows 
  $1-{\rm Exp}(-\tau_{\rm CII})$.}
  \label{Fig:BruteForceCII_inter_80voxel}
\end{figure}

Each map has been derived on a spatial grid with a grid-spacing 
$d_{\rm grid}\ll\Delta s'$ {\changed (see Sect.~\ref{Sect:EnsembleStatistics}). 
The large maps, for instance Fig.~\ref{Fig:BruteForceCII_Tb}, contain $n_{\rm grid} = 1741\times 1741$ grid points
while the small maps (see Figs.~\ref{Fig:BruteForceCO3_tau} and \ref{Fig:BruteForceCO16_Tb}) 
contain $n_{\rm grid} = 1184\times 1184$ grid points.
For each map the ensemble-averaged quantities are calculated using}
\begin{align}
\label{Eq:BruteForceI}
\langle I \rangle_{\rm grid}&=\frac{1}{n_{\rm grid}}\sum_{k=0}^{n_{\rm grid}}\, I_k\, \\
\label{Eq:BruteForceTau}
\langle {e^{-\tau}}\rangle_{\rm grid}&=\frac{1}{n_{\rm grid}}\sum_{k=0}^{n_{\rm grid}}\, e^{-\tau_k}\, 
\end{align}
{\changed where the index $k$ denotes different gridpoints. 
The resulting $\langle I \rangle_{\rm grid}$ and $\langle {e^{-\tau}}\rangle_{\rm grid}$} 
can be compared to the result{\changed s from} the ``binomial'' approach 
(Eqs.~\ref{Eq:IAv_centre} and \ref{Eq:TauAv_centre}).

In the presented comparison we account for only one velocity bin with 
centre-velocity ${\rm v}_i={\rm v}_1={\rm v}_{\rm sys}$. This velocity bin contains all 
clumps of the ensemble and Eq.~\ref{Eq:Nji} simplifies to 
\begin{equation}
 \Delta N_{j, 1}=N_j\, .
\end{equation}
Furthermore, due to ${\rm v}_i={\rm v}_{\rm obs}$, Eqs.~\ref{Eq:Ixi} and \ref{Eq:tauxi} simplify to 
\begin{align}
\label{Eq:IAv_centre}
 I_{x_1}({\rm v_{obs}})  = &\sum_{j=1}^{n_M} k_{j,1}\, \overline{I_{j,\, {\rm line}}}\\
 \label{Eq:TauAv_centre}
 \tau_{x_1}({\rm v_{obs}})= &\sum_{j=1}^{n_M} k_{j,1}\, \overline{\tau_{j,\, {\rm line}}}\,
\end{align}
giving the line-centre intensity and optical depth for each combination of clumps. 
As we do not have to sum up contributions from different centre-velocities, the 
ensemble-averaged quantities are provided by 
Eq.~\ref{Eq:IAvi} and \ref{Eq:tauAvi},
for $i=1$. These will be compared to the results from the probabilistic approach.

{\changed For the [C{\sc ii}] line the ensemble-averaged line intensity and optical 
depth have been derived for 30 different realisations of the ensemble including 20 
small and 10 large maps. The results are summarised in Figs.~\ref{Fig:BruteForce_statistics_CII_Tb} 
and \ref{Fig:BruteForce_statistics_CII_tau} for the intensities and optical depths, 
respectively. As expected, the averaged values calculated for single realisation in 
Figs.~\ref{Fig:BruteForce_statistics_CII_Tb} and \ref{Fig:BruteForce_statistics_CII_tau}} 
show some scatter, {\changed which is} larger for the smaller maps {\changed (this effect is better visible in 
Fig.~\ref{Fig:BruteForce_statistics_CII_inter} where the size-difference between the maps is larger).
The scatter shows us how much single representations of the ensemble, which might exist in
molecular clouds, can differ from the mean value derived with the probabilistic approach.}
The black line{\changed s} give the {\changed line intensity or optical depth} averaged over the 
{\changed 30 results of the individual representations}. {\changed Within this} calculation
the maps got different weights, {\changed proportional to} their respective {\changed sizes}.
The red, dashed line{\changed s} show the {\changed ensemble-averaged quantities} for 
the same ensemble, calculated {\changed using} the {\changed probabilistic} approach. 
{\changed For [C{\sc ii}] the two approaches agree within 2.6\% for line intensities and 
optical depths}.

The reliability of the approach presented here depends on the size of 
the calculated maps and on the number of representations of the 
ensemble to be averaged over. However, a statistical difference 
remains between simulating one very large map or several smaller maps 
with the same total size. This difference can be understood from
the different [C{\sc ii}] maps: while for the large maps with (for 
example) $N_{n_M}=2$ a clump at the highest mass point can overlap 
with a clump of the same kind, this overlap is not possible for the 
small maps with $N_{n_M}=1$.
However, Fig.~\ref{Fig:BruteForce_statistics_CII_Tb} shows that this 
difference is negligible for the dense ensembles in this work, due to 
their small area filling factors. For the interclump medium, which has a higher area filling factor, the 
situation is different (see below).
 
For each CO line 10 small maps have been analysed. {\changedTwo 
Fig.~\ref{Fig:BruteForceCO3_tau} shows such a map with the optical 
depth of the CO~$3-2$ transition given on the colour scale.} As expected the CO~$3-2$ optical 
depth is highest for lines of sight intersecting with the dense cloud 
cores. The related statistical overview is shown in Fig.~\ref{Fig:BruteForce_statistics_CO3_tau}.
We find that the two approaches agree within 3.7\% (3.5\%) for the optical depths (line intensities). 
Possibly, increasing the number of samples could improve this result. 

For the CO~16-15 line we present an ``intensity map'' in Fig.~\ref{Fig:BruteForceCO16_Tb}
and the related statistical overview in Fig.~\ref{Fig:BruteForce_statistics_CO16_I}.
The map (Fig.~\ref{Fig:BruteForceCO16_Tb}) shows the 5th realisation of the ensemble from  
Fig.~\ref{Fig:BruteForce_statistics_CO16_I}. Here, the large clump is sitting at the lower 
edge of the map and is partly cut away causing the small ensemble-averaged intensity (and 
optical depth). However, this situation is part of the ``normal'' statistics. For CO~$16-15$ 
the agreement between the two approaches is excellent (which is coincidental due to the rather 
low number of sample maps), we find deviations between the two approaches of 0.004\% and 0.8\% 
for line intensities and optical depths, respectively.

In addition to the ensemble of dense clumps we test one ``typical'' 
interclump-ensemble. This ensemble contains clumps at one mass point 
($10^{-2}$~M$_\sun$), a total ensemble mass of 
$0.5\times 0.00173$~M$_\sun$ per $(0.01\, {\rm pc})^2$ projected 
surface area, and an ensemble averaged density of 
$1.91\times 10^4$~cm$^{-3}$. It has an area filling factor
{\changedTwo ($N_1 \pi (R_1)^2/(\Delta s)^2$, i.e.~not accounting for 
the fact that clumps do overlap)} of about 0.8. The clumps are lying on 
a gridpoint of the KOSMA-$\tau$ model grid, making interpolations 
unnecessary. One representation of this ensemble is shown in 
Fig.~\ref{Fig:BruteForceCII_inter} where the colour scale gives the 
[C{\sc ii}] line (centre) intensity. The map originally contained 100 
identical clump, a few clumps have been cut away during the removal of 
the edges of the map. Note that for this ensemble the clumps at mass 
point $10^{-2}$~M$_\sun$ are about a factor six larger compared to the 
previous ensemble, due to the reduced density. The [C{\sc ii}] 
intensities of different realisations of the ensemble, for maps that 
contained initially 10 or 100 clumps, are shown in
Fig.~\ref{Fig:BruteForce_statistics_CII_inter}. The comparison to the 
probabilistic approach\footnote{For the interclump medium the 
probabilistic approach yields an ensemble-averaged [C{\sc ii}] line 
centre intensity of 23.68~K for $N_{n_M=1}=10$ and of 23.57~K for 
$N_{n_M=1}=100$.} shows that the deviation between the two results 
lies below 1\% for line intensities and optical depths.

The excellent agreement between the two approaches for the 
interclump ensemble indicates that averaging over the projected 
surface (see Sect.~\ref{Sect:KOSMA}) hardly introduces any error. 
Furthermore, the interpolation between the line \emph{profiles} 
$I_{\rm line}(p)$ and $\tau_{\rm line}(p)$, as needed for the 
ensemble of dense clumps during the analysis with the direct 
approach, does not cause large deviations. All compared results are 
found to agree within the statistical scatter, which is quantified 
by the standard deviations of the scatter of the ensemble-averaged 
quantities, calculated with the direct approach.

{\changedTwo In addition to the ensembles used for the statistical
analysis in this section, we have added
Fig.~\ref{Fig:BruteForceCII_inter_80voxel}, which shows clumps based on 
the parameters of the interclump medium of model 6j (see 
Table~\ref{Tab:models}), with a number surface density corresponding to 
80 voxels along a line of sight. This figure is further discussed in Sect.~\ref{Sect:Discussion_interclump}.}

\section{\changed Numerical handling of the radiative transfer}
\label{appendix:radTrans}

A solution of the equation of radiative transfer, based on linear 
approximations of the emission and absorption coefficient, 
has been presented in Sect.~\ref{Sect:radTrans}. Here, we discuss the 
numerical handling of the integral in 
Eq.~\ref{Eq:linearApproxRadTrans} in the KOSMA-$\tau$-3D code.
The numerical integrations in this section were performed using 
Wolfram 
Mathematica\footnote{\url{http://www.wolfram.com/mathematica/}}.

Before we solve the integral in Eq.~\ref{Eq:linearApproxRadTrans} we need to distinguish three different cases: 
\begin{enumerate}
\item If no absorption takes place between two pixels, i.e.~$k_0=0$ and $k_1=0$, the equation of radiative 
transfer (Eq.~\ref{Eq:radtrans1}) 
 reduces to 
 \begin{equation}
  {\rm d}I= \epsilon\, {\rm ds} = (e_0+e_1 s)\,  {\rm ds}
 \end{equation}
 and integration between 0 and $\Delta s$ yields
  \begin{equation}
  I = e_0 \Delta s + \frac{1}{2} e_1 (\Delta s)^2 + I_{\rm bg}
 \end{equation}
 where $I_{\rm bg}$ indicates the incident background emission.
Note that this case (i.e.~an infinite source function and hence an infinite
 excitation temperature)
cannot occur in any physical source. However, in the code it is needed if there are
``holes'' in the set-up i.e.~if voxels on a line of sight are not occupied
by an ensemble (at a specific velocity) or for
artificial sources which have been constructed in a way
that voxels emit at a specific frequency without absorbing it. 
Practically, in the code, the condition $|k_0\ \Delta s|<10^{-10}$ and $k_1=0$ has been 
used for this case, avoiding errors due to numerical inaccuracies.
\item If the absorption coefficient does not change between two pixels, i.e.~$k_1=0$ but $k_0\neq 0$, 
Eq.~\ref{Eq:linearApproxRadTrans} reduces to
\begin{equation}
I={\rm e}^{-k_0\Delta s}\left[ \int_0^{\Delta s}(e_0+e_1\, s'){\rm e}^{k_0 s'}{\rm d}s' + I_{\rm{bg}}\right]\, 
\label{Eq:case2}
\end{equation}
and numerical integration yields
\begin{equation}
I={\rm e}^{-k_0\Delta s}\left[\left(\frac{e_0k_0 + e_1(k_0\Delta s - 1)}{k_0^2} \right){\rm e}^{k_0\Delta s} - 
\left(\frac{e_0k_0-e_1}{k_0^2}\right) + I_{\rm{bg}}\right]\, .
\end{equation}
Practically, in the code, the condition $k_0>10^3\ k_1\Delta s$ has been used for this case.
\item The third case is the general case which is always used if $k_1\neq 0$. Numerical integration
of Eq.~\ref{Eq:linearApproxRadTrans} yields
\begin{equation}
I=\tilde{I}+I_{\rm{bg}}\cdot{\rm e}^{-k_0 \Delta s- \frac{1}{2}k_1 (\Delta s)^2} 
\end{equation}
with
\begin{align}
\label{Eq:Itilde}
\tilde{I}= \frac{e_1}{k_1}\bigg\lbrack 1-&{\rm e}^{-k_0 \Delta s- \frac{1}{2}k_1 (\Delta s)^2} \bigg\rbrack 
- (e_0\ k_1 - e_1\ k_0)\frac{1}{k_1^{3/2}}\sqrt{\frac{\pi}{2}}\times \\\nonumber
&{\rm e}^{-\frac{(k_0+k_1\Delta s)^2}{ 2k_1}}\bigg\lbrack \operatorname{erfi}\Big(\frac{k_0}{\sqrt{2 k_1}} \Big) - \operatorname{erfi}\Big(\frac{k_0+k_1 \Delta s}{\sqrt{2 k_1}} \Big) \bigg\rbrack\, 
\end{align}
where $\operatorname{erfi}(...)$ denotes the imaginary Error Function.
$\tilde{I}$ is further discussed below.
\end{enumerate} 
Implementation, runtime and precision of the imaginary error functions in Eq.~\ref{Eq:Itilde} 
are problematic 
for large function values. In the code this is avoided by the following rearrangements and substitutions:\\
The function $\tilde{I}$, Eq.~\ref{Eq:Itilde}, can be written as
\begin{align}
\tilde{I}=& \frac{e_1}{k_1}\bigg\lbrack 1-\exp(-k_0 \Delta s- \frac{1}{2}k_1 (\Delta s)^2) \bigg\rbrack - \\\nonumber
&\frac{e_0\ k_1 - e_1\ k_0}{k_1}\sqrt{\frac{\pi}{2|k_1|}}\bigg\lbrack\exp(a^2-b^2)\tilde{E}(a)-\tilde{E}(b)\bigg\rbrack\, 
\end{align}
with
\begin{align}
&a := {\frac{k_0}{\sqrt{2k_1}}}\quad\mbox{and}\\\nonumber
&b := {\frac{k_0 + k_1\, \Delta s}{\sqrt{2k_1}}}\, .
\label{Eq:ab_kgr0}
\end{align}
The function $\tilde{E}$ has been constructed in a way that (subtractions between) large numbers are avoided.
It is different for $k_1>0$ and $k_1<0$. 
Furthermore, it can be approximated for large and small function values.
For $k_1>0$ it is given by
\begin{equation}
\tilde{E}(x) =\begin{cases}
\frac{2x}{\sqrt{\pi}}, \quad & \text{ if } x<0.01  \\
\frac{1}{\sqrt{\pi}x}, \quad & \text{ if } x>8.0 \\
\exp(-x^2)\operatorname{erfi}(x), \quad & \text{ else }
\end{cases}
\label{Eq:Ereal}
\end{equation}
for $k_1<0$ (and consequently imaginary a and b) it is given by
\begin{equation}
\label{Eq:Eimag}
\tilde{E}(x) =\begin{cases}
1-\frac{2|x|}{\sqrt{\pi}},             \quad & \text{ if } x=-i|x| \text{ with } |x|<0.01  \\
\frac{1}{\sqrt{\pi}|x|},               \quad & \text{ if } x=-i|x| \text{ with } |x|>8.0  \\
1+\frac{2|x|}{\sqrt{\pi}},             \quad & \text{ if } x=+i|x| \\
\exp(-x^2)\operatorname{erfc}(i\, x)                                  \\
\quad =\exp(|x|^2)\operatorname{erfc}(|x|),\quad & \text{ else }\, .
\end{cases}
\end{equation}
For $0.01\leq |x|\leq 8.0$ both functions, Eqs.~\ref{Eq:Ereal} and \ref{Eq:Eimag},
have been tabulated. The code interpolates linearly between the tabulated values.
Note that maser lines ($k_0+k_1\Delta s<0$, i.e.~$x=+i|x|$) are treated in 
linear approximation. Hence, the code should not be used for strong maser lines. 
For all other $x$, including weak maser lines ($x=+i|x|$ with $|x|<0.1$), the relative error made by 
interpolation or approximation of $\tilde{E}$ is less than one percent.

\section{FUV flux at the ionisation front}
\label{appendix:FUVatIF}

\citet{Jansen_1995} state that the radiation field incident on the Orion Bar 
corresponds to an enhancement over the average interstellar radiation field, 
$\chi_0$, of a factor $\approx4.4\cdot10^4$. Other authors give similar values, 
\citet{Marconi_1998} estimate a flux of $1-3\cdot10^4$ times the average interstellar 
field. \citet{Arab_2012, YoungOwl_2000, Walmsley_2000} used $1-4\, G_0$ at the IF 
with $G_0$ being the Habing field \citep{Habing_1968}\footnote{The {\changed ratio} 
between the Draine field $\chi_0$ and the Habing field $G_0$ {\changed(both 
integrated over the FUV range) is $\chi_0/G_0\approx 1.71$ \citep{Draine_1996}.}}.

Here, we have re-estimated the FUV flux at the IF, originating from $\Theta^1$~Ori~C,  
based on synthesised stellar spectra provided by 
\citet{Martins_2005}\footnote{{\changed A}vailable online: \url{http://www.mpe.mpg.de/~martins/SED.html}}.
Different spectra from their sample have been investigated, for stars having effective 
temperatures between 35000 and 39540 K which covers the spectral classes from O7V to O6V.
Different authors \citep{Stahl_2008, Pellegrini_2009, Arab_2012} obtained varying results 
for the spectral class of $\Theta^1$~Ori~C, {\changed all falling} in the range between
O6 {\changed and} O7.

The selected spectra have been integrated in the FUV range, 2066 to 911~\AA\ (6 to 13.6~eV), 
and the resulting flux at the position of the Orion Bar has been computed. For this calculation, 
the distance between $\Theta^1$~Ori~C and the Orion Bar has been assumed to be 0.223~pc, 
i.e.~equal to the projected distance (neglecting a possible offset in radial direction which 
is not precisely known, {\changed see} Sect.~\ref{Sect:geometry}). The calculated fluxes 
fall between 17 and 373~erg~s$^{-1}$~cm$^{-2}$ ($0.63\cdot10^4$~$\chi_0 - 13.8\cdot10^4$~$\chi_0$),
covering the range of values discussed above. 

{\changed The FUV flux representative} for an O6.5 star ($T_{\rm{eff}}=36826$~K, \citealt{Martins_2005})
{\changed is found to be} 38~erg~s~s$^{-1}$~cm$^{-2}$ or $1.4\cdot 10^4\,\chi_0$ at the IF.
{\changed The flux of $4.4\cdot10^4\,\chi_0$ from \citetalias{Hogerheijde_1995}} is best 
reproduced by the star from the sample with $T_{\rm{eff}}=37760$~K, which indicates the 
spectral class O6.5V \citep{Martins_2005}. {\changed The calculated FUV fluxes make strong} 
FUV absorption inside the H{\sc ii} region between star and PDR {\changed improbable} 
in agreement with the lack of dust observed in the cavity. Dust must have been blown out by the 
strong stellar winds.

\section{\changed Cylindrical models}
\label{appendix:cylindrical}

\begin{table}
\caption{\changed Overview over cylindrical models.}    
\label{Tab:cyl}   
\centering
\begin{tabular}{c c c c c}
\hline\hline
Name & parameters adopted from & $\chi_{\rm I}^2$& $\chi_{\rm off}^2$ & $\chi_{\rm tot}^2$ \\ 
\hline
   Cyl. 1 & 1z       & 602 & 120 & 722  \\ 
   Cyl. 2 & 1A       & 674 & 201 & 875 \\    
   Cyl. 3 & 2b/2e/2h & 336 & 79  & 415  \\    
   Cyl. 4 & 6b       & 419 & 66  & 485  \\    
   Cyl. 5 & 6j       & 515 & 59  & 574  \\    
   Cyl. 6 & 6k       & 648 & 74  & 722  \\    
   Cyl. 7 & 6l       & 825 & 50  & 875  \\     
   \hline
   \end{tabular}
\end{table}

\begin{figure*}
\centering   \includegraphics[width=17cm]{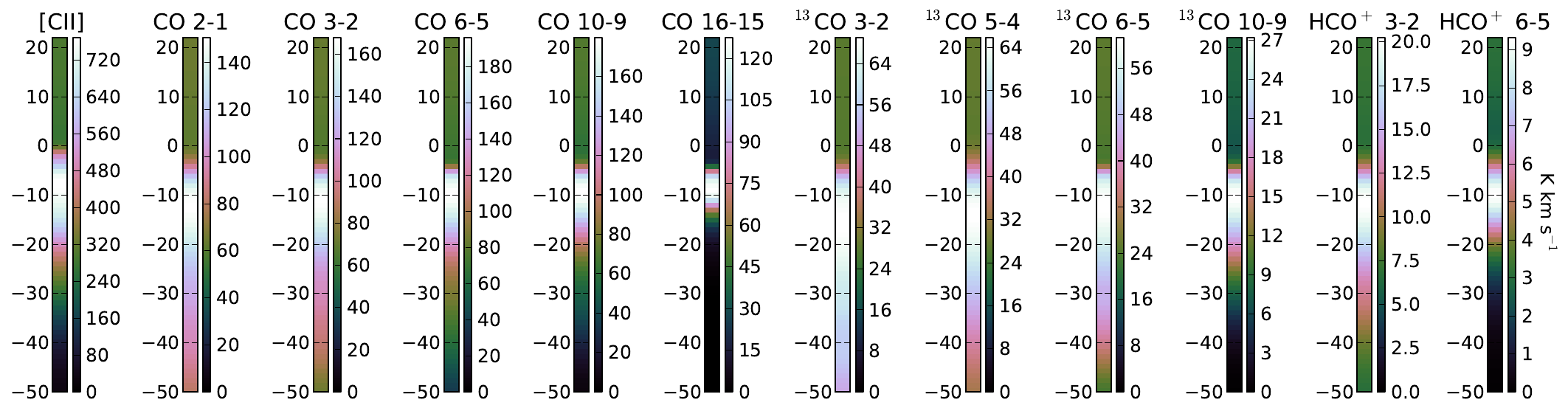}
     \caption{Simulated cuts perpendicular to the Orion Bar, based on model Cyl.~5
     (see Table~\ref{Tab:cyl}). Each colour scales gives the line integrated {\changedTwo intensity} of 
     the transitions indicated above the respective cut.
     }
     \label{Fig:cuts_cyl5}
\end{figure*}

\begin{figure}
  \resizebox{\hsize}{!}{\includegraphics{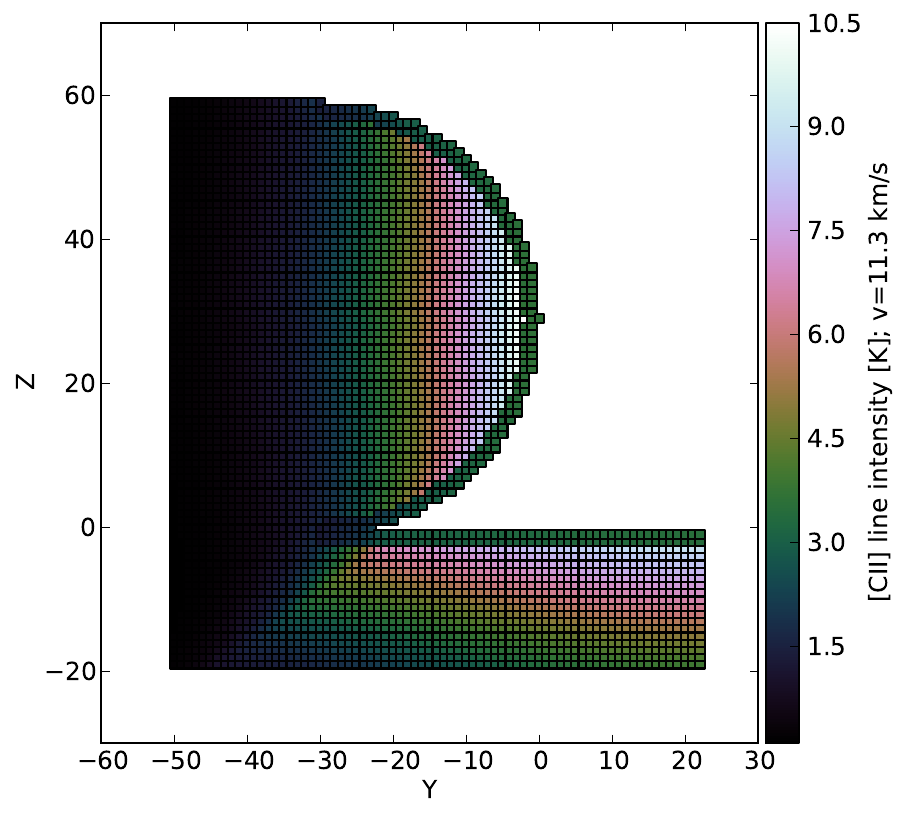}}
  \caption{\changed Cut through the cylindrical Orion Bar model Cyl.~4. For each voxel 
  the colour scale gives the [C{\sc ii}] line intensity of dense clumps and interclump medium, 
  at the line centre (at 11.3~\kms{}). The illuminating star 
  $\Theta^1$~Ori~C is located at $\lbrack 0, 22.3, 30\rbrack$.
  }
  \label{Fig:cyl_CII_intensity}
\end{figure}

\begin{figure}
  \resizebox{\hsize}{!}{\includegraphics{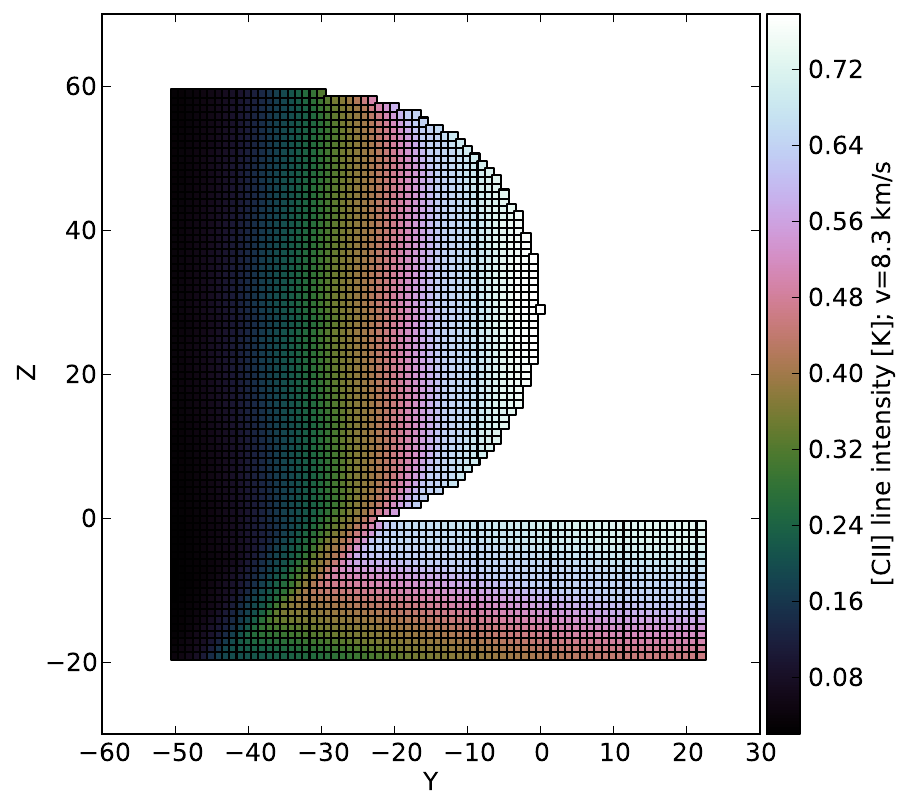}}
  \caption{Same as Fig.~\ref{Fig:cyl_CII_intensity} but plotted for 
  the outer line wing (at 8.3~\kms{}). Effectively, only the 
  interclump medium contributes at this velocity.}
  \label{Fig:cyl_CII_intensity_v8p3}
\end{figure}

As discussed in Sect.~\ref{Sect:OriModel} we have also tested models where the Orion Bar has
a cylindrical shape (see Figs.~\ref{Fig:cyl_CII_intensity} and \ref{Fig:cyl_CII_intensity_v8p3}). 
The tests of this geometry are based on the assumption 
that the sets of parameters that provide the best simulation results for the 
\citetalias{Hogerheijde_1995} geometry are also good initial guesses for the cylindrical model. 
We use a cylindrical model with a deep cavity, $d_{\rm cavity}=0.6$~pc, and hence a radius 
for the cylinder of 0.3~pc. The {\changedTwo centre} point of the cut through the cylinder is 
$\lbrack 0, -30, 30\rbrack$ and the illuminating source is located at $\lbrack 0, 22.3, 30\rbrack$ 
(see Fig.~\ref{Fig:cyl_CII_intensity}). The inclination parameter $\alpha'$ is not needed for a cylinder.
Table~\ref{Tab:cyl} gives an overview over some cylindrical models. The second column of the 
table refers to the corresponding model from Table~\ref{Tab:models} for 
the remaining parameters, i.e.~the composition of the clump ensembles, $I_{\rm UV}$ and $d_{\rm clumps}$. 

Figures~\ref{Fig:cyl_CII_intensity} and \ref{Fig:cyl_CII_intensity_v8p3} show cuts through 
the cylindrical model Cyl.~4, which is an inhomogeneous model with $d_{\rm clumps}=0.02$~pc. The 
colour scales give the [C{\sc ii}] line intensity emitted by the dense clumps and the
interclump medium of the respective voxel at a specific velocity, Fig.~\ref{Fig:cyl_CII_intensity} 
shows the line-centre intensity (at 11.3~\kms{}) and Fig.~\ref{Fig:cyl_CII_intensity_v8p3} 
the emission in the line wing (at 8.3~\kms{}). The Fig.~\ref{Fig:cyl_CII_intensity} 
illustrates that at the line 
centre velocity the [C{\sc ii}] emission is dominated by the dense clumps, that
only start two voxels below the cloud surface. Due to the higher velocity
dispersion of the interclump medium (see Sect.~\ref{Sect:line_profiles}), 
the  8.3~\kms{} channel in Fig.~\ref{Fig:cyl_CII_intensity_v8p3} is dominated 
by the [C{\sc ii}] emission
from the interclump medium, providing the  highest line intensities in the 
voxels that are closest to the illuminating 
source. The observable stratification pattern follows from the radiative 
transfer and the beam convolution of these pictures.

In general, for cylindrical models the $y$-offsets where the line
integrated intensity peaks for the different transitions are shifted 
deeper into the cloud (into the negative $y$-direction) compared to 
the corresponding cavity-wall models. 
For all models listed in Table~\ref{Tab:cyl} the [C{\sc ii}] line
integrated intensity peak (i.e.~the reference positions for the
$y$-offsets) is shifted by one to three pixels deeper into the PDR 
compared to the cavity-wall model with the same parameters. As an 
example  Fig.~\ref{Fig:cuts_cyl5} shows the simulated cuts of model 
Cyl.~5. Here the [C{\sc ii}] peak lies at $y=-0.08$~pc, while the peak 
appears at $y=-0.06$~pc in model 6j. Hence, to provide a stratification 
pattern, the $y$-offsets of all other transitions need to be shifted 
even deeper into the cloud in the cylindrical models. This is hardly
observed. The line intensity is dominated by the column density
along the line of sight given by the cylindrical geometry, less
by changing composition or excitation conditions.
%
For the cylinder the lines of sight through the compound close 
to $y=0$ are shorter than the lines of sight through the cloud
material in the cavity-wall {\changedTwo set-up} (assuming small $\alpha'$). 
Therefore, the peaks are moved away from the edge of the 
cloud in the cylindrical model, to positions where the lines 
of sight through the compound and hence the column densities are 
larger. This leads to the higher $y$-offsets.

Depending on the specific model set-up, the $\chi_{\rm off}^2$ can 
decrease or increase when we switch from the cavity-wall models to the 
cylindrical geometry. However, this comparison is ambitious, 
as the $\chi_{\rm off}^2$ of the \citetalias{Hogerheijde_1995}-models 
also depend on the inclination angle $\alpha'$. Similar to the 
\citetalias{Hogerheijde_1995}-models we find that the fit of the 
stratification pattern is better when an inhomogeneous model 
(i.e.~models Cyl.~4 to Cyl.~7) is used, compared to the homogeneous 
set-ups (models Cyl.~1 to Cyl.~3). Non of the homogeneous models can 
reproduce any stratification for the CO~$10-9$ transition. Models 
Cyl.~5, 6 and 8 do provide peak positions $\Delta y_i<0$ relative to 
the [C{\sc ii}] peak (see Sect.~\ref{Sect:ModelAssessment}) for all 
transitions. However, the $\Delta y_i$ scatter around the observed 
values, complicating the fine-tuning of the stratification pattern. Non 
of the tested models reaches the $\chi_{\rm off}^2$ of our 
best \citetalias{Hogerheijde_1995}-models.

With the parameters from Table~\ref{Tab:models} the cylindrical 
geometry provides always systematically lower line integrated 
intensities for all transitions than the cavity-wall model, 
resulting in a deteriorated fit (in terms of the 
$\chi_{\rm I}^2$). This reduction results from a combination of 
two effects: (a) shorter lines of sight and hence reduced column 
densities at the front of the cylinder, (b) less FUV flux 
available for the excitation when the peaks are shifted 
deeper into the cloud (see above). This already shows the 
fundamental problem of cylindrical models. 

Consequently, we find that the fits based on the tested 
cylindrical model do not reach the quality (in terms of the 
$\chi_{\rm tot}^2$) of the \citetalias{Hogerheijde_1995}-
models. Of course these results could be improved if 
step-by-step fine-tuning of the parameters is performed, 
however, fine-tuning of the stratification patter is expected to 
be difficult due to the scattering of the $\Delta y_i$ of the 
different transitions. An improvement of the fit of the line 
integrated intensities might be possible if the mass per voxel 
in the dense clumps is drastically increased, leading to larger 
molecular column densities close to $y=0$. A cylinder with less 
curvature might also provide better results.

\end{document}